\documentclass[11pt]{article}

\usepackage{natbib}
\usepackage[usenames]{color}

\usepackage{graphicx}

\newcommand\sech{{\rm \;sech}}
\newcommand\D{\partial}
\newcommand\gt{{\rm \;>\;}}
\newcommand\lt{{\rm \;<\;}}
\newcommand\lesim{\lower.5ex\hbox{$\; \buildrel < \over \sim \;$}}
\newcommand\gesim{\lower.5ex\hbox{$\; \buildrel > \over \sim \;$}}
\def\simgt{\lower.5ex\hbox{$\; \buildrel > \over \sim \;$}}
\def\simlt{\lower.5ex\hbox{$\; \buildrel < \over \sim \;$}}
\def\gsim{\lower.5ex\hbox{$\; \build/rel > \over \sim \;$}}
\def\lsim{\lower.5ex\hbox{$\; \buildrel < \over \sim \;$}}
\newcommand\HI{H{\small I}}
\newcommand\HII{H{\small II}}
\newcommand\kms{km s$^{-1}$}
\newcommand\Ha{H$\alpha$}

\newcommand\msun{M$_{\odot }$}

\newcommand{\ie}{{\it i.e.}}
\newcommand{\eg}{{\it e.g.}}
\newcommand{\etal}{{et al.}}
\newcommand\aj{{\it Astron.~J.\ }}%
\newcommand\araa{{\it Annu.~Rev.~Astron.~Astrophys.\ }}%
\newcommand\annap{{\it Ann.~Ap.\ }}%
\newcommand\apj{{\it Ap.~J.\ }}%
\newcommand\apjs{{\it Ap.~J.~Suppl.\ }}%
\newcommand\assl{{\it Astrophys.~Space~Sci.~Lib.\ }}%
\newcommand\aap{{\it Astron.~Astrophys.\ }}%
\newcommand\aapr{{\it Astroph.~Astrophys.~Rev.\ }}%
\newcommand\aaps{{\it Astron.~Astrophys.~Suppl.\ }}%
\newcommand\aspcs{{\it Astron.~Soc.~Pac.~Conf.~Ser.\ }}%
\newcommand\mnras{{\it MNRAS\ }}%
\newcommand\na{{\it New~Astron.\ }}%
\newcommand\pasp{{\it PASP\ }}%
\newcommand\ssr{{\it Space~Sci.~Rev.\ }}%
\newcommand\nat{{\it Nature\ }}%
\newcommand\bain{{\it Bull.~Astron.~Inst.~Netherlands\ }}%
\begin{document}

\markboth{van der Kruit \&\ Freeman}{Galaxy disks}

\begin{center}
{\Huge {\bf Galaxy disks}}
\vspace{1cm}

{\Large {\bf P.C. van der Kruit}}\\
{Kapteyn Astronomical Institute, University of Groningen,\\
P.O. Box 800, 9700 AV Groningen, the Netherlands
(vdkruit@astro.rug.nl)}

\vspace{0.5cm} 

{\Large {\bf K.C. Freeman}}\\
{Research School of Astronomy and Astrophysics, Australian National
University,\\
Mount Stromlo Observatory, Cotter Road, ACT 2611, Australia
(kcf@mso.anu.edu.au)}
\end{center}
\vspace{1cm}


\vspace{0.5cm}

\begin{abstract}
The disks of disk galaxies contain a substantial fraction of 
their baryonic matter and angular momentum, and much of the evolutionary activity 
in these galaxies, such as the formation of stars, spiral arms, bars and rings 
and the various forms of secular evolution, takes place in their disks.  The 
formation and evolution of galactic disks is therefore particularly important 
for understanding how galaxies form and evolve, and the cause of the variety 
in which they appear to us.   Ongoing large surveys, made possible by new 
instrumentation at wavelengths from the ultraviolet (GALEX), via optical (HST 
and large groundbased telescopes) and infrared (Spitzer) to the radio are 
providing much new information about disk galaxies over a wide range of redshift.  
Although progress has been made, the dynamics and structure of stellar disks, 
including their truncations, are still not well understood. We do now have plausible 
estimates of disk mass-to-light ratios, and estimates of Toomre's $Q$ parameter 
show that they are just locally stable.  Disks are mostly very flat and sometimes 
very thin, and have a range in surface brightness from canonical disks with a 
central surface brightness of about 21.5 $B$-mag arcsec$^{-2}$ down to very 
low surface brightnesses. It appears that galaxy disks are not maximal, except 
possibly in the largest systems. Their \HI\ layers display warps whenever \HI\ 
can be detected beyond the stellar disk, with low-level star formation going on 
out to large radii. Stellar disks display abundance gradients which flatten at
 larger radii and sometimes even reverse. The existence of a well-defined baryonic 
(stellar + \HI) Tully-Fisher relation hints at an approximately uniform baryonic 
to dark matter ratio. Thick disks are common in disk galaxies and their 
existence appears unrelated to the presence of a bulge component; they are old, 
but their formation is not yet understood.  Disk formation was already advanced at 
redshifts of $\sim 2$, but at that epoch disks were not yet quiescent and in 
full rotational equilibrium. Downsizing (the gradual reduction with time in 
the mass of the most actively star-forming galaxies) is now well-established. 
The formation and history of star formation in S0s is still not fully understood.

The final version of this chapter as it will appear in Annual Reviews of 
Astronomy \&\ Astrophysics, vol. 49 (2011), will have fewer figures and a somewhat shortened 
text.
\end{abstract}

\newpage 

\thispagestyle{empty}
{\small
\tableofcontents
}

\newpage

\section{INTRODUCTION AND OVERVIEW}

\subsection{Historical Background}\label{sec:history}
\label{sect:history}

Disks are the most prominent parts of late-type spiral galaxies. The disk of
our own Milky Way Galaxy stretches as a magnificent band of light from horizon 
to horizon, particularly from a dark site at southern latitudes, as in the
Astronomy Picture of the Day for January 27, 2009 \citep{pac09}. Its
appearance with the Galactic Center high in the sky is reminiscent of the
beautiful edge-on spiral NGC 891 \citep{hs99}, which can be regarded as a
close twin to our Galaxy \citep{vdk84}. What might in hindsight be called the
first study of the distribution of stars in a galaxy disk is William
Herschel's famous cross-cut of the `Sidereal System' based on his `star
gauges' \citep[{\it ``On the Construction of the Heavens''};][]{her85}, from 
which he concluded that the distribution of stars in space is in the form of 
a flattened system with the sun near its center. His counts, on which he based 
this `section' of the system, were performed along a great circle on the sky almost
perpendicular  to the Galactic equator, crossing it at longitudes 45$^\circ$
and 225$^\circ$ and missing the poles by 5$^\circ $. Comparison to modern star 
counts shows that Herschel counted stars consistently down to magnitude
$V\sim15$  \citep{vdk86}. It was the first description of the flattened nature 
of galactic
disks.  

The next major step in the study of the distribution of stars in the Milky Way
was that of Jacobus Kapteyn \citep{kvr20, kap22}. In spite of his deduction that
interstellar extinction must have the effect of reddening of starlight with
increasing distance, \citet{kap09a, kap09b, kap14} was unable to establish the
existence of interstellar absorption in a convincing way and was led to ignore
it. As a result he ended up with a model for what we now know to be the
Galactic disk that erroneously had the Sun located near its center. The
proceedings  {\it `The Legacy of J.C. Kapteyn'} \citep{vdkvb00} has a number
of interesting studies of Kapteyn and astronomy in his time.  

Even before galaxies were established to be `Island universes', spiral
structure was discovered in 1845 by William Parsons, the Third Earl of
Rosse. His famous drawing of M51 appears in many textbooks, popular literature
and books on the history of astronomy. The importance of the disk and the
development of spiral structure were the basis for the classification scheme
that John Reynolds and Edwin Hubble developed; the background and  early
development of galaxy classification was described by \citet{san05}. Eventually 
the concept of Stellar Populations, first proposed by \citet{baa44}, led to the 
famous Vatican Symposium of 1957 \citep{ocon58}, where a consistent picture was 
defined to interpret the  presence of disk and halo populations in the context 
of the structure and formation of galaxies. 

{ Not long after that, the collapse picture of \citet{els62} provided 
a working picture for the formation and evolution of the Galaxy and by implication 
of galaxies in general. It was modified by \citet{sz78} to include extended evolution, 
whereby the outer globular clusters originated and underwent chemical evolution in 
separate fragments that fell into the Galaxy after the collapse of the central halo 
had been completed. The basic discrete two-component structure of the edge-on galaxy 
NGC 7814 led \citet{vdks82b} to deduce that there are two discrete epochs of star 
formation, one before and the other after virialization of the spheroid and the 
formation of the disk. 

Two related important developments in understanding the properties of galaxies and 
their formation were the discovery of dark matter halos, and the appreciation of the
role of hierarchical assembly of galaxies. The concepts of hierarchical assembly
were already around in the early 1970s, and became widely accepted at a landmark 
symposium on the {\it Evolution of Galaxies and Stellar Populations} at Yale 
University in 1977 \citep{tl77}. The discovery of dark matter halos (see below) 
led to the two-stage galaxy formation model of \citet{wr78}, in which 
hierarchical clustering of the dark matter took place under the influence of gravity, 
followed by collapse and cooling of the gas in the resulting potential wells. The 
Hubble Space Telescope made possible the high resolution imaging of galaxies at high 
redshift, and showed directly that merging and hierarchical assembly are significant 
in the formation of massive galaxies.  

The significance of internal secular evolution for the evolution of disks has become
clear in recent years. The presence of oval distortions, bars and spiral structure 
can have a profound effect on the changing structure of disks, as has been 
extensively reviewed by \citet{kk04} and \citet{kor07}.}

The rotation of galaxies was discovered early in the twentieth century. 
For a historical introduction, see \citet{vdka78}, \citet[][Ch 10]{gkvdk90}
and \citet{sf01}, documenting the first derivation of rotation velocities as a
function of radius using optical absorption lines \citep{pea14}, emission
lines \citep{bab39} and \HI\ \citep{vdhrvw57,arg65}, all in the Andromeda Nebula
M31. The subject developed into the extensive mapping of the velocity fields of
disk galaxies in the optical and \HI, which eventually led to the discovery of
flat rotation curves and the existence of dark matter  \citep[see \eg\ reviews
by][]{vdka78, fg79, rob08}. 

{ Quantitative surface photometry of disk galaxies to study their 
structure and luminosity distributions began with the work of \citet{rey13} 
for the bulge of M31: for a review, see \citet{gkvdk90}, Ch. 5. Photometry of 
the much fainter disks came later, and revealed the exponential nature of the 
radial surface brightness distributions.} This was first described 
in a Harvard thesis, using observations of M33 \citep{pat40}: the data appear 
as fig.~10 in the review by  \citet{dvauc59a}.  He had undertaken in the late 
fifties a systematic survey of the light distributions in nearby spirals, 
particularly in M31 and M33 \citep{dvauc58, dvauc59b} and established the 
universal `exponential disk' description of the radial light distribution in 
galactic disks. At about the same time, \citet{holm58} completed a survey of 
the diameters of 300 galaxies from micro-photometer tracings of photographic 
plates in two colors. This heroic effort was the culmination of work started 
much earlier \citep{holm37}.

The exponential nature of the radial surface brightness distributions of disks
was discussed in detail by \citet{kcf70}, who noted that many of the larger
spirals had a remarkably small range in the extrapolated central (face-on)
surface brightness around  21.65 $B$-mag arcsec$^{-2}$. This result still
holds for classical spiral galaxies.  The exponential surface brightness
distribution of starlight in disks was complemented by the observations of the
vertical light distribution in edge-on spirals. The distribution could be
approximated very well by an isothermal sheet \citep[][but for practical
purposes an exponential can be used as well]{cam50} with a scaleheight that
- surprisingly - is to an excellent approximation independent of galactocentric
radius  \citep{vdks81a,vdks81b, vdks82a}. 

\citet{kcf70} also noted that for a self-gravitating exponential disk 
{ the expected rotation curve} peaks at 2.2 scalelengths and then declines. A decline at
2.2 scalelengths was however not observed in the rotation data for NGC 300 
and M33 at the time. { This 1970 paper appears to be the first 
indication from rotation curve analysis that the rotation curve is not  
determined by the mass distribution in the disk alone, but requires a
contribution to the amplitude of the rotation curve from an extended 
distribution of invisible matter. Subsequent observations of rotation curves 
eventually led to the concept of dark halos in individual galaxies 
\citep[{\eg\ }][]{fg79,rob08}.}

An important concept in the analysis of rotation curves is that of
`maximum disk', introduced by \citet{cf85} and \citet{vabbs85}. In this
concept, because the $M/L$ ratio of the disk is unknown, the contribution of
the disk mass to the rotation curve is taken to be as large as permitted by
the observed rotation curve. This means in practice (see below) that the
amplitude of the rotation curve from the disk itself is about 85\%\ of the
observed amplitude \citep{sac97}. In principle an independent measurement of
the disk mass distribution can be obtained from hydrostatic considerations,
comparing the thickness and velocity dispersion of the stars,  as was
pioneered for the Galaxy by \citet{kap22} and \citet{oort32}, or the \HI\ gas
\citep{vdk81}. { \citet{smg02} have reviewed Modified Newtonian Dynamics as an 
alternative to dark matter.}

Within disks, the star formation history was studied first in our Galaxy in the
solar neighborhood. The stellar initial mass function IMF { (the 
statistical distribution of stellar massses during star formation)}
was derived first by \citet{sal59}.  \citet{ms59} defined the `Schmidt-law' 
for the rate of star formation as a function of the density of the ISM, { in 
which the rate of star formation is proportional to the square of the local gas 
density}; see reviews by \citet{krou02a} and \citet{ken98} for subsequent refinements. 
Studies of the chemical evolution in the local disk identified the `G-dwarf problem' 
in the `Simple Model' of chemical evolution \citep{ms63}. { In this simple model 
the chemical evolution is followed in a galactic disk starting as pure gas 
with zero metallicity and without subsequent inflow or outflow; then the 
result is a much higher fraction of low-metallicity, long-lived stars as G-dwarfs
than is observed in the solar neighborhood. This can be rectified by extensions
of the model;} see e.g. the review by \citet{bt80}.  The basic models for chemical
evolution were able to represent the radial gradients in metal abundance in
the gas of disk galaxies \citep{ls73} in terms of the extent to which star
formation and chemical enrichment has proceeded  \citep[\eg][for
M81]{gs87}. The mean metal abundance of stars that formed over the lifetime of
a disk approaches that of the abundance of the gas at the time of disk
formation plus an `effective yield' (the net production of heavy elements,
modified by effects of zero-metal inflow or enriched gas outflow). In this
`simple model with bells and whistles' \citep{mou84} it follows that, while
gas consumption is still proceeding, the abundance gradients in the stars will
be in principle shallower than that in the gas. 

{ Reviews of Stellar Populations include those by \citet{king71}, \citet{san86}, \citet{bah86}, \citet{kcf87} and \citet{gwk89}. An IAU Symposium 
in  1994 on the subject of Stellar Populations \citep{vdkg95} includes a historical 
session. For a recent review on the structure and evolution of the Galaxy, see \citet{fbh02}.}

The integral properties of galaxies and the systematics of their 
distribution have been used as a tool towards understanding galaxy formation and 
the origin of the variety among them. Chief among these relations are those between 
the morphological type and properties of their stellar and gas content, such as
\HI\ content and integrated color \citep{rh94}.  These latter properties were
convincingly interpreted as a measure of the process of depletion of the
interstellar gas in star formation and the rate of current star formation
relative to that averaged over a galaxy's lifetime \citep{ss72,ssb73,lt78},
which then correlate with galaxy type.  The Tully-Fisher relation \citep{tf77}
provides a tight correlation of rotation velocity and integrated luminosity,
although it still is not clear why it is so tight when the rotation velocity
is determined not only by the mass in the stars that provide the luminosity 
but also by the dark matter halo.  

We should mention here the discovery of low surface brightness galaxies, 
which was anticipated by the work of \citet{dis76}. \citet{dp83} showed that the 
observed range in central surface brightness of galaxy disks (and also of 
elliptical galaxies) is severely restricted by the necessity for them to stand 
out against the background sky; the exponential nature of disks naturally 
restricted samples to a small range in central surface brightness comparable to 
the value first noted by \citet{kcf70}. This selection effect had been described 
earlier in qualitative and more general terms by \citet{arp65}. Many low surface 
brightness galaxies are known today, although it appears that the bright limit 
of the surface brightnesses seen by \citet{kcf70} is not an effect of observational 
selection \citep{as79,bf93}

\subsection{Setting the Scene}

This brief description of the historical development of our subject already
indicates that a comprehensive treatment of all aspects of galaxy disks 
is beyond the scope of
a single chapter in Annual Reviews. Topics that we will not review in detail
include radio continuum studies and magnetic fields
\citep{vdka76,con92,beck08}, AGNs and black holes in the centers of galaxies
\citep{kr95,ff05,petal07}{ , spiral structure \citep{toom77}, bars
\citep{sell10b} and secular evolution \citep{kk04}.}
Also we will not review issues related to physical
or chemical processes in the ISM. We refer the reader to the proceedings of
some recent symposia that concentrate on disks in galaxies, including {\it
  `The dynamics, structure and history of galaxies'} \citep{dcj02},  {\it
  `Island Universes: Structure and evolution of disk galaxies'}
\citep{dejong07}, {\it `Formation and Evolution of Galaxy  Disks'}
\citep{fc08}, {\it `Unveiling the mass: extracting and interpreting galaxy
  masses'} \citep{cdj09}\footnote{For this symposium on the occasion of Vera
  Rubin's 80th birthday, there will be no printed proceedings; electronic
  versions of presentations or posters are available through the conference
  website.} and {\it `Galaxies and their Masks'} \citep{bef10}. 

Despite the correlations between overall properties, there are galaxies with
very similar properties but very different morphologies. M33 and the Large
Magellanic Cloud provide an example. Both galaxies have very similar central
surface brightness ($\sim21.2$ $B$-mag arcsec$^{-2}$), scalelength ($\sim$ 1.6
kpc), integrated magnitude ($\sim -18.5$ in $B$), $(B-V)$ color ($\sim$ 0.51),
IRAS luminosity ($\sim1.0\times10^{8}$ L$_\odot$), \HI\ mass
($\sim$9.5$\times10^{8}$ M$_\odot$) and rotation velocity ($\sim 90-100$ \kms)
\citep[see][Ch. 10]{gkvdk90}.  Our point is that these two systems differ
significantly only in morphological classification and nothing else. The
detailed structure of a galaxy, its morphology and spiral structure, may be
determined by external properties such as environment or may even be
transient, so that during the lifetime of the systems there might have been
periods when M33 looked very much like the LMC and {\it vice versa}.  

In the final section, we will discuss the origin of S0 galaxies . Originally 
introduced by Hubble as a transition class between elliptical and spirals, 
they were believed to be systems that had quickly used all of their remaining 
gas. The Sa systems had sufficient gas left after completion of disk formation  
to support star formation at a low level up to the present, while later types 
had more gas left and were able to form stars more vigorously up till now 
\citep{sfs70}. Alternative theories were suggested, involving the stripping of 
gas from existing spirals by collisions \citep{sb51} or intergalactic gas \citep{gg72}. 
Assuming that it is unlikely that we are living at the time just when within 
a few Gyr all spirals will `run out of gas', \citet{ltc80} argued that gas must 
be replenished in normal spirals but not in S0's.

\section{SURVEYS}

Surveys provide the basis of much of the observational studies of disk galaxies. In 
the past, major surveys were very time consuming. For example, the {\it Hubble Atlas of 
Galaxies} \citep{san61} which provided the basic source list for much of the past 
work on nearby galaxies,  was the culmination of decades of photography of galaxies 
by Hubble, Sandage and others in order to survey the variety of morphologies among 
galaxies. For many years,  the \citet{hms56} survey was a main source of galaxy 
redshifts and magnitudes; it was the result of 20 years of observations at Mount 
Wilson, Palomar and Lick, and was only surpassed decades after its publication.   
The advent of dedicated, automated survey telescopes, multi-object spectrographs, 
and high-resolution imaging and spectroscopic space facilities has transformed our 
ability to make surveys of galaxies.   In  this section, we will give a brief overview 
of surveys, currently or recently undertaken, that are relevant to studies of  disks 
in galaxies as discussed in later sections of this review.  

Kinematic surveys aimed at the dynamics of (stellar) disks (see \S~3) using integral 
field spectrographs include DiskMass \citep{bvsawm10} (146 nearly face-on galaxies for 
which \Ha\ velocity fields have already been measured, and a subset of 46 galaxies with 
stellar velocities and velocity dispersions) and PINGS\footnote{{\it PPAK IFS Nearby 
Galaxies Survey}; www.ast.cam.ac.uk/research/pings/html/.} \citep{pings10}, which will 
provide 2-dimensional spectroscopy in 17 nearby galaxies. For these surveys, the data 
are supplemented by extensive observations at other wavelengths. 

Surveys specifically designed to gather detailed information on the properties
of disks in galaxies usually involve samples of nearby galaxies that are not
statistically complete but are designed to cover the range of morphological
types.  \HI\ surveys of individual galaxies are often complemented by optical or near-IR 
surface photometry to aid the analysis of their rotation curves (see \S~4).  A first 
such survey of spiral galaxies, combining imaging (three-color
photographic surface photometry) at optical wavelengths and mapping of
distributions and kinematics of \HI, was made by \citet{wvdka86}. This {\it 
Palomar-Westerbork Survey of northern spiral galaxies} included only 16
galaxies, but required 64 observing periods of 12 hours with the Westerbork
Synthesis Radio Telescope (WSRT) and 42 dark nights at the Palomar 48-inch Schmidt. 
This was extended substantially in the WHISP survey\footnote{{\it Westerbork observations 
of neutral Hydrogen in Irregular and SPiral 
galaxies}; www.astro.rug.nl/$\sim$whisp.} \citep{vdh02,nvdhssva05} of a sample of a
few hundred galaxies.  THINGS\footnote{{\it The \HI\ Nearby Galaxy 
Survey}; www.mpia.de/THINGS/.} \citep{things08} 
is the most detailed recent uniform set of
high-resolution and high-sensitivity data on 34 nearby disk galaxies available
at this time; data were taken with the Very Large
Array VLA. A special section, devoted to THINGS, appeared in the December 2008 issue of 
the {\it Astronomical Journal}.  Another major survey of nearby
galaxies is SINGS \citep[{\it Spitzer Infrared Nearby Galaxies Survey};
sings.stsci.edu][]{sings03}. This is a comprehensive imaging and spectroscopic study of 
75 nearby galaxies in the infrared. 

Other surveys provide large samples of galaxy data of various kinds, in different 
wavelength regions. Images of galaxies in two UV bands from the {\it Galaxy Evolution 
Explorer} GALEX \citep{galex05} survey are particularly useful for estimating the recent 
star formation history of galaxies. In the optical B-band, the {\it Millennium Galaxy 
Catalogue}\footnote{see www.eso.org/$\sim$jliske/mgc.} comes from a $37.5$ deg$^2$ medium-deep 
imaging survey of galaxies in the range $13 < B < 24$, connecting the local and distant 
universe.  The 6dF and 2DF Galaxy Redshift Surveys\footnote{www.aao.gov.au/local/www/6df  
and msowww.anu.edu.au/2dFGRS.} and the SDSS\footnote{{\it Sloan Digital Sky Survey}; 
www.sdss.org.} \citep{sdss00} provide vast samples of optical galaxy redshifts and 
spectroscopic properties related to their star formation history (see \S~5).    
The 2MASS\footnote{{\it Two Micron All Sky Survey}; www.ipac.caltech.edu/2mass.} 
\citep{2mass06} 
gives integrated near-IR photometry for a very large sample of galaxies, and also 
relatively shallow near-IR images for the brighter galaxies.  High-resolution deep 
imaging in the near and mid-infrared over a wide redshift range is provided by the 
{\it Spitzer Space Telescope} mission : see \citet{shw08} for a recent summary of 
extragalactic studies. Large \HI\ surveys are of interest for studies of the \HI\ mass 
function in the universe, and also for scaling laws (see \S~6). For example, the HIPASS 
survey\footnote{{\it \HI\ Parkes All-Sky Survey}; www.atnf.csiro.au/research/multibeam/release} 
gives integrated \HI\ data for 
galaxies south of declination $+25^\circ$ out to velocities of 12,700 \kms.  

Two major surveys are using HST to study resolved stellar populations in nearby galaxies. 
ANGST\footnote{{\it ACS Nearby Galaxy Survey Treasury}; www.nearbygalaxies.org.}
\citep{detal09} 
establishes a legacy of uniform multi-color photometry of resolved stars  for a 
volume-limited sample of nearby galaxies. GHOSTS\footnote{{\it Galaxy Halos,  Outer disks, 
Substructure, Thick disks and Star clusters}; 
www-int.stsci.edu/$\sim$djrs/ghosts} \citep{djetal07a} is imaging several edge-on galaxies 
with a range in masses to study their stellar populations.  These population studies 
are important for understanding the star formation history in galaxies (see \S~5 and 
\S~7).  For more nearby population studies, SEGUE\footnote{{\it Sloan Extension for Galactic 
Understanding and Exploration}; www.sdss.org/segue/.} and RAVE\footnote{{\it  RAdial Velocity 
Experiment}; www.rave-survey.aip.de/rave/.} focus on kinematic and chemical surveys of 
very large samples of stars in the Galactic disk and halo.

For studies of disk galaxies at high redshift,  the Hubble (Ultra-)Deep Fields
\citep{hdfn96,hdfs00,hudf06} and the GOODS and COSMOS surveys have been been very 
influential.   The GOODS\footnote{{\it Great Observatories Origin Deep 
Survey}; www.stsci.edu/science/goods.} \citep{goods03} survey  involves two fields centered on the 
Hubble Deep Field North and the Chandra Deep Field South and  combines deep observations 
from NASA's Great Observatories, Spitzer, Hubble, and Chandra, ESA's  Herschel and 
XMM-Newton, and from the most powerful ground-based facilities such as Keck, VLT, 
Gemini and Subaru. The {\it Cosmological Evolution Survey} COSMOS 
(cosmos.astro.caltech.edu) covers a two square degree  equatorial field with a similar 
range of facilities, aimed at probing the formation and evolution of galaxies with 
cosmic time (see \S~8).

Astronomy profits enormously from new facilities, and this is equally true for our 
subject of disk galaxies.  For the future, we look forward to new insights from major 
facilities.  In the submillimeter and radio,  Herschel and the  Atacama Large 
(sub-)Millimeter Array ALMA will revolutionize studies of star formation and the 
interstellar medium in disk galaxies.  The LOw Frequency ARray LOFAR, its southern 
MWA counterpart, the MeerKAT and ASKAP pathfinder arrays and ultimately the SKA itself 
will have a profound impact on studies of the formation of galaxies and the structure 
of disk galaxies. Current deep HST surveys (see for example candels.ucolick.org) and 
surveys to come with the James Webb Space Telescope JWST, will bring new insights into 
the properties of disk galaxies and their assembly.   For studies of the stucture and 
evolution of the Milky Way,  the Gaia mission will give astrometric data of unparalled precision. Combined with panoramic surveys like those planned with Pan-Starrs, SkyMapper and LSST, it will help us to understand the structure and genesis of the different components of our disk galaxy.

\section{STELLAR DISKS}

\subsection{Luminosity Distributions}

\subsubsection{Exponential disks}

The structure and general properties of stellar disks have previously been
reviewed by us \citep[\eg][]{vdk02,kcf07}. As mentioned in the introduction,
the radial distribution of surface brightness in the disks of { face-on
or moderately} inclined  galaxies can be approximated by an exponential: 
$I(R) \propto \exp(-R/h)$.  Before we  discuss the three-dimensional distribution 
we first review work on the exponential disks in { such}  galaxies. Fits to actual 
surface photometry result in two parameters, the radial scalelength $h$ and the 
(extrapolated and corrected to face-on) central surface brightness $\mu_{\circ}$, 
both as a function of photometric band. The determination of these parameters can in 
general be done in a reasonably reliable way from component separations
\citep{kor77}; \citet{sb87} and \citet{bf95} showed from realistic simulations that 
one-dimensional and two-dimensional bulge-disk separations do return input
values for bulge and disk parameters very well. Nevertheless, independent determinations 
of scalelengths of the same galaxies in the literature give results that differ with a 
standard deviation of 20\%\ \citep{kvdk91}.   In his CCD study of exponential disks in 
a sample of bright UGC galaxies, \citet{cou96} also stresses the pitfalls, and cautions 
that comparison of central  surface brightnesses and scalelengths is complicated by the 
subjective nature of their measurement. { We note that older fits adopt the $R^{1/4}$ 
law $I(R) \propto \exp (R^{-1/4})$ for the bulge, whereas most authors now use the 
S\'ersic (1963)-profiles $I(R) \propto \exp (R^{-1/n})$, with the S\'ersic index $n = 1$ 
for the exponential disk and $n = 4$ for the $R^{1/4}$-law. In the context of 
two-component decompositions of radial surface brightness distributions, we note that 
the flat pseudo-bulge structures discussed by \citet{kk04} have values for $n$ of about 
2.5.}

The original publication on exponential disks of \citet{kcf70} used
observations in the $B$-band. In that paper the distribution of the two
parameters was discussed, finding an apparent constancy of $\mu_{\circ}$ for
about 75\% of the sample and that disk galaxies have scalelengths with a wide
range of values (predominantly small in later-type galaxies). We already noted
in \S~\ref{sect:history} that the apparent constancy of central surface
brightness is seriously affected by observational selection
\citep{arp65,dis76,dp83}, leading to the conclusion that there must be many
lower surface brightness galaxies. However, the upper limit is believed to be
real \citep{as79,bf93,dj96b}.  

With the advent of large datasets of surface photometry (such as from the
SDSS), it has become possible now to study large samples of galaxies. For
example, \citet{gad09} has collected $g$, $r$ and $i$-band images of a
representative sample of nearly 1000 galaxies from the SDSS and decomposed
them into bulges, bars and disks.  \citet{pt06} and \citet{pzpd07} have used
SDSS data to determine radial luminosity distributions and look for radial truncations 
(see \S~\ref{sect:truncations}). \citet{fabhp10} determined scalelengths, using an 
automatic technique, for over 30,000 galaxies in five wavelength bands, together with 
indices for asymmetry and concentration.
{ Comparison with the overlap with the sample of \citet{gad09} shows in general terms
good agreement \citep[see fig.~1 of ][which concerns the same sample]{fat10}. 
\citet{fabhp10} form sub-samples for which reliable morphological types or central velocity dispersions are available. As before, the average scalelength ($3.8\pm2.0$ kpc) is independent of morphological type and is very similar in the optical bands ($g$, $r$, $i$ and $z$).  In the $u-$band, they find a mean scalelength of $5\pm3$ kpc. Galaxies of smaller mass ($10^9$ to $10^{10}$ M$_{\odot}$) have smaller scalelengths ($1.5\pm0.7$ kpc) than larger mass ($10^{11}$ to $10^{12}$ M$_{\odot}$) 
galaxies ($5.7\pm1.9$ kpc). The distributions in this study have {\it not}
been corrected for sample selection.}

{ It is possible to study the bi-variate distribution function of the disk
parameters. It is
most important for such studies that the sample is complete with respect to
well-defined selection criteria and that the distribution in the
($\mu_{\circ}$,$h$) plane is corrected for the effect of these selection
criteria  \citep[following the prescriptions of][]{dp83}. This was first done in 
\citet{vdk87}, later at various optical and
near-infrared colors by \citet{djvdk94} and \citet{dj96a,dj96b,dj96c}
and more recently by \citet{fat10} using the large SDSS \citet{fabhp10} sample. 
The study of the distribution
of parameters in this plane reveals important results that bear on the
formation models of disks.}

\begin{figure}[t]
\begin{center}
\includegraphics[height=99mm]{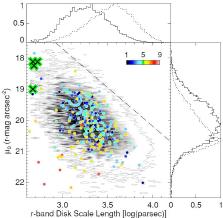}
        \caption{ The bivariate distribution function of face-on central 
surface brightness and scalelength of galaxy disks in a sample of almost 30,000
galaxies taken from SDSS, corrected for selection effects. Superposed are the 282 most 
reliable data as colored points (coding for revised Hubble type) and a few disky
ellipticals as green crosses. The dashed line shows a
slope of 2.5 corresponding to a constant disk luminosity. The distributions
on the right and top are as observed (dotted) and after correcting for sample 
selection (solid).
\citep[From ][]{fat10}.
}
\label{fig:bivariate1}
\end{center}
\end{figure}

\begin{figure}[t]
\begin{center}
\includegraphics[width=160mm]{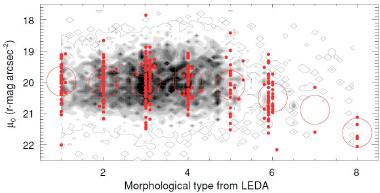}
        \caption{ The distribution of face-on central surface brightness
for the same sample as in fig.~\ref{fig:bivariate1}
as a function of Hubble type, corrected for selection and again with the 282 most 
reliable determinations as points. The open circles show the average surface brightness 
for each type. \citep[From ][]{fat10}.
}
\label{fig:bivariate2}
\end{center}
\end{figure}

The distribution in the $(\mu _{\circ}, \log h)$ diagram 
{ (Fig.~\ref{fig:bivariate1})} shows a broad band
running from bright, large disks to faint, small disks
\citep{vdk87,dj96b,gdb01,fat10}. { \citet{gdb01} find that there is a morphological 
type dependence in this plane: among low surface brightness galaxies \citep[central
surface brightness more than 1 mag fainter 
than the 21.65 $B$-mag arcsec$^{-2}$ mean value of ][]{kcf70}, the early-type spiral 
galaxies have large scalelengths (larger than 8-9 kpc), while the late-type spirals
have smaller scalelengths.} Further, \citet{dj96b} finds that the scale
parameters of disks and bulges are correlated at all morphological types, but
are not correlated themselves with Hubble type. On the other hand, low surface
brightness galaxies are usually of late Hubble type. He also concludes that
the bulge-to-disk ratio is not correlated with Hubble type, nor is 
the disk central surface brightness. The significant parameter that does
correlate with morphological type is the effective surface brightness of the
bulge. Color information shows that, within and among galaxies, low surface
brightness corresponds to bluer colors \citep{dj96c}. This results from the
combined effect of mean stellar age and metallicity and 
not from dust reddening and implies significant mass-to-light ratio variations.
{ Fig.~\ref{fig:bivariate2} shows the modern version of fig.~5 of \citet{kcf70}, 
corrected for volume selection effects. There is still a mean value (but with a 
large scatter around it), that does not depend on morphological type, except for the
later ones.}

What properties of galaxies do correlate with $h$ and $\mu_{\circ}$?
\citet{cou07} collected surface photometry of 1300 galaxies and determined the
photometric parameters from either one-dimensional bulge-disk decompositions
of the surface brightness profile or using the so-called `marking-the-disk'
method, where the extent of the exponential disk profile is judged by eye.  
They also find some variation of central surface brightness or scalelength as
a function of morphological type, with earlier types having fainter surface
brightness and larger scalelengths, but the effects are marginal. In addition,
they find well-defined relations between luminosity, scalelength and rotation
velocity, but the slopes show a definite dependence on morphological type and
a small but significant dependence on the wavelength band. The scalelengths in
the $I$-band correlate with integrated luminosity and rotation velocity (see
eqn.~(\ref{eqn:courteau}) and fig.~\ref{fig:scaling} below). In summary,
although $h$ has no strong dependence on morphological type, it is clearly
larger in the mean for more luminous and more massive galaxies.  

There is some argument concerning the scalelength of the disk of our own Galaxy.  
If $V_{\rm rot}=220$ km s$^{-1}$, then the expected scalelength of the Galactic disk
would be 4.4 kpc with a one-sigma range between 3.6 and 6.6 kpc \citep[see
slide 19 in][]{vdk09}. These values would be larger if $V_{\rm rot}=250$ km s$^{-1}$
\citep{reid09}.  The often quoted values of 2.5 to 3.0 kpc
\citep{sac97,freu98,hpcfl07,yhpbcsz09,rmrs09} put the Galaxy outside the
one-sigma range of scalelengths for its rotation speed. A value more like 4.5
kpc \citep{vdk08,vdk09} \citep[or even the probably too large $5.5\pm1.0$ kpc
from the Pioneer 10 photometry alone;][]{vdk86}  
would be more typical for our Galaxy. On the other hand, \citet{hpcfl07} argue that 
our Galaxy is exceptional in many aspects: however their adopted low value for the
scalelength of the disk is a major contributor to this conclusion. 

The origin of the exponential nature of stellar disks is still
uncertain. \citet{kcf70, kcf75} already pointed out that the distribution of
angular momentum in a self-gravitating exponential disk resembles that of the
uniform, uniformly rotating sphere \citep{mes63}. This is also true for an
exponential density distribution with a flat rotation curve \citep{gun82,
  vdk87}. A model in which the disk collapses with detailed conservation of
angular momentum \citep{fe80} would give a natural explanation for the
exponential nature of disks and maybe even their truncations (see
below). However, bars or other non-axisymmetric structures may give rise to
severe redistribution of angular momentum; nonaxisymmetric instabilities and
the secular evolution of disks and their structural parameters may be
important \citep{dmcmwq06}. 
 
Before leaving the subject of luminosity distributions, we will briefly
address the issue of low surface brightness (LSB)  disks. Often these have
central (face-on) surface brightnesses that are 2 magnitudes or more fainter
than the canonical 21.65 $B$-mag arcsec$^{-2}$ of 
\citet{kcf70}. Traditionally these are thought to be galaxies with low (gas)
surface densities, in which the star formation proceeded slowly. Analysis of
available data (\HI\ rotation curves, 
colors and stellar velocity dispersions) led \citet{dbm97} to argue that LSB
galaxies are not described well by models with maximum disks (see below). LSB
galaxies appear to be slowly evolving, low density, dark matter dominated
systems. The star formation in low surface brightness disks can now be studied
with GALEX by directly mapping their near-UV flux.  \citet{wmbetal09} combined
such data with existing \HI\ observations and optical images from the SDSS for
19 systems. Comparison with far IR-data from Spitzer shows that there is very
little extinction in the UV, consistent with the fact that LSB galaxies appear
to have little dust and molecular gas 
\citep[see \eg][]{dbvdh98a,dbvdh98b}. The star formation rate in LSB galaxies
lies below the extrapolated rate as a function of gas surface density for high
surface brightness galaxies, 
implying a lower mean star formation efficiency in LSB systems. This may be
related to the lower density of molecular gas.  

\begin{figure}[t]
\begin{center}
\includegraphics[width=135mm]{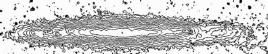}
\includegraphics[width=135mm]{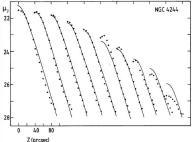}
        \caption{ The surface brightness distribution in the edge-on, pure disk
galaxy NGC 4244. At the top the isophotes in blue light at 0.5 mag 
arcsec$^{-2}$ intervals. At the bottom vertical
$z$-profiles at a range of distances from the center after averaging over
the four equivalent quadrants. The curves are those for an isothermal
sheet with $n$=1 in eqn.~(\ref{eqn:vdksdisk}).
\citep[After ][]{vdks81a}
}
\label{fig:isothermal_disk}
\end{center}
\end{figure}

\subsubsection{Three-dimensional distributions}

We now turn to the three-dimensional distribution. 
The vertical distribution of luminosity within a galactic disk  can be
modelled to a first approximation with an isothermal sheet \citep{cam50} with
a scaleheight that is independent of galactocentric distance \citep{vdks81a}.  
{ This is a surprising observational result. We will discuss its possible 
origin in \S~\ref{sect:scaleheight}.} 
In a more general form { the luminosity density distribution} 
can be written as
\citep{vdk88}
\begin{equation}
L(R,z)=L(0,0) {\rm \;e}^{-R/h} \sech ^{2/n} \left( {{n z} \over
{2 h_{\rm z}}} \right),
\label{eqn:vdksdisk}
\end{equation}
This ranges from the isothermal distribution ($n=1$: $L(z) \propto
\sech ^2(z/z_{\circ})$ with $z_{\circ}=2h_{\rm z}$) to the exponential 
function ($n=\infty $: $L(z) \propto \exp (-z/h_{\rm z})$), 
as was used by \citet{whf89,whf90},
and allows for the more realistic case that the
stellar distribution is not completely isothermal in the vertical direction. 
{ The uncertainty resulting from what the detailed vertical distribution of
stellar mass really is in a disk can be estimated by taking a realistic range
in the parameter $n$, as we have done for example in eqn.~(\ref{eqn:hydrostat})
below.}
{ Fig.~\ref{fig:isothermal_disk} shows the fits of this distribution projected
in edge-on orientation to the surface brightness distribution in
the pure-disk, edge-on galaxy NGC 4244, for
the case of the isothermal sheet ($n$=1). The outer profile does not fit, because
the truncation (see \S~\ref{sect:truncations})
has not been taken into account.}
From actual fits in $I$ and $K'$ \citet{dgpvdk97} found
\begin{equation}
{2 \over n}=0.54 \pm 0.20
\end{equation}
for a sample of edge-on galaxies.
A detailed study by \citet{dgp97} has shown that the constancy of the vertical
scaleheight $h_{\rm z}$ is accurate in disks of late-type spirals, but
in early type galaxies $h_{\rm z}$ may increase { outward by} 
as much as 50\%\ per scalelength $h$.

The distribution of the scale parameters is most easily studied in edge-on
galaxies. Following the work by \citet{vdks81a, vdks81b, vdks82a}, an
extensive sample of edge-on galaxies was studied by \citet{dg98} and
re-analysed by \citet{kvdkdg02}. The results on the distribution of scale
parameters can be summarized as follows. Both scalelength $h$ and scaleheight
$h_{\rm z}$   
correlate well with the rotation velocity of the galaxy: \eg\ for the scaleheight
\begin{equation}
h_{\rm z}=(0.45 \pm 0.05)\, (V_{\rm rot}/100\ {\rm km\  s}^{-1}) -
(0.14 \pm 0.07)\ {\rm kpc}
\end{equation}
with a scatter of 0.21 kpc. This relation is important as it can be used to
make a statistical 
estimate of the thickness of disks in galaxies that are not seen edge-on. The
correlation between $h$ and $V_{\rm rot}$ is comparable to that found by
\citet{cou07}. The flattest galaxies (largest ratio of $h$ and $h_{\rm z}$)
appear to be those with late Hubble type, small rotation velocity and faint
(face-on) surface brightness. Among galaxies with large \HI\ content, a large
range of flattening is observed, becoming smaller with lower \HI\ mass. 
The flattest disks occur among galaxies with about $10^{10}$ M$_{\odot}$ in
\HI. We will return to this subject in \S~\ref{sect:superthin} when we discuss
`super-thin' galaxies.

\subsection{Stellar Kinematics, Stability and Mass}\label{sec:stellardyn}

\subsubsection{Vertical dynamics}

We will first turn to the dynamics of stellar disks in the vertical ($z$) direction. 
At the basis of the analysis of the vertical dynamics of a stellar disk
we have the Poisson equation for the case of axial symmetry
\begin{equation}
{\D K_{\rm R} \over \D R} + {K_{\rm R} \over R} + {\D K_{\rm z} \over \D z} =
-4\pi G \rho (R,z),
\label{eqn:Poisson}
\end{equation}
where $K_{\rm R}$ and $K_{\rm z}$ are the gravitational { force} components. 
At small $z$, the first two terms on the { left} are equal to
$2(A - B)(A + B)$ \citep[\eg][]{oort65, kcf75} and this is zero for a flat 
rotation curve.\footnote{The Oort constants are $A=\textstyle{1\over2} 
\{V_{\rm rot}/R - 
(dV_{\rm rot}/dR)\}$ and $B = -\textstyle{1\over2} \{V_{\rm rot}/R + 
(dV_{\rm rot}/dR)\}$, so $A + B = -(dV_{\rm rot}/dR)$.}
So we have
\begin{equation}
{dK_{\rm z} \over dz} = -4\pi G \rho (z).
\label{eqn:plane}
\end{equation}
This is the plane-parallel case and flat rotation curves do make this
an excellent approximation at low $z$ \citep{vdkf86}. The Jeans
equation then becomes
\begin{equation}
{ d \over {d z}} \left[ \rho (z) \sigma_{\rm z}^2 (z) \right]
= \rho (z) K_{\rm z}.
\end{equation}
Combining these gives \citep[\eg][]{vdk88} 
\begin{equation}
\sigma_{\rm z}(R) = \sqrt{c \pi G \Sigma (R) h_{\rm z}},
\label{eqn:hydrostat}
\end{equation}
where the velocity dispersion $\sigma_{\rm z}$ is now the velocity dispersion 
integrated over all $z$ (corresponding to the second moment of the
distribution observed when the disk is seen face-on) and the constant $c$ 
varies between 3/2 for an exponential  
[$n=\infty$ in eqn.~(\ref{eqn:vdksdisk})] to 2 for an isothermal 
distribution ($n=1$).
Eqn.~(\ref{eqn:hydrostat}) is the equation for hydrostatic equilibrium that
relates the vertical distribution of the stars and their 
mean vertical velocity dispersion to the  distribution of mass;
this principle was used already by \citet{kap22} and \citet{oort32} to derive
the mass density in the solar neighborhood.
If the mass-to-light ratio $M/L$ is constant with radius, the exponential
radial distribution and the constant scaleheight imply through hydrostatic
equilibrium that the vertical velocity dispersion $\sigma_{\rm z}(R)$ of the 
old stars in the disk should { be proportional to} the square-root of the surface
density $\Sigma$ or  as an exponential with galactocentric radius, 
but with an e-folding of twice the scalelength. 

The mass-to-light ratio $M/L$ is a crucial measure of the contribution
of the disk to the rotation curve and the relative importance of disk
mass and dark matter halo in a galaxy. An often used hypothesis
is that of the `maximum disk' (see also \S~\ref{sect:rotcur}), in which 
the disk contribution to  a galaxy's rotation curve is maximized in the sense 
that the amplitude of the disk-alone rotation curve is made as large as the
observations allow. Using hydrostatic equilibrium, we may estimate $M/L$ and
obtain information on whether or not disk are maximal of sub-maximal. This can
in principle be done from eqn.~(\ref{eqn:hydrostat}) by measuring the velocity
dispersion in a face-on galaxy and using a statistical estimate of the
scaleheight. 

The measurement of stellar velocity dispersions in disks, which has to be
done using stellar absorption lines in the optical or near-IR, is seriously
hampered by the low surface brightness. In 1984 \citet{vdkf84} made the
first successful measurements of stellar velocity dispersion in the 
face-on spirals
NGC 628 and 1566.\footnote{At the same time and independently, \citet{kor84a,kor84b} 
succeeded in measuring stellar
velocity dispersions in the disks of two S0 galaxies with more or less
the same aim; we will discuss this in \S~\ref{sect:S0}.}
This work was followed by more detailed observations by \citet{vdkf86} for 
NGC 5247 (inclination about 20$^{\circ}$), where the prediction 
was verified: the e-folding length of $\sigma_{\rm z}$ was 2.4$\pm$0.6 photometric
scalelengths, the predicted value of 2.0 being well within the
uncertainty. Many studies have since shown that 
$\sigma_{\rm z}$ decreases with galactocentric radius \citep[\eg][and 
references therein]{bot93,kvdkf04, kvdkf05, kvdk05}. 
\citet{gkm97} found in NGC 488 that
the kinematic gradient was comparable to the photometric gradient, which they
attributed to the fact that the scalelength should really be measured in
$K$-band to represent the stellar distribution. The same authors
found in NGC 2985, that these scalelengths were indeed as expected from a constant
$M/L$. There is certainly support from stellar dynamics that 
in general there are no substantial gradients in mass-to-light ratios in disks.
We will come back to this below in the context of photometric models and 
stellar composition and ages in disks.

Two recent developments are making an impact on this issue. The first is 
the use of integral field units that enable a more 
complete sampling of the disks. The DiskMass Project 
\citep{vbsaw07,wbvas08,bvsawm10,bvwasm10} aims at mapping the stellar 
vertical velocity dispersion in 46 face-on or moderately inclined spiral galaxies. 
This will provide a kinematic 
measurement of the mass surface density of stellar disks. The final results 
have not yet appeared in the literature, but recent conference presentations
show that the `kinematics follows the light', i.e. the velocity dispersions
drop off according to the rule described above. Also the actual values
indicate relatively low mass-to-light ratios and disk masses that are well
below those required for maximum disk fits.

Similarly, the use of planetary nebulae (PNe) as test particles in the disks 
\citep{hcfv08,hc09a,hc09b} of five face-on spirals method allows
the velocity dispersion of these representative stars of the old disk population
to be measured out to much larger radii (see also \S~\ref{sect:S0}). 
In general the findings are similar:
except for one system, the $M/L$ is constant out to about three radial
scalelengths of the exponential disks. Outside that radius, the velocity
dispersion stops declining and becomes flat with radius. Possible 
explanations proposed for this behavior include an increase 
in the disk mass-to-light ratio, an increase in the importance of the thick 
disk, and heating of the thin disk by halo substructure. They also find that 
the disks of early type spirals have higher values of $M/L$ and are closer to 
maximum disk than later-type spirals.

In summary, the vertical dynamics of stellar disks show that in general
the velocity dispersions of the stars falls off with an e-folding length double 
that of the exponential light distribution,  as required for a constant 
$M/L$, while for the majority of disks the inferred mass-to-light ratios
are almost certainly lower than required in the maximum disk hypothesis.

\subsubsection{Stellar velocity dispersions in the plane}

The stellar velocity dispersions {\it in} the plane are more complicated
to determine from observations. The  radial and tangential components are 
not independent, but governed by the local Oort constants\footnote{For small 
deviations from circular motions around the galactic center, the stellar 
orbit may be described by a small epicycle superposed on the circular motion 
around the galactic center. The
frequency in the epicycle is $\kappa=2\sqrt{-B( A - B)}$ and its axis
ratio $\sqrt{-B/(A - B)}$ \citep{oort65}. The ratio between the two 
velocity dispersions derives from the shape of the epicycle.}
\begin{equation}
{ \sigma_{\theta} \over \sigma_{\rm R}}=\sqrt{ -B \over {A - B}}.
\label{eqn:disps}
\end{equation}
For a flat rotation curve $A=-B$ and this ratio is 0.71.
In highly inclined or edge-on systems the dispersions can be measured
both from the line profiles and the asymmetric drift equation 
\begin{equation}
V^2_{\rm rot} - V^2_{\theta}=\sigma^2_{\rm R} \left\{ {R \over h} -
R {\D \over {\D R}} \ln (\sigma_{\rm R}) - \left[ 1 - { B \over {B - A}}
\right] \right\},
\end{equation}
where the circular velocity $V_{\rm rot}$ can be measured with sufficient 
accuracy from the gas (optical emission lines or \HI\ observations), which
have velocity dispersions of order 10 \kms\ or less and have therefore 
very little asymmetric drift. 

{ The stability of a galactic disk to local axisymmetric disturbances depends on the 
the (stellar) radial velocity dispersion $\sigma_{\rm R}$, the epicyclic frequency
$\kappa$, and the local mass surface density $\Sigma$. Toomre's (1964) 
criterion is}
\begin{equation}
Q={ {\sigma_{\rm R} \kappa} \over {3.36 G \Sigma}}.
\end{equation}
On
small scales local stability results from a Jeans-type stability, where 
tendency to collapse under gravity is balanced by the kinetic energy in random
motions, but only up to a certain (Jeans) scale. On scales
larger than some minimum radius, shear as a result of galactic
differential rotation provides stability. In the Toomre $Q$-criterion
this smallest scale is just equal to the (maximum) Jeans scale, so that
that local stability exists on {\it all} scales. According to \citet{too64},
local stability requires $Q \gt 1$. Numerical simulations suggest that galaxy
disks are on the verge
of instability \citep{hohl71,sc84,as86,mmb97,bot03}, having stellar
velocity dispersions that are slightly larger than Toomre's critical
velocity dispersion. The simulations suggest $Q=$1.5--2.5.

The first attempt to measure these in-plane velocity
dispersion components was by \citet{vdkf86} for the highly inclined 
galaxy NGC 7184. They fitted their data using two different assumptions
for the radial dependence of the radial velocity dispersion: one 
that the axis ratio of the velocity ellipsoid (between the vertical and radial
dispersion) is the same everywhere, and the other that the Toomre $Q$ is 
constant with radius. Both assumptions worked well; over the observed range 
of one or two scalelengths from the center, the two assumptions correspond to 
similar variations \citep[see][page 196]{gkvdk90}.

\begin{figure}[t]
\begin{center}
\includegraphics[width=80mm]{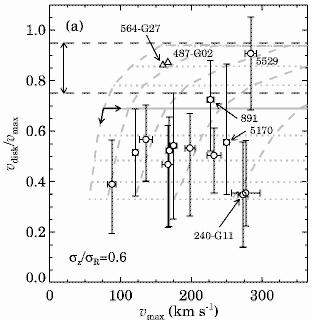}
        \caption{Stellar disk velocity dispersion, measured at one scalelength
in edge-on galaxies  versus the
        maximum rotational velocity. The gray lines indicate the relation
        $\sigma_{\rm R}(h) =$ (0.29$\pm$0.10) $V_{\rm rot}$
        \citep{bot93}. \citep[From][]{kvdkf05}}
\label{fig:7_bottema2}
\end{center}
\end{figure}

More extensive observations by \citet{bot93} on a sample of 12 galaxies
\citep[including the Milky Way Galaxy from][]{lf89}
resulted in the discovery of a relation between a fiducial value of the
velocity dispersion (either the vertical one measured at or 
extrapolated to the center or the radial velocity dispersion  
at one scalelength) and the integrated luminosity or
the rotation velocity. Luminosity and rotational velocity are equivalent through the
Tully-Fisher relation. This has been confirmed by \citet{kvdk05} and 
\citet{kvdkf05}, see fig.~(\ref{fig:7_bottema2}). 
The relation\footnote{The formal fit to the sample in
\citet{kvdkf05} is $\sigma_{\rm R|1h}=(-2 \pm 10) + (0.33 \pm 0.05) 
V_{\rm rot}$; the equation above gives the relation adopted from all 
available studies.} is
\begin{equation}
\sigma_{{\rm z}|0}=\sigma_{\rm R|1h}=(0.29 \pm 0.10) V_{\rm rot}.
\label{eqn:bottema}
\end{equation}
It extends to very
small dwarf galaxies, \eg\ UGC 4325 with a velocity dispersion
of 19 \kms\ still falls on the relation \citep[][chapter 7]{swat99}. 
The scatter in this relation is not random but appears 
related to other properties. Galaxies with 
lower velocity dispersions 
have higher flattening, lower central surface brightness or dynamical 
mass ($4 h V^2_{\rm rot}/G$) to disk luminosity ratio.

The linear $\sigma-V_{\rm rot}$ relation follows from straightforward
arguments, presented in \citet{gkvdk90}, Ch. 10 
\citep[see also][]{bot93, vdkdg99}.
We evaluate properties at one radial scalelength ($R=1h$) without
using subscripts to indicate this. Using the definition for Toomre $Q$ 
for a flat rotation curve, so that $\kappa=\sqrt{2}V_{\rm rot}/R$, 
and eliminating $h$
using a Tully-Fisher relation $L_{\rm disk}
\propto \mu _{\circ} h^{2} \propto V_{\rm rot}^{4}$ results in
\begin{equation}
\sigma _{\rm R} \propto Q {{\mu _{\circ} (M/L)_{\rm disk} h}
\over V_{\rm rot}} \propto Q \left( M \over L \right)_{\rm disk} \mu
_{0}^{1/2} V_{\rm rot}.
\label{eqn:botrel}
\end{equation}
This shows that, when $Q$ and $M/L$ are constant among galaxies, the Bottema 
relation follows, with indeed the proviso that galaxy disks with lower
(face-on) central surface brightness $\mu _{\circ}$ at a given value of
$V_{\rm rot}$ have  
lower stellar velocity dispersions than given by the mean $\sigma-V_{\rm rot}$
relation. 

\subsubsection{Origin of the constant scaleheight}
\label{sect:scaleheight}

The origin of the constant scaleheight of stellar disks --or of the fall-off
of stellar vertical velocity dispersion such to precisely compensate for
the decline in surface density-- is not obvious. If the evolution of the
stellar velocity dispersions (the `heating of the disk') is similar at
all radii and if it evolves to a radial velocity dispersion such that
the disk is just stable everywhere, we may expect 
$\sigma_{\rm z}/\sigma _{\rm R}$ (`the axis ratio of the
velocity ellipsoid') and Toomre $Q$ to be independent of galactocentric
radius. This would however imply that $\sigma_{\rm z} \propto (R/h) \exp
(-R/h)$. Although this is not all that much different from the
exponential decline $\exp(-R/2h)$
that follows from eqn.~(\ref{eqn:hydrostat}) between
say one and three scalelengths\footnote{The reason why the two analyses
of the measurements of velocity dispersion in \citet{vdkf86} and by
Bottema et al. \citep[see references in][]{bot93} both gave good
fits.}, it is significantly different at larger radii. In fact, the
simple assumption would result in $h_{\rm z} \propto (R/h)^2
\exp(-R/h)$,
which is far from constant over the range $R=0$ to $R=5h$.

It is necessary to look first at the evolution of random stellar motions in
disk before we can proceed. There are three general
classes of models for the origin of the velocity
dispersions of stars in galactic disks.  The first, going back to
\citet{ss51}, is scattering by irregularities in the
gravitational field, later identified with the effects of Giant
Molecular Clouds (GMC's).  The second class of models can be traced back
to the work of \citet{bw67}, who suggested transient
spiral waves as the scattering agent; this model has been extended by
\citet{cs85}. More recently,
the possibility of infall of satellite galaxies has been recognized as a
third option \citep[\eg][]{vw99}.

{ Almost all of the observational information about the evolution of velocity 
dispersion with age in galactic disks comes from the solar neighborhood, and we must 
stress that this information remains quite insecure.  While much of the earlier work 
invokes the results of \citet{wie77}, which indicates a steady increase of stellar 
velocity dispersion with age, some of the more recent observational studies indicate 
that the velocity dispersion increases with age for only 2 to 3 Gyr, and then 
saturates, remaining constant for disk stars of older age 
\citep[see \eg\ ][]{egn93, kcf91, sbmk08}.  
The observational situation regarding disk heating is 
far from certain, and this in turn must reflect on the various theories of disk heating. 
}

In the solar neighbourhood the ratio of the radial and vertical velocity
dispersion of the stars $\sigma _{\rm z}/\sigma _{\rm R}$ is usually
taken as roughly 0.5 to 0.6 \citep{wie77, gdg90, db98, mig00}, although values
on the order of 0.7 are also found in the literature \citep{wmpsa77,
mrs91}.  The value of this ratio
can be used to test predictions for the
secular evolution in disks and perhaps distinguish between the
general classes of models.  \citet{lac84} and \citet{vil85} have
concluded that the Spitzer-Schwarzschild mechanism is not in agreement
with observations; the predicted time dependence of the
velocity dispersion of a group of stars
as a function of age disagrees with the observed age
-- velocity dispersion relation (see also Wielen, 1977), while it would
not be possible for the axis ratio of the velocity ellipsoid 
$\sigma_{\rm z}/\sigma _{\rm R}$ to be less than about 0.7 
\citep[but see][]{ikm93}.

\citet{jb90} argued that it is likely that the dynamical
evolution in the directions in the plane and that perpendicular to it
could have proceeded with both mechanisms contributing, but in different
manners.  Scattering by GMC's would then be responsible for the vertical
velocity dispersion, while scattering from spiral irregularities would
produce the velocity dispersions in the plane.  The latter would be the
prime source of the secular evolution with the scattering by molecular
clouds being a mechanism by which some of the energy in random motions
in the plane is converted into vertical random motions, hence
determining the thickness of galactic disks.
The effects of a possible slow, but significant accretion of
gas onto the disks over their lifetime has been studied by
\citet{jen92}, who pointed out strong effects on the time dependence of
the vertical velocity dispersions, in particular giving rise to enhanced
velocities for the old stars. On the other hand, \citet{hf00, hf02} conclude
that observations such as the radial dependence of stellar velocity
dispersions in the Milky Way Galaxy by \citet{lf89} can be reproduced if
scattering occurs by a combination of massive halo objects (black holes) and
GMC's. \citet{db98} conclude that 
spiral structure is probably a major contributor to disk heating.
More recently, \citet{mq06} suggested from 2D simulations 
that multiple patterns of spiral structure could cause strong variations of
stellar velocity dispersions with galactocentric radius, which has not been
observed.  Our conclusion is that there still is much uncertainty about the
process of heating of the (thin) disk. Some of this uncertainty is due to
uncertainty in the observational relation between stellar ages and velocity
dispersions, because stellar ages are so difficult to measure. 

Theoretical arguments suggest that a constant axis ratio of the
velocity ellipsoid is a fair approximation in
the inner parts of galaxy disks \citep{ca92, fvcd02}. An observational
argument for the approximate constancy of the velocity anisotropy
is provided by the ages and kinematics of 182 F and G dwarf stars in
the solar neighbourhood \citep{egn93}. This indicates that the
anisotropy was set after an early heating phase and, although the
Galaxy has probably changed much over its lifetime, has remained
constant throughout the life of the old disk \citep{kcf91}.

So, where does this leave us with respect to the origin of the
constant scaleheight?  As long as there is no detailed understanding of 
the evolution of the velocity dispersions as a function of galactocentric 
radius, we cannot even begin to address this in a meaningful way. A constant
stability parameter $Q$ and a constant axis ratio of the velocity ellipsoid
$\sigma_{\rm z}/\sigma_{\rm R}$  do give an approximate constant thickness
over the inner few scalelengths, but this fails at larger radii.

\begin{figure}[t]
\begin{center}
\includegraphics[width=80mm]{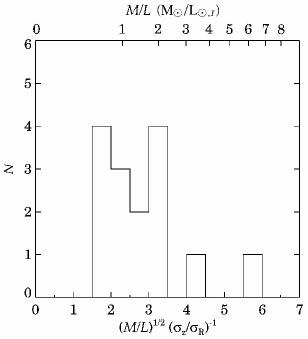}
   \caption{Histogram of the product $\sqrt{M/L_I}\
        (\sigma_{\rm z}/\sigma_{\rm R})^{-1}$ from stellar kinematics
in edge-on galaxies. Except for two outliers
        the distribution of $\sqrt{M/L_I}\ (\sigma_{\rm z}/\sigma_{\rm
        R})^{-1}$ is rather narrow. The outliers are ESO 487-G02 and 
564-G27; data for these galaxies are less complete than for the other
ones. Along the top we show the values of
        $M/L_I$ implied by $\sigma_{\rm z}/\sigma_{\rm R}=0.6$. 
\citep[From][]{kvdkf05}}
\label{fig:7_histogram}
\end{center}
\end{figure}

\subsubsection{Mass distributions from stellar dynamics}

The stellar velocity dispersions can still be used to derive information on the
disk mass distribution. For a self-gravitating
disk which is exponential in both the radial and vertical direction,
the vertical velocity dispersion goes as \citep[cf.][]{vdk88}:
\begin{equation}
\sigma_{\rm z}(R, z) = \sqrt{\pi G h_{\rm z} (2 -
e^{-z/h_{\rm z}}) (M/L) \mu_{0}}\ e^{-R/2 h},
\label{eqn:sigma_z}
\end{equation}
Assuming a constant (but unknown) axis ratio of the velocity ellipsoid 
$\sigma_{\rm z}/\sigma_{\rm R}$, the radial velocity dispersion becomes
\begin{equation}
\sigma_{\rm R}(R,z) = \sqrt{\pi G h_{\rm z} (2 -
e^{-z/h_{\rm z}}) (M/L) \mu_{0}}\
\left( {{\sigma_{\rm z}} \over {\sigma_{\rm R}}} \right)^{-1}\ 
e^{-R/2 h}.
\label{eqn:sigma_R}
\end{equation}
The distribution of the products $\sqrt{M/L_I}\ (\sigma_{\rm
z}/\sigma_{\rm R})^{-1}$, deduced from this equation in the \citet{kvdkf05} 
sample is
shown in fig.~\ref{fig:7_histogram}. { This sample of edge-on galaxies 
has a range of Hubble types from Sb to Scd, absolute $I$-magnitudes between 
-23.5 and -18.5, and a range in rotation velocities from 89 to 274 \kms.}
Thirteen of the fifteen disks
have 1.8 $\simlt \sqrt{M/L_I}\ (\sigma_{\rm z}/\sigma_{\rm R})^{-1}
\simlt$ 3.3. The values of the outliers may have been overestimated
\citep[see][]{kvdkf05}. Excluding these, the
average is $\left<\right.\sqrt{M/L_I}\ (\sigma_{\rm z}/\sigma_{\rm
R})^{-1}\left.\right>=$ 2.5$\pm$0.2 with a $1\sigma$ scatter of 0.6. 
The near constancy of the product can be used with { mass-to-light ratios}
based on stellar population synthesis models to estimate the axis
ratio of the velocity ellipsoid. Conversely, the upper scale of
fig.~(\ref{fig:7_histogram}) indicates that a typical $M/L$ in the
$I$-band of a galactic stellar disk is of order unity and for the 
majority systems lies between 0.5 and 2. 

It is possible to relate the axis ratio of the velocity ellipsoid to the
flattening of the stellar disk, i.e. the ratio of the radial exponential
scalelength and the vertical exponential scaleheight \citep{vdkdg99}. In the
radial direction, the velocity dispersion { at one scalelength can be
written using the definition of Toome $Q$ as 
$\sigma _{\rm R,h} \propto Q \Sigma(h) h / V_{\rm rot}$, where
 a flat rotation curve has been assumed. At this radius of one scalelength the 
hydrostatic equation gives $\sigma _{\rm z} \propto \sqrt{\Sigma (h) 
h_{\rm z}}$. Eliminating $\Sigma (h)$ between these two equations then gives 
\begin{equation}
\left( {{\sigma _{\rm z}} \over {\sigma _{\rm R}}} \right)^{2}_{\rm h}
\propto {1 \over Q} {{h_{\rm z}} \over h}
\end{equation}
If $Q$ is constant within individual disks, then the disk flattening depends 
directly on the axis ratio of the velocity ellipsoid.}

Eqn.~(\ref{eqn:botrel}) shows that when $Q$ and $M/L$ are constant among 
galaxies, galaxy disks with lower (face-on) central surface brightness 
$\mu _{\circ}$ have
lower stellar velocity dispersions.
Combining eqn.~(\ref{eqn:botrel}) with the hydrostatic
equilibrium eqn.~(\ref{eqn:hydrostat}) and using eqn.~(\ref{eqn:bottema})
gives \citep{kvdkf05, vdkdg99}
\begin{equation}
{h \over h_{\rm z}} \propto Q \left( {\sigma _{\rm R} \over \sigma
_{\rm z}} \right) \sigma _{\rm z}^{-1} V_{\rm rot} 
\propto Q \left( {\sigma _{\rm R} \over \sigma _{\rm z}} \right).
\label{eqn:B1}
\end{equation}
The observed constancy of $\sqrt{M/L}\ (\sigma_{\rm z}/\sigma_{\rm R})^{-1}$
implies that the flattening of the disk $h/h_{\rm z}$ is proportional
to $Q\sqrt{M/L}$.

\begin{figure}[t]
\begin{center}
\includegraphics[width=80mm]{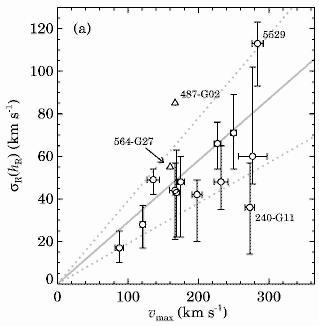}
\caption{The contribution of the disk to the amplitude
of the rotation curve $V_{\rm disk}/V_{\rm rot}$. 
for a sample of 15 edge-on galaxies as a function
of the rotation velocity itself. The horizontal dashed lines are the limits
of $0.85 \pm 0.10$ from \citet{sac97}, which would indicate
Fmaximal disks. The axis ratio of the velocity 
ellipsoid is assumed to be 0.6. The grey lines correspond to collapse
models of \citet{dal97}; { dashed lines connect models of
        the same total mass ($\log_{10}(M_{\rm tot}) = 10-13$ in steps
        of 0.5) and dotted lines connect models with the same spin
        parameter (logarithmically spaced, separated by factors of
        0.2 dex, with the solid line at $\lambda = 0.06$). The
        arrows indicate the direction of increasing $M_{\rm tot}$ and
        $\lambda$.} The two { galaxies}
without error bars are the same { ones} as 
the outliers in fig.~\ref{fig:7_histogram}.
\citep[From][]{kvdkf05}}
\label{fig:7_maximumdisk1}
\end{center}
\end{figure}

\begin{figure}[t]
\begin{center}
\includegraphics[width=75mm]{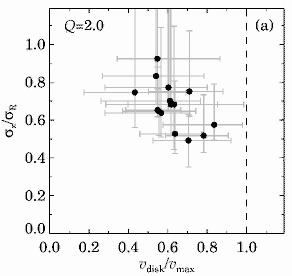}
\includegraphics[width=75mm]{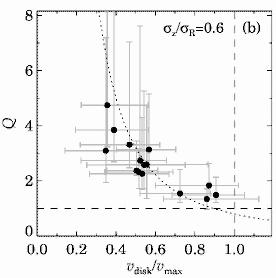}
        \caption{Stellar dynamics parameters for edge-on galaxies.
(a) The axis ratio of the velocity ellipsoid as
a function of $V_{\rm disk}/V_{\rm rot}$ for $Q$=2.0. (b)
$V_{\rm disk}/V_{\rm rot}$ as a function of $Q$ for an assumed axis ratio of
the velocity ellipsoid of 0.6.  \citep[From][]{kvdkf05}}
\label{fig:7_maximumdisk2}
\end{center}
\end{figure}

For a { self-gravitating} exponential disk, the { expected rotation curve peaks
at 2.2 scalelengths. The ratio of this peak of the} rotation velocity of
the disk to the maximum rotation velocity of the galaxy
($V_{\rm disk}/V_{\rm rot}$) is
\begin{equation}
\frac{V_{\rm disk}}{V_{\rm rot}}=\frac{0.880\ (\pi G\,
\Sigma_{0}\, h)^{1/2}}{V_{\rm rot}}.
\label{eqn:7_freeman}
\end{equation}
Using eqn.~(\ref{eqn:hydrostat}) and eqn.~(\ref{eqn:bottema}) this can be
rewritten as 
\begin{equation}
\frac{V_{\rm disk}}{V_{\rm rot}}=(0.21 \pm 0.08) \sqrt{h \over {h_{\rm
z}}}.
\end{equation}
So we can estimate the disk contribution to the rotation curve from
a statistical value for the flattening 
\citep[see also][]{bot93, bot97, vdk02}. For the sample of \citet{kvdkdg02} this
then results in  $V_{\rm disk}/V_{\rm rot}=$ 0.57$\pm$0.22 (rms scatter).
In the dynamical
analysis of \citet{kvdkf05}, the ratio $V_{\rm disk}/V_{\rm rot}$
is known up to a factor $\sigma_{\rm z}/\sigma_{\rm R}$ and
distance-independent. The derived
disk contribution to the observed maximum for the same sample
rotation is on average $V_{\rm disk}/V_{\rm rot}=$ 0.53$\pm$0.04, 
with a $1\sigma$ scatter of 0.15. Both estimates agree well.

In the maximum disk hypothesis,  $V_{\rm disk}/V_{\rm rot}$ will be 
a bit lower than unity to allow a bulge contribution 
and let dark matter 
halos have a low density core. A working definition that has been
adopted generally is
$V_{\rm disk}/V_{\rm rot}=$ 0.85$\pm$0.10 \citep{sac97}. Thus, at least for
this sample, the average spiral has a submaximal disk. Note
that eqn.~(\ref{eqn:7_freeman}) strictly applies to a razor-thin
disk. For a disk with a flattening of $h/h_{\rm z} \simeq$ 10
the radial gravitational force is weaker, leading to decrease of about
5\%\ in $V_{\rm disk}/V_{\rm rot}$ \citep{vdks82a}. Taking the gravity of
the gas layer and dark matter 
halo into account would yield a 10\%\ effect,
also in this direction. So, these effects work in the direction of making the
disks more sub-maximal.

The values obtained from stellar dynamics are illustrated in 
Figs.~\ref{fig:7_maximumdisk1} and \ref{fig:7_maximumdisk2}. { The 
measurement of stellar velocity dispersions can been used to derive the disk
surface density at some point (e.g. one scalelength) up to a factor 
$(\sigma _{\rm z}/\sigma _{\rm R})^2$, but can be estimated also
from the velocity dispersion for an
assumed value of $Q$. Comparing the two gives then an estimate of the axis 
ratio of the velocity ellipsoid. In fig.~\ref{fig:7_maximumdisk2} 
on the left $Q$ is assumed 2.0 
and on the left the velocity anisotropy is assumed to be 0.6 and then 
a value for $Q$ results.}  Most galaxies are not `maximum-disk'. The
ones that may be maximum disk have a high surface density
according to fig.~\ref{fig:7_histogram}. From the panels we also note
that disks that are maximal appear to have more anisotropic velocity
distributions or are less stable according to Toomre $Q$. We will
return to the maximum disk hypothesis below (\S~\ref{sect:rotcur}).

\subsection{Age Gradients and Photometric $M/L$ Ratios}

{ Colors contain information on the history of star formation, as can be
studied in the context of integrated colors of galaxies, pioneered by 
\citet{ssb73} and \citet{lt78}, and described in much detail by 
\citet{bt80}, and also  as a function of radius in a galaxy disk.}
Observing and interpreting color gradients in galactic disks is not
straightforward. Obviously one needs accurate photometry for unambiguous
interpretation in terms of stellar synthesis and star formation
histories. However, the effects of age\footnote{It should
be noted that in discussions of these subjects the property `stellar age'
is usually the mean age of all stars derived as a luminosity-weighted average, 
further weighted by the
star formation rate over the lifetime of the disk,
and should not be confused with the age of the oldest stars.}
and metallicity
are difficult to separate.
Dust absorption is also a
major factor, often making degenerate the effects of stellar age and
metallicity on the one hand and extinction and reddening by dust on the other. 
Fig.~7 of \citet{lt78} is instructive. It shows a sequence of population
synthesis models in the two-color ($U$ - $B$) {\it vs.} ($B$ - $V$) diagram,
with ages of $10^{10}$ years and star formation histories ranging from initial 
burst to constant with time. The effects of age, metallicity and absorption,
and even changes in the IMF, shift the models in very similar directions! 

\citet{wvdka86} were the first to undertake a systematic survey of the
luminosity, color and \HI\ distributions in a well-defined set of spiral
galaxies. The surface photometry was based on photographic plates and,
although the data did show color gradients, \citet{wev84} was not
sufficiently confident to conclude that these were significant. In 
hindsight this was not justified: a
detailed comparison by \citet{kgb87} with later CCD-photometry of \citet{kent87}
for three systems showed deviations of at most 0.2 magnitudes in the radial
profiles down to 26 $r$-magnitudes arcsec$^{-2}$.  Although common wisdom
holds that old photographic surface photometry is not reliable, at least
some of it certainly is.

A comprehensive study of the broadband optical and near-infrared colors
in a sample of 86  disk galaxies was performed by 
\citet{djvdk94} and \citet{dj96a,dj96b,dj96c}. These studies 
established the existence of
color gradients both within and among galaxy disks, fainter surface
brightness systematically corresponding to bluer colors. It was also found 
that the degeneracies between dust absorption, stellar age and metallicity
can be broken to a large extent by
use of a set of photometric bands from the blue ($B$-band) to the near
infrared ($K$) and it was concluded 
from 3D radiative transfer models that dust extinction cannot be the major 
cause of the observed  gradients. The color gradients 
must be the result of significant
differences in star formation history, whereby the outer regions are
younger and of lower metallicity than the central parts. The lack
of suitable stellar population models made it impossible to quantify the
trends, although the extreme variations predicted by the models of
\citet{rl76} seemed outside the range of possibilities offered by the
observed color gradients. 

\citet{pdg98} used $(I - K)$ colors in edge-on galaxies away from the
central planes to derive a dust-free near-IR color-magnitude relation for spiral
galaxies. The slope of this relation is steeper for spirals than for
elliptical galaxies.
This is most likely not a result of vertical abundance
gradients, but of average age with height. The surprising thing
is that the scatter in this relation is small, possibly even only due to
observational uncertainty. Average stellar age must be an important 
contributor to variations in broadband colors. 

\citet{bdj00} made an important step forward by using maximum-likelihood
methods to match observed colors with stellar population synthesis
models, using simple star formation histories. These showed that spiral
galaxies almost all have significant gradients in the sense that the
inner regions are older and more metal rich than the outer regions. The
amplitude of these gradients is larger in high-surface brightness
galaxies than in low-surface brightness ones, and the progress of 
star formation (as evidenced by  decreasing age and increasing
metallicity) depends primarily on the surface brightness (most clearly
in the $K$-band) or surface density. The local surface density seems to
shape the star formation history in a disk more strongly than the
overall mass of the galaxy.

These models can also be used to derive values and gradients of the
mass-to-light ratio $M/L$ in and among disks. This was done by
\citet{bdj01} under the assumption
of a universal initial mass function (IMF). They conclude that their relative
trends in $M/L$ with color are robust to uncertainties in the stellar
populations and galaxy evolution models. Corrections for dust extinction
are not critical in the final determination of the stellar masses. They
also find that limits on the 
$M/L$ ratios derived from maximum disk fits to rotation
curves \citep[for galaxies in the Ursa Major cluster
by][]{verh01, vs01} 
match their $M/L$'s well, providing support for the universality
of the  IMF and the notion that at least some high surface brightness galaxies
are close to maximum disk. The variations in $M/L$ span a factor between
3 and 7 in the optical and about 2 in the near-infrared. 

The IMF provides the normalisation of the $M/L$ 
through the numbers of low-mass stars, but
the slope of the relation between color and $M/L$ is largely independent
of what models are used or what IMF is adopted \citep{djb09}. The
Salpeter IMF gives too massive a normalisation \citep{bdj01}, which can
be remedied by using a `diet' Salpeter IMF \citep[i.e. deficient in
low-mass stars such that it has only 70\%\ of the mass for the same
color;][]{bdj01}, or adopting an IMF that is itself more deficient in
low-mass stars \citep{ken83, krou01, cha03}. 
\citet{krou02a} has rather convincingly  argued that the IMF is universal to the
extent that its variations are smaller than would follow from the
expected varying conditions on the basis of elementary considerations.
{ \citet{bcm10} have recently concluded
that ``there is no clear evidence that the IMF varies strongly and 
systematically as a function of initial conditions after the first 
few generations of stars''.}

Default models, produced by adopting a
declining star formation rate, the population synthesis models of \citet{bc03} 
and the IMFs listed above, give consistent estimates of $M/L$ \citep{djb09}.
In fact, the $M/L_{I}$ values implied in fig.~(\ref{fig:7_histogram}) 
on the top-axis (derived for an axis ratio of the velocity ellipsoid of 0.6)
are 0.2 dex lower than from \citet{bdj01} but, as \citet{djb09} point out,
the axis ratio of the velocity ellipsoid scales with the {\it square} of 
$M/L$. 
The conclusion is that the determination of mass-to-light ratios from
broadband colors is reliable and robust in a relative sense, but that
there are still some uncertainties in the normalisation resulting from
imprecise knowledge of the faint part of the IMF.

\subsection{Global Stability, Bars and Spiral Structure}

Local stability of stellar disks has already been discussed in relation to
local stellar velocity dispersions, Toomre's $Q$ and the secular evolution
(`heating') of disks. We will say a few words here about global stability,
bars in galaxies and spiral structure. Much of these subjects has been covered
recently in the reviews, such as that on dynamics of galactic disk by
\citet{sell10} and for 
the case of bars in relation to pseudo-bulges by \citet{kk04}.  

Global stability of disks has been a subject ever since numerical simulations
became possible, starting about 1970 \citep[\eg][]{mpq70, hohl71}. Criteria for
stability were formulated empirically by \citet{op73} and \citet{eln82}.
In the latter criterion the halo stabilizes the disk; the criterion is in
terms of `observables'
\begin{equation}
Y=V_{\rm rot} \left( {h \over {G M_{\rm disk}}} \right) ^{1/2} \simgt
1.1,
\label{eqn:eln}
\end{equation}
where the disk is assumed exponential with scalelength $h$ and total mass
$M_{\rm disk}$. Since the rotation velocity $V_{\rm rot}$ is related to the
total mass, it is a criterion that relates to the relative mass in disk and
halo. It can be rewritten to say 
that within the radial distance from the center corresponding to the edge of
the disk, the dark matter halo
contains up to 60--70\%\ of the total mass \citep{vdkf86}. Such galaxies are in
fact sub-maximal. \citet{sell10} concludes that these criteria are only
necessary for disks that have no dense centers, since central concentrations
of mass in disks themselves could also provide global stability. It was shown
already some decades ago \citep{kal87} that halos are not very efficient in
stabilising disks acompared to budges.

\begin{figure}[t]
\begin{center}
\includegraphics[width=160mm]{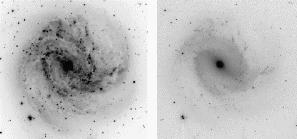}
        \caption{M83 in blue light at the left and in the $K$-band on the
          right. The bar is much more obvious in the near-IR. 
(Unpublished images by Park \& Freeman).
}
\label{fig:M83}
\end{center}
\end{figure}

We will not discuss the formation of bars in galaxies, as this subject has
been covered in detail by \citet{kk04} in
relation to pseudo-bulges,  and by \citet{sell10}. We do want to stress the
fundamental point that the incidence of bars is much larger than traditionally
thought; a typical fraction that figured in previous decades --although
admittedly for strongly barred galaxies as in \citet{san61}-- was of the order
of a quarter to a third. Current estimates are much higher; 
\citet{sheth08} found in the COSMOS field that in the local Universe
about 65\%\ of luminous spiral galaxies are barred. This fraction is a strong
function of redshift, dropping to 20\%\ at a redshift of 0.8. The {\it 
Spitzer Survey of Stellar Structure in Galaxies} S$^4$G \citep{sheth10} aims
among others at studying this in the near IR.
As an example, we show in fig.~\ref{fig:M83} a blue and near-IR image of the
large spiral M83. Although it appears mildly barred in the optical, it is
clear that in $K$-band the bar is very prominent and extended. 

Throughout the previous century much attention has been paid to the matter 
of the formation and 
maintenance of spiral structure. It was extensively reviewed by \citet{toom77, toom81}. 
Spiral structure in itself is unquestionably an
important issue (see the quote to Richard Feynman in the introduction in 
Toomre's review), as it is so obvious in galaxy 
disks and appears to play a determining role in the evolution of disks through
the regulation of star formation and therefore the dynamical, photometric and
chemical evolution. 
We will not discuss theories of spiral structure itself as
progress in this area has recently been somewhat slow. 
We refer the reader to the contributions of \citet{kn79}, \citet{sc84}, 
\citet{eel03} and \citet{sell08,sell10,sell10b}. Spiral structure is often
related to gravitational interaction between galaxies;
for interactions and subsequent merging see the work of \citet{tt72}, 
\citet{schwei86} and \citet{bar88}.

\subsection{The Flatness of Disks}

The inner disks of galaxies are often
remarkably flat. For the stellar disks this can be studied in
edge-on systems by
determining the centroid in the direction perpendicular to the major
axis at various galactocentric distances \citep[\eg][chapter 5]{ssbf90,
fpbmss91, dg97}. These studies were aimed at looking for warps in the
outer parts of the stellar disks (see below), but it is obvious from the
distributions that in the inner parts the systematic deviations are
very small.

\begin{figure}[t]
\begin{center}
\includegraphics[width=160mm]{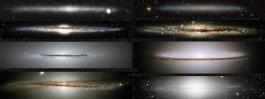}
        \caption{Selected images of edge-on disks and dustlanes
from various public Web-galleries.
Top: `Superthin' galaxies  IC 5249 \citep[from the 
{\it Sloan Digital Sky Survey},][]{vdkjvkf01}
and UGC 7321 (cosmo.nyu.edu/hogg/rc3/UGC\_732\_irg\_hard.jpg);
second row: NGC 4565 
(www.cfht.hawaii.edu/HawaiianStarlight/AIOM/English/2004/Images/Nov-Image2003-CFHT-Coelum.jpg) 
and NGC 891 (www.cfht.hawaii.edu/HawaiianStarlight/Posters/ 
NGC891-CFHT-Cuillandre-Coelum-1999.jpg);
third row: NGC 5866 (heritage.stsci.edu/ 2006/24/big.html) and
M104 (heritage.stsci.\-edu/2003/28/big.html);
bottom row: ESO 510-G013 (heritage.stsci.edu/2001/23/big.html) and
NGC 7814 
(www.cfht.hawaii. edu/HawaiianStarlight/English/Poster50x70-NGC7814.html).
}
\label{fig:edge-ons}
\end{center}
\end{figure}

The evidence for the flatness of stellar disks is more compelling when
we look at the flatness of the layers of the ISM. First look at the
dustlanes. In fig.~\ref{fig:edge-ons} we collect some images of
edge-on disk galaxies. At the top are two `super-thin' galaxies (which
we discuss further in \S~\ref{sect:superthin});
the disks are straight lines to within a few percent. The same holds for
the dustlanes in NGC 4565 (allow for the curvature due to its imperfectly
edge-on nature) and  NGC 891. Again the dustlanes indicate flat layers
to a few percent. In the third row the peculiar structure of NGC 5866
has no measurable deviation from a straight line, while for the Sombrero
Nebula the outline of the dustlane  fits very accurately to an ellipse. 
In the bottom row, NGC 7814 (right) is straight again to within a few
percent, but NGC 5866 is an example of a galaxy with a large warp in the
dust layer.

The \HI\ kinematics provide probably the strongest indications for
flatness. In three almost completely face-on spirals (NGC 3938, 628 and
1058), \citet{vdksh82, vdksh84} and \citet{shvdk84} 
found that the residual velocity
field after subtraction of that of the systematic rotation shows no
systematic pattern and had an r.m.s. value of only 3 - 4 \kms. So,
vertical motions must be restricted to only a few \kms (a few pc per Myr). 
Then, even a vertical oscillation with a period equal to the
typical time of vertical oscillation of a star in the Solar Neighborhood
($10^7$ years) or of the rotation of the Sun around the Galactic Center
($10^8$ years) would have an amplitude of only ten to a hundred pc.
Again, this is of order a few percent or less of the diameter of a galaxy like
our own. The absence of such residual patterns shows that the \HI\
layers and, since by implication they are more massive, 
the stellar disks must be
extraordinarily flat, except maybe in their outer regions. This
obviously does not hold for galaxies that are or have recently been in
interaction. 

Recently, \citet{mu08a, mu08b} found evidence for a pattern of corrugation
in the disk of the edge-on galaxy IC 2233. The excursion of the plane
is most pronounced in younger tracer populations, such as \HI\ or Young
stars. The amplitude is up to 250 pc (compared to a radius of order 10
kpc). The older disk shows much less of an effect. IC 2233
is a relatively small galaxy (rotation velocity about 100 \kms) and
has extensive star formation;  it appears that the effect is related
to the process of star formation. 

\subsection{`Superthin' Galaxies}
\label{sect:superthin}

We have indicated above that the flattening of the stellar disk 
$h_{\rm z}/h$ is smallest for systems that are of late Hubble type,
small rotation velocity and faint (face-on) surface brightness. It is of
interest then to look more closely at systems at this extreme end of 
the range of flattening; such systems are referred to as `superthin'. Of
course, the ones we can identify are seen edge-on. A prime example is the
galaxy UGC 7321, studied extensively by \citet{mgvd99}, \citet{mat00}, 
\citet{mw03}, \citet{um03} and \citet{bmj10}. 
This is a very low surface brightness galaxy (its face-on $B$-band 
central surface brightness would be $\sim$23.4 mag arcsec$^{-2}$) with a
scalelength of about 2 kpc, but a projected vertical scaleheight of only
150 pc. There is evidence for vertical structure: it has a color
gradient (bluer near the central plane) and appears to consist of two
components. Rotation curve analysis \citep{bmj10, obfvdk10d} 
indicates that it has a large amount
of mass in its dark matter 
halo compared to the luminous component. Its \HI\ is
warped in the outer parts, starting at the edge of the light
distribution. Extended \HI\ emission is visible at
relative high $z$ (more than 2 kpc out of the plane). \citet{pbld03}
argue that the deviation in the light profile in the central regions
and the shape of the isophotes point at a presence of a large bar. 

Another good example of a superthin galaxy is IC 5249 \citep{byun98,
abc99, vdkjvkf01}. This also is a low surface brightness galaxy with
presumably a small fraction of the mass in the luminous disk. However,
the disk scaleheight is not small (0.65 kpc). It has
a very long radial scalelength (17 kpc); its faint surface brightness 
$\mu_{\circ}$ then causes only the parts close to the plane to be easily
visible against the background sky, while the long radial scalelength
assures this to happen over a large range of $R$. Therefore it appears
thin on the sky. The flattening $h_{\rm z}/h$  is 0.09 (versus 0.07 for
UGC 7321). The stellar
velocity dispersions are similar to those in the Solar Neighborhood;
disk heating must have proceeded at a pace comparable to that in the
Galaxy. 

The flattest galaxies appear not only very flat on the sky, but have indeed
very small values of $h_{\rm z}/h$. However, these two examples show
that the detailed structure may be different. Superthin galaxies do
share the property of late type, faint face-on surface brightness and
small amounts of luminous disk mass compared to that in the dark 
matter halo. 

\citet{kau09} has reviewed the observations of `flat and superthin'
galaxies, especially in view of the fact that these late-type, bulgeless
systems present challenges to models of disk galaxy formation within
the hierarchical growth context of $\Lambda$CDM. These pure
disk systems have low surface brightness blue structures with low
angular momenta, that may have formed with a lower frequency of  merging
events than disk galaxies with bulges and thick disks. 
In large and giant galaxies the question of the frequency of the
presence of a `classical' bulge has been addressed by \citet{kdbc10}.
They find that giant, pure disk galaxies are far from rare and their
existence presents a major challenge to formation pictures with histories of
merging in an hierarchical clustering scenario.

\subsection{Warps in Stellar Disks}
\label{sect:warps}
 
In their outer parts, stellar disks have deviations from both the plane of
the inner parts (warps) as well as deviations from the extrapolated
exponential surface brightness distributions (truncations). We will discuss 
these phenomena in turn.

First we turn to warps in the outer parts of stellar disks. Studies
referred to above \citep{ssbf90,fpbmss91,dg97,rbcjv02} have indicated
that most, if not all, disks display warps in their very outer
parts, often up to 0.5$h_{\rm z}$ or more. Recently, \citet{sdjh09} have
studied edge-on galaxies observed with the {\it Spitzer Space
Telescope} in the 4.5$\mu$ band.\footnote{This paper contains also a rather
complete inventory of publications concerning warps; optical, near-IR as
well as \HI}. Out of 24 galaxies they found evidence for warps in 10.  The 
radius of the onset of the warp indicates that
there must also be a moderate amount of flaring, in order to match the
response to the indicated mass distribution from the light distribution
and rotation curve. The warp onset is asymmetric and the more so in
small scalelength systems. The reason for this is not clear, but could
point to asymmetries in the dark matter distribution. The warp profiles 
shown in their figures reinforces the point made above about the flatness of disks; 
in the inner parts the deviations from a straight line are exceedingly 
small (only a percent or less of the radial extent). Theoretical
work related to warps and dynamics in stellar disk has recently been 
reviewed by \citet{sell10}, in the context of collective global
instabilities, bending waves, bars and spiral structure. 

Sometimes optical warps are very pronounced, such as in the so-called 
`Integral Sign' galaxy UGC 3697 \citep{bbs67, ann07}. There have recently been
a number of statistical studies \citep[\eg][]{sd01, ap06} from large
samples, that contain more and less isolated systems. 
The conclusions are that strong warps are probably all a result of interactions,
while at least a fraction may arise from accretion of gaseous material.
An important point to note is that even isolated galaxies show signs of
accretion. Beautiful examples have recently been presented in
much detail, including NGC 5907 \citep{mbh94, md08}, NGC 4736 \citep{ti09},
NGC 4013 and NGC 5055 \citep{md09}. In NGC 5055 the brightest part of the
faint loops have been registered also in the photographic surface photometry
of \citet{vdk79} in two colors; it appeared definitely red and presumably
dominated by older stars. These relatively
isolated systems appear to show signs for recent accretion events, which therefore 
must be common. Of course, much is known now about substructure in the halo
of our Galaxy \citep{ah08} and M31 \citep{af07} and the evidence for continuing
accretion that this provides, but that is beyond the scope of this review.

\subsection{Truncations}
\label{sect:truncations}

\begin{figure}[t]
\begin{center}
\includegraphics[width=160mm]{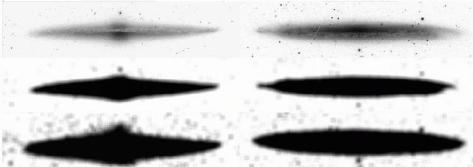}
\caption{NGC 4565 and NGC 5907 at various light levels. These have been
produced from images of the {\it Sloan Digital Sky Survey}, which were clipped at 
three different levels (top to bottom) and turned into two-bit
images and subsequently smoothed  
\citep[see][for an explanation of the details]{vdk07}. Note that the
disks grow significantly along the minor axes but not in radial extent.}
\label{fig:truncations}
\end{center}
\end{figure}

Truncations in stellar disks were first found in edge-on galaxies, where the
remarkable feature was noted that the radial extent did not grow 
with deeper photographic exposures \citep{vdk79}. Especially,
when a bulge was present the minor axis did grow considerably
on IIIa-J images compared to the Palomar Sky Survey IIa-D exposures in
contrast to the major axes that did not grow at all. Detailed surface
photometry \citep{vdks81a,vdks81b} confirmed the presence of these truncations 
in the four brightest, edge-on, disk-dominated galaxies in the northern sky, 
NGC 891, 4244, 4565 and 5907. For the last two we illustrate 
this phenomenon of truncation in  fig.~\ref{fig:truncations}. 
The truncations appear very sharp, although of course not 
infinitely.\footnote{In fact, the statement in  \citet{vdks81a} reads:
{\it ``This cut-off is very sharp with an e-folding of less than about
1 kpc''}, based on the spacing of the outer isophotes.}
Sharp outer profiles are actually obtained
after deprojecting near-IR observations of edge-on galaxies
\citep[\eg][]{flo06}. \citet{fry98}, using CCD surface
photometry, and \citet{djetal07b}, from HST star counts, show
that the disk of NGC 4244 has a sharp truncation,
occurring over only about 1 kpc.

Various models have been proposed for the origin of truncations.
In the model by \citet{rl76}, the truncations are the current extent of 
the disks while they are growing from the
inside out from accretion of external material.
This predicts larger age gradients across disks than are observed 
\citep{dj96b}. Another possibility is that
star formation is inhibited when the gas surface (or space?) density
falls below a certain threshold for local stability \citep{fe80,ken89,js04}.
The Goldreich--Lynden-Bell criterion for stability of gas layers
gives a poor prediction for the truncation radii \citep{vdks82a}.
Another problem is that the rotation curves of some galaxies, \eg\ NGC 5907
and NGC 4013 \citep{sc83,bot96}, show features
near the truncations that indicate that the {\it mass}
distributions are also truncated.
\citet{js04} predicts an anti-correlation between $R_{\rm max}/h$ and $h$,
which is not observed.

Obviously, the truncation corresponds to the maximum 
in the specific angular momentum distribution of the present disk, 
which would correspond to that in the protogalaxy \citep{vdk87}
if the collapse occurs with detailed conservation of specific
angular momentum \citep{fe80}. As noted above, if the protogalaxy starts 
out as a  \citet{mes63} sphere with uniform density and
angular rotation in the force field of a dark matter 
halo with a flat rotation curve, a roughly exponential disk
results. This disk has then a truncation at about 4.5 scalelengths, so 
this hypothesis provides at the same time an explanation for the exponential 
nature of disk as well as for the occurrence of the truncations.
On the other hand it is possible that substantial redistribution of angular
momentum takes place, so that its distribution now is
unrelated to the initial distribution in the material that formed the disks.
Bars may play an important role in such redistribution, as suggested by \citet{dmcmwq06} and \citet{erw07}.
In fact a range of possible agents in addition to bars, such as density waves,
heating and stripping of stars by bombardment of dark matter sub-halos, has
been invoked \citep{djetal07b}.
\citet{rdsqkw08,rdqsw08} have studied the origin of truncations or breaks
in the radial distributions in stellar disks as related to a rapid drop in star
formation and include the effects of radial migration of stars.
Observations of stellar populations and their ages in the regions near
the truncation \citep{yrd10} have been used to provide evidence that migration of
stars is a significant phenomenon in the formation and evolution of
stellar disks. Finally, there are models
\citep{bat02,flo06} in which a magnetic force breaks
down as a result of star formation so that stars escape.
The evidence for sufficiently strong magnetic fields
needs  strengthening.

\begin{figure}[t]
\includegraphics[width=160mm]{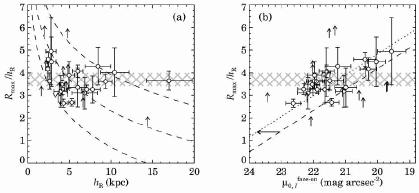}
\caption{Correlations of $R_{\rm max}/h$ with scalelength $h$ and
face-on central surface brightness $\mu _{\circ,{\rm fo}}$
for a sample of edge-on galaxies.
 The cross-hatched regions show the prediction from
a collapse model as in  \citet{vdk87} and \citet{dal97}; 
the dotted and dashed lines show 
predictions from the star formation threshold model of \citet{js04} for 
three different values of the disk mass. \citep[from][; see there for 
details of the models.]{kvdk04b}}
\label{fig:truncs}
\end{figure}

\citet{kvdk04b} derive correlations of the ratio of the cut-off 
radius $R_{\rm max}$ to the radial
scalelength $h$ with $h$ itself and with the face-on central surface
brightness $\mu _{\circ,{\rm fo}}$ (fig.~\ref{fig:truncs}).
$R_{\rm max}/h$ does not depend strongly on $h$, but is somewhat less
than the 4.5 predicted from the collapse from a simple Mestel-sphere.
There is some correlation between $R_{\rm max}/h$ and $\mu _{\circ, {\rm fo}}$,
indicating approximately constant disk surface density at the truncations,
as possibly expected in the star-formation threshold model.
But this model predicts an anti-correlation between $R_{\rm max}/h$ and
$h$ \citep{js04}, which is not observed.
The maximum angular momentum hypothesis predicts that
$R_{\rm max}/h$ should not depend on $h$ or $\mu _{\circ, {\rm fo}}$ and
such a model therefore requires some redistribution of angular momentum in the
collapse or somewhat different initial conditions.

The truncation in the
stellar disk in our Galaxy has been identified using star counts in a 
number of surveys \citep{rcm92,rrecbfg96} at a galactocentric radius
of 14 to 15 kpc.  Typical values for the ratio between the truncation radius
and the radial scalelength $R_{\rm max}/h$ are 3.5 to 4 (see 
fig.~\ref{fig:truncs}), so that the Galaxy's scalelength is expected on this
basis to be 3.5 to 4 kpc. 

Due to line-of-sight integration, truncations should more difficult to
detect in face-on galaxies than in edge-on ones.
The expected surface brightness at 4 scalelengths is
about 26 $B$-mag arcsec$^{-2}$, only a few percent of the dark sky.\footnote{
This surface brightness is close to that often associated with
the `Holmberg diameters' \citep{holm58}, which are often assumed to be 
diameters at 26.5 $B$-mag arcsec$^{-2}$ and corrected for inclination.
For a discussion of the history of Holmberg radii, see the appendix in
\citet{vdk07}.  Contrary to common belief, they are defined in terms of
photographic density (rather than a well-defined surface brightness), in two
bands (photographic and photovisual rather than the $B$-band) and \underline{not}
corrected for inclination.\label{footnote:Holmberg}} 
In face-on
galaxies like NGC 628 \citep{shvdk84,vdk88} an isophote map shows that 
the outer contours have a much smaller spacing than the inner ones.
The usual analysis uses an inclination and major axis determined from
kinematics if available, (otherwise these are estimated from the
average shape of isophotes) and then determines an azimuthally averaged
radial surface brightness profile. But this will smooth out any truncation
if its radius is not exactly constant with azimuthal angle.
At this level spiral disks are indeed often lopsided, as is seen
from the $m=1$ Fourier component of surface brightness 
maps \citep{rz95,zr97}, presumably as a result of interactions and merging. 
The effects are nicely illustrated in the study of NGC 5923
by \citet{pdla02} (their fig.~9), which has isophotes in polar coordinates.
The irregular outline shows that some smoothing will occur contrary
to observations in edge-on systems. Unless special care is taken we will always find a (much) less sharp truncation in face-on than in edge-on systems.

{ Although we will not discuss oval distortions of disks here because they
were reviewed by \citet{kk04}, we emphasize that they are potentially important
for studies of truncations and are also of intrinsic interest. Oval distortions 
in unbarred galaxies can have significant kinematical and dynamical effects.}  

\citet{pt06} studied a sample of moderately inclined systems
through ellipse-fitting of isophotes in SDSS data.
They distinguish three types of profiles:
{\it Type I}: no break;
{\it Type II}: downbending break;
{\it Type III}: upbending break.
\citet{pzpd07} have reported that the same profiles occur in edge-on
systems; however, of their 11 systems there was only one for each of 
the types I and III. 

Various correlations have been reviewed by \citet{vdk09}. In general, the
edge-on and face-on samples agree in the distribution of $R_{\rm max}/h$; 
however the fits in moderately inclined systems result in small values of the 
scalelength. A prime example of a Type III profile in \citet{pt06} is 
NGC 3310, which is a well-known case of an merging galaxy \citep{vdk76, ks01}.
{ In fact, on close examination of their images at faint levels, many of the Type III 
systems show signs of outer distortions, presumably due to interactions.}

There is a good correlation between the truncation radius $R_{\rm max}$ 
and the rotation velocity \citep{vdk08}. On average a galaxy like our own 
would have an $R_{\rm max}$ of 15 - 25 kpc (and a scalelength of 4 - 5 kpc). 
It is of interest to compare this correlation to the case of NGC 300
\citep{bh05}, which has no truncation even at 10 scalelengths from
the center ($R_{\rm max} \gt$ 14.4 kpc), and therefore is an example of a type I
disk in the terminology of \citet{pt06}. Despite showing no sign of truncation
down to a very faint surface brightness level, we note that its lower limit
to $R_{\rm max}$ is still consistent with the observed correlation between
$R_{\rm max}$  and $V_{\rm rot}$ in edge-on systems (NGC 300 has a rotational velocity 
of $\sim105$ \kms, which would give an $R_{\rm max}$ of 8 - 15 kpc and an $h$ of 2 - 4 
kpc ). NGC 300 could be interpreted as having an unusually small $h$ for its 
$V_{\rm rot}$ rather than an unusually large $R_{\rm max}$ for its scalelength. 
 
These examples show that at least some of the type III galaxies could  arise 
from the effects of interactions and merging, and type I systems could at least 
partly be disks with normal truncation radii, but large $R_{\rm max}/h$ and small
scalelengths, so that their truncations occur at much lower surface brightness.

\subsection{Nuclei of Pure-Disk Galaxies}

Late type pure disk galaxies are commonly nucleated, with central nuclear 
star clusters { \citep[e.g. M33: ][]{kmc93}}. For a sample studied by \citet{wal05}, the dynamical masses of the nuclear clusters are in the range $8\times10^5$ to $6\times10^7$ \msun. These star clusters usually lie within 
a few arcsec of the isophotal centers of the galaxies \citep[\eg][]{bok02}.
{ How are the nuclei able to locate accurately the centers of 
the apparently shallow central potential wells of their exponential disks?  
The reason may be that the center of the gravitational field of an exponential disk is 
actually well-defined: the radial gradient of its potential does not vanish at its 
center, so the force field defines the center of the disk to within a fraction of the 
scaleheight of the interstellar medium of the disk, which is of order 100 pc.}

Structurally the nuclear star clusters are much like Galactic globular 
clusters \citep{bok04}. Their stellar content is however very different. 
The light of the nuclear star clusters is typically dominated by a 
relatively young star population \citep{ros06, kdbc10}, but the young 
population provides only a few percent of the stellar mass. They have 
an underlying older population with an extended history of episodic star 
formation \citep{wal06}. This episodic star formation may come from gas 
funnelled in to the center of the galaxy by local torques. 

AGNs are rare or absent from these nuclei of pure disk galaxies 
\citep{sat09}. For the nucleus of the nearby system M33, with a total 
nuclear mass of about $2\times10^6$ \msun, \citet{geb01} were able to 
derive an upper limit of $1500$ \msun\ for a supermassive black hole within 
the nuclear star cluster.

These nuclear star clusters, { with their episodic and
extended star formation history, are interesting in their possible relation 
to some of the Galactic globular clusters, such as the massive cluster 
$\omega$ Centauri which also shows evidence of an extended history of 
episodic star formation \citep[\eg][]{bel10} and an inhomogeneous 
distribution of heavy element abundances. This is unusual in globular 
clusters.}  Based on chemical evolution arguments, \citet{sz78} proposed that the 
Galactic globular clusters originated in small satellite galaxies which were accreted 
long ago by the Milky Way. The small galaxies are 
tidally disrupted but the globular clusters survive.  \citet{fre93} and 
\citet{bok08} argued that { at least some of} the globular clusters may 
have been the nuclei of such satellite systems.

\section{HI DISKS}

\subsection{HI Distributions, Kinematics and Dynamics}

The study of the distribution of \HI\ in samples of { more than a few}
disks in galaxies has been possible only since the advent of aperture
synthesis measurements of the 21-cm line. Early observations with single disk
instruments could be made only for the very nearest systems,
notably the Andromeda Nebula \citep[in particular][]{rw75}. Observations
with the necessary angular resolution started with the Owens Valley
Two-Element Interferometer \citep[\eg][]{rs71}, the Half-Mile Telescope
at Cambridge \citep[\eg][]{bfww71} and the Westerbork Synthesis Radio
Telescope \citep[\eg][]{agvw73}. This early work up to about 1977 has
been reviewed in \citet{vdka78}, although mostly in the context of
kinematics.

These studies revealed distributions of the \HI\ in most cases to be
much more extended than the stellar disks and often warped away from the
plane of the stellar disk beyond the boundaries of the light
distribution, both in edge-on \citep{san76} and moderately
inclined systems \citep{rs71, bos81a,bos81b}. The most important finding was
that  
the rotation curves at these radii remained flat (see 
\S~\ref{sec:history}). 

\begin{figure}[t]
\begin{center}
\includegraphics[width=160mm]{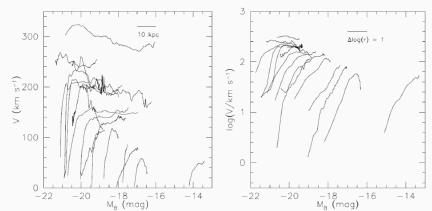}
        \caption{The rotation curves from {\it The \HI\ Nearby Galaxies
            Survey} THINGS, plotted in linear units in the left panel and 
in logarithmic units in the right panel. The origin of the rotation curves 
has been shifted according to their absolute luminosity as indicated on the 
horizontal axis { such that that the innermost point of each
rotation curve is plotted at the $M_B$ value of the galaxy}. 
The bar in the respective panels indicates the radial scale.
\citep[From ][]{blok08}.}
\label{fig:rotcurves}
\end{center}
\end{figure}

Since then many observations of disk galaxies have been taken. The first 
extensive survey, including comparison with optical surface photometry, was 
done by \citet{wvdka86}. More recently the extensive {\it Westerbork \HI\ Survey 
of Spiral and irregular
galaxies} WHISP \citep{ksva96, vdhvas01, nvdhssva05, svavdhs02, swb02, grsk02,
nvdhssva07} was made. The most advanced survey at this stage is {\it The \HI\
Nearby  
Galaxies Survey} THINGS \citep{things08,blok08}, which provides very
high-resolution \HI\ maps and rotation curves and has been analysed by
comparison with 3.6 $\mu$m data from the {\it Spitzer Infrared Nearby 
Galaxies Survey} SINGS. Fig.~\ref{fig:rotcurves} shows the rotation curves
from THINGS. The details of the rotation curves correlate with the
absolute magnitude of the galaxies, as was first described by 
\citet{bro92}. Luminous galaxies have
rotation curves that rise steeply, followed by a decline and
an asymptotic approach to the flat outer part of the curve;
low-luminosity galaxies show a more gradual increase, never
quite reaching the flat part of the curve over the extent of their
\HI\ disks. { Although in some curves the rotation velocity does decrease
at large radii,} none of the { galaxies shows an decline in their rotation curves
that can unambiguously be associated with a cut-off in the mass distribution
so that in no case has the rotation curve been traced to the limit of
the dark matter distribution}
\citep{blok08}.

We comment on a number of properties of the surface density distribution
of the neutral hydrogen first. The total content, especially relative to
the amount of starlight $M_{\rm HI}/L$ (a property that is distance
independent) is well-known to correlate with morphological type 
\citep[\eg][]{rh94}, while in later types the more luminous galaxies
contain relatively less \HI\ \citep{vs01}. 
The \HI\ diameter compared to the optical diameter (at 25 $B$-mag
arcsec$^{-2}$) is about 1.7 with a large scatter, but does not depend on 
morphological type or luminosity, according to \citet{br97}. However,
there are very good correlations between \HI-mass and - diameter, 
$\log M_{\HI}$ and $\log D_{\HI}$, both with slopes of about 2. 
This implies that the \HI\ surface density averaged over the whole \HI\ disk
is constant from galaxy to galaxy, independent of luminosity or type.
Also note that there is a relatively well-defined maximum surface density 
in disks of galaxies observed (at least with resolutions of the current 
synthesis telescopes), which amounts to about 10 $M_{\odot}  {\rm pc}^{-2}$ 
\citep{wvdka86} .

Many systems have asymmetries in their \HI-morphologies or -kinematics.
Often this is in the form of lopsidedness in the surface density distributions. 
This can to some extent already be seen in the asymmetries in integrated 
profiles;  studies have claimed that up to 75\%\ of galaxies are asymmetric or 
lopsided \citep{mvdg98, hvzhrm98, sssva99, nvdhssva05} at some detectable
level.  
Lopsidedness appears to be independent of whether or not the galaxy is
isolated,  
so interactions or mergings 
cannot always be invoked as its origin. It is suggested that it is either
an intrinsic property of the disks or is induced by asymmetries in the 
dark matter distribution \citep{nsl01}.

The extraction of kinematic data from the raw observations of moderately 
inclined systems has often been discussed. 
The basics are summarized in
\citet{vdka78}, \citet{bos81a}, \citet{wvdka86} and \citet{kgb87}. 
The results are a radial distribution of \HI\ surface
brightness, a rotation curve and a radial distribution of velocity
dispersion; from these one can in principle make maps of residuals
compared to these azimuthal averages. The derivation of the \HI\ velocity
dispersions is easiest in face-on spirals, where there is no gradient
in the systematic motions across a telescope beam \citep{vdksh82, vdksh84,
shvdk84}. In moderately inclined systems it is possible to measure the 
velocity dispersion, provided that the angular resolution 
and signal-to-noise are sufficiently 
high and that the effects of \HI\ layer thickness can be separated from 
that of random motions in the observed profiles \citep[see][]{sick97}.

For edge-on systems the procedure is more complicated as a result of
the line-of-sight integration. Various methods have been devised,
initially only to derive the rotation curve and the radial distribution
of \HI\ surface brightness, later in some cases also the flaring
(the increasing thickness as a function of galactocentric radius) 
of the \HI\ layer  \citep[\eg][]{sa79, vdk81, sof86, rbft85, mfb92,
ts02, grsk02, um03, kvdkdb04, kvdk04a}. 
Recently \citet{olling96a, olling96b} and 
\citet{obfvdkb10, obfvdk10b, obfvdk10c} 
have also fit for the \HI\ velocity dispersion.
The paper by \citet{obfvdk10b} provides a detailed description of the 
various methods and a discussion on the relative merits and pitfalls.

In general, for larger disk galaxies, the radial distributions show 
(sometimes in addition to a central depression) a radial surface density 
that falls off more slowly than that of the starlight, a rotation curve 
that rises to a maximum and stays at that level, and a velocity dispersion 
of 7 -- 10 km s$^{-1}$, often near the higher end of this range in the inner 
regions and in spiral arms and in the lower range in the outer parts and 
inter-arm regions \citep{shvdk84, vdksh84, sick97}. 

The velocity dispersion of the \HI\ can be measured only in
cases where there is a negligible gradient in the overall
radial velocity over the beam of the radio telescope. Early
determinations have been made in our Galaxy;  \citet[\eg][]{vw67} found 7
km s$^{-1}$ for the Solar Neighborhood and \citet{emer76} found $12\pm1$
\kms\ from aperture synthesis observations using the Half-Mile Telescope at
Cambridge. For larger angular size galaxies, the use of face-on galaxies
(as judged from the narrowness of their integrated \HI\ profiles) is
required to fulfill this condition. It was done in much detail on three
galaxies that are only a few degrees from face-on \citep[NGC 3938, 628 and
1058;][]{vdksh82,vdksh84,shvdk84}. They found that the
dispersions over the optical extent of the disks were rather constant, 
with no significant decline with radius at
$7-10$ km s$^{-1}$. Only in NGC 628 was a systematic difference seen betwen 
the inter-arm region (at the lower end of the range quoted) and the spiral
arms themselves (the higher end). 

The most recent study of galaxies { that are not very close to face-on,}
is {\it The \HI\
Nearby Galaxy Survey} (THINGS) by \citet{tamb09}, where references to
other earlier work can be found. They do find significant declines of
velocity dispersion with radius in their sample, but there appears a 
characteristic value of $10\pm2$ km s$^{-1}$ at the outer extent of the 
star-forming part of the disk.
Inward of this, the dispersion correlates with indicators of star
formation, suggesting that the supernovae associated with this star
formation activity are driving the turbulence, although magnetorotational
instabilities may add significantly to the random motions as well. In
edge-on galaxies, \citet{obfvdkb10} and \citet{obfvdk10b,obfvdk10c} made
detailed analyses to retrieve the radial distributions of \HI\ surface
density, rotation velocity and velocity dispersion, finding that there
is significant structure, but not much systematic radial decline. 
This study involved \HI\ rich, dwarf systems, quite dissimilar from the 
SINGS sample. Typical dispersions are 6.5 to 7.5 km s$^{-1}$, increasing with
the  amplitude of the rotation curve. { In the solar neighborhood the velocity
dispersion of OB stars is comparable to that in the \HI, so stars are born with 
that same amount of random motion which then increases as a result of secular 
evolution. The same will happen in other spirals, but in dwarf galaxies the 
stellar velocity dispersions are similar to the ones reported in \HI. This may 
suggest that no significant secular evolution of random motions occurs in 
dwarf systems.}

\begin{figure}[t]
\begin{center}
\includegraphics[width=78mm]{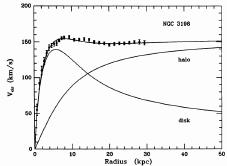}
\includegraphics[width=78mm]{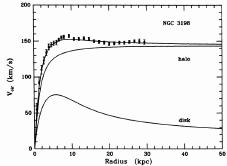}
        \caption{ The rotation curve decomposition for NGC 3198.
The shape of the curve for the dark matter halo is set by a core radius that determines
the initial rise and tends asymptotically to that of a mass density distribution 
that falls off $\propto R^{-\gamma}$ with $\gamma$ close to 2. The shape of the 
disk curve is that of the exponential disk with a scalelengh from the light 
distribution. On the left the `maximum disk' fit that maximizes the amplitude 
of the disk rotation curve and on the left an arbitrary fit with a disk mass 0.3
times that of the maximum disk. \citep[From][]{vabbs85}
}
\label{fig:maximum_disk}
\end{center}
\end{figure}

\subsection{Dark Matter Halos}
\label{sect:rotcur} 

The discovery of dark matter halos has been described very briefly in 
\S~\ref{sec:history}, where also the concept of the `maximum-disk hypothesis'
(see fig.~\ref{fig:maximum_disk})
was introduced, in which the contribution of the stellar disk to the rotation 
curve is taken to be as large as possible,  consistent with the observations
\citep{cf85,vabbs85,vas86}. As mentioned above, this contribution has in
practice  
an amplitude at its maximum within the range $0.85 \pm 0.10$ of the observed
maximum 
rotation, following \citet{sac97}. The evidence from stellar dynamics 
(see \S~\ref{sec:stellardyn}) is that the majority of galaxies 
have disk masses significantly below maximum disk, except for 
some galaxies with the highest surface brightness.

Recent reviews of disk masses in galaxies and implications for
decompositions of rotation curves into contributions from dark and
baryonic matter have been presented by \citet{vdk09} and 
\citet{mg09} at the Kingston symposium. These reviewers  
take the view that
most galaxy disk are sub-maximal, except possibly those  with the highest 
surface brightness and surface density. In contrast { to these studies that
involve mostly non-barred galaxies}, we note 
that \citet{wsw01} and \citet{pff04} find from detailed fluid dynamical gas flows 
in some barred galaxies that their disks are close to maximal. \citet{debs00}
argue from the observed rapid rotation of the bars of barred galaxies that
the stars are dominating the gravitational field in the inner
regions of these galaxies.

\begin{figure}[t]
\begin{center}
\includegraphics[width=130mm]{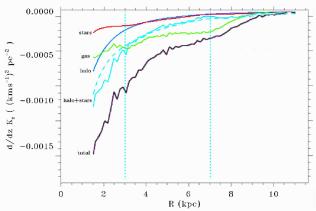}
        \caption{The vertical gradients of the force field for various
          components in the edge-on 
          galaxy UGC 7321. For an explanation see the text.  
The measured halo flattening was $q=1.0\pm0.1$. 
The vertical dotted (cyan) lines show the radial regime used for the fit.
\citep[From ][]{obfvdk10d}}
\label{fig:7321}
\end{center}
\end{figure}

The thickness of the gas layer in a disk galaxy can be used to measure the surface
density of the disk. It has been known for a long time 
that this layer in our Galaxy is `flaring' \citep[for a recent discussion see][]{kk09}.
Assume that the vertical density distribution of the exponential stellar disk is locally 
isothermal ($n=1$ in eqn.~(\ref{eqn:vdksdisk})).
If the \HI\ velocity dispersion is
$\langle V_{\rm z}^{2} \rangle _{\rm HI}^{1/2}$ and isotropic,
and if the stars dominate the gravitational field, then the \HI\ layer  has a
full width at half maximum (to \lsim 3\%) of 
\begin{equation}
W_{\rm HI} = {1.7 \langle V_{\rm z}^{2} \rangle _{\rm HI}^{1/2}} \left[
{z_{\circ}} \over {\pi G(M/L) \mu_{\circ }} \right] ^{1.2} e^{R/2h}.
\label{eq:flaring}
\end{equation}
So, if the \HI\ velocity dispersion is independent of radius, the \HI\ layer 
increases exponentially in thickness with an e-folding length $2h$.
This was first derived and applied to \HI\ observations of NGC 891 by
\citet{vdk81} to demonstrate that the dark matter
indicated by the rotation curve did indeed not reside in the disk but in a more 
spherical volume. Using the photometry in \citet{vdks81b} and the
observed \HI\ flaring \citep{sa79} 
and taking a gaseous velocity dispersion of 10 km
s$^{-1}$ resulted in a rotation curve of the disk alone with a maximum 
of $\sim140$ \kms. 
A smaller value for $\langle V_{\rm z}^{2} \rangle _{\rm
HI}^{1/2}$ would, according to eqn.~(\ref{eq:flaring}), result in a smaller
value for the inferred $M/L$ for the same \HI\ thickness
and therefore to a lower value for the maximum disk-alone rotation. The 
observed maximum rotation velocity is $225\pm 10$ km s$^{-1}$, which indicates
that NGC 891 is not maximum disk.  In a similar analysis, \citet{olling96b}
inferred for NGC 4244  
that the maximum disk-alone rotation velocity is between 40 and 80\%\ of the
observed rotational velocity. 

Another important property of dark matter halos for understanding galaxy
formation is their three-dimensional shape. There are various ways to address 
this issue, of which the use of the flaring of the \HI\ layer is the most prominent. 
It was first developed by \citet{olling95} and subsequently applied 
to the nearby edge-on dwarf galaxy NGC 4244 \citet{olling96a,olling96b}. He found 
the dark halo of this galaxy to be highly flattened.
{ Observations of the kinematics in polar ring galaxies provide estimates of the 
potential gradients in the two orthogonal planes of the galaxies. These galaxies can 
potentially give useful information about the shapes of their dark halos \citep{sac99}, 
although it is possible that special halos are needed to host these rare polar ring 
systems.  The resulting halo flattenings derived for polar ring galaxies and also for 
streams in the halo of the Galaxy \citep{ah04} range from a few tenths to unity.}

A more extensive survey of a sample of eight late-type, \HI-rich dwarfs
in which the dark halo appears to dominate the gravitational field even in the
disk, was undertaken by \citet{obfvdkb10}. The basic premise of the approach
is that the radial gradient of the dark matter halo force $\D K_{\rm R}/\D R$
can be 
measured from the rotation curve after correction for the contribution of the
stellar disk and its ISM, while the flaring of the \HI\ layer
together with a measurement of the velocity dispersion of the \HI\ provides 
a measure of the vertical gradient $\D K_{\rm z} /\D z$. 
The ratio of the two force gradients is related to the flattening of the
dark halo (assumed spheroidal), as measured by the axis ratio $c/a$. The
derivation of the necessary properties from the 
\HI\ observations have been presented in \citet{obfvdk10b,obfvdk10c}.  The method 
relies on two assumptions concerning the \HI\ velocity dispersion, that it is 
isotropic (since we measure in an edge-on galaxy the line-of-sight dispersion
component parallel to the plane and use in the analysis the vertical component) 
and isothermal with $z$. High signal-to-noise observations in edge-on and
moderately inclined galaxies are needed.

The first galaxy for which this analysis has been completed is UGC 7321
\citep{obfvdk10d}. The halo density distribution was modelled as a pseudo-isothermal 
spheroid \citep{srjf94} and the
disk as a double exponential determined from $R$-band surface photometry. After
allowing for the gravitational field of the gas layer, the hydrostatic equations
gives the $M/L$ of the disk and a value for the flattening of the dark matter halo. 
For UGC 7321, the $M/L$ in the disk is very low: the contribution of the
stellar disk to the radial force is small and is far below `maximum disk' in
this low surface brightness galaxy \citep[][reached the same
conclusion]{bmj10}. 
The best fits of the force gradients $\D
K_{\rm z}/\D z$  
from the various components are shown in fig.~\ref{fig:7321}. The vertical gradient
of the total vertical force field derived from the hydrostatics of the flaring
gas layer  
(for stars + gas + halo) is shown in black as a function of radius. After
taking off the known contribution from the gas layer, the blue curve labelled
`halo + stars' shows the remaining contribution from the halo + stars. This is
well modelled for $R > 3$ kpc by the adopted contributions from the stars and
the dark halo model (dashed curve).  Note that the shape of the observed and
fitted curves are remarkably 
similar for $R > 3$ kpc.  For this best fit, the dark halo model is spherical. 
More analysis is required for the rest of the sample to draw firm conclusions. 

In the Galaxy the flaring of the disk can be studied in much more detail. An
extensive study has been performed recently by \citet{kdkh07} based on the
Leiden/Argentina/Bonn all-sky survey of Galactic \HI\ \citep[see
also][]{kk09}. In this study they map the Galactic \HI\ layer and its flaring
and warp in much detail and fit the observations  
with models containing a dark matter disk as well as a dark matter halo.
Their best fitting model has a local disk surface density of 52.5 \msun\
pc$^{-2}$ with a scalelength of 2.5 to 4.5 kpc. This corresponds to a
maximum `disk-alone' rotation velocity of respectively 200 to 130 km
sec$^{-1}$. So again, if the Galaxy has a normal scalelength ($\sim 4$ kpc) for 
its rotation velocity, it is far below maximum disk, but if the scalelength is 
2.5 kpc it is close to maximum disk.  A complete analysis of the flaring \HI\ 
disk in terms of Galactic dark matter distribution yields evidence for a 
significant dark matter disk with a large scalelength of order 8 kpc and in 
addition a dark matter ring at 13 to 18 kpc. This ring could be the remnant of
a  
merged dwarf galaxy.

The distribution of the density in the inner parts of dark halos has
been a subject of much attention, as a consequence of the Cold Dark Matter
paradigm \citep{bfpr84,bffp86}, in which structure grows hierarchically
with small objects collapsing first and then merging into massive objects.
Cosmological $n$-body simulations based on $\Lambda$CDM \citep{dc91,nfw96,nfw97}
have long predicted that the inner density profile of the dark
matter halo ($\rho \propto r^{\alpha}$) would have an exponential slope
$\alpha$ of about $-1$ (a {\it `cusp'}), while observations
seemed to suggest a slope near zero (a {\it `core'}). The high spatial resolution 
of the THINGS data \citep{blok08} are well suited to investigate this
matter. They find that, for massive disk-dominated
galaxies, all halo models appear to fit equally well, while
for low-mass galaxies a core-dominated halo is clearly preferred over a 
cusp-like halo. This cusp-core controversy
(with $\alpha$ assuming somewhat different values) is a
long-lasting hot item in the study of \HI-rich, dwarf galaxies and low
surface brightness systems, where the
contribution of the dark matter to the rotation curve is large even in the
inner regions. Recently, \citet{db10} has summarised the situation in
such systems, concluding that the problem is still unsolved, even with
the use of current high resolution rotation curves. This issue has potentially 
important implications for galaxy formation models in $\Lambda$CDM.
                                                                   
\subsection{Outer \HI\ and Warps}
\label{sec:warps}

The warping of the outer parts of the neutral hydrogen layer of
our Galaxy has been known for a long time. It was discovered
independently in early surveys of the Galactic \HI\ in the north by
\citet{bur57} and in the south by Kerr and Hindman \citep[see][]{kh57}. In
external spiral galaxies the first indication came from the work of
\citet{rlw74}, when they obtained aperture synthesis
observations of the \HI\ in M83. The distribution and the velocity field
of the \HI\ both showed features that could be interpreted as a warping of
the gaseous disk in circular motions in inclined rings. This later
became known as the `tilted-ring' description for `kinematic
warps' and many more galaxies have been shown to have such deviations
using this method \citep[\eg][]{bos81a,bos81b}. The case that this warping occurs
in many spiral galaxies was strengthened by the early observations of
edge-on systems. \citet{san76} was the first to perform such
observations and showed that the \HI\ in four out of five observed edge-ons
(including NGC 4565 and NGC 5907)  displayed strong deviations
from a single plane. \citet{san83} discussed these warps in somewhat
more detail and in particular noted that in the radial direction the \HI\
surface densities often display steep drop-offs followed by a `shoulder' or
`tail' at larger radii.

The most extreme (`prodigious') warp in an edge-on system was observed by
\citet{bsvdk87} \citep[see also][]{bot95, bot96} in NGC 4013. 
\citet{gar01,grsk02} presented 21-cm observations of a sample of 26
edge-on galaxies in the northern hemisphere. This showed that \HI-warps
are ubiquitous; the authors state that {\it ``all galaxies
that have an extended \HI-disk with respect to the optical are
warped''}. Studies of possible warps in the stellar
disks have also been made; for recent results see \S~3.7; 
although there is evidence for
such stellar warps in most edge-on galaxies, the amplitude is very small
compared 
 to what is observed in the \HI.

The origin of warps has been the subject of extensive study and has been
reviewed for example by \citet{grsk02}, \citet{ss06}, \citet{bin06}
and \citet{sell10}. Although none of the models is
completely satisfactory, most workers seem to agree that it has
something to do with a constant accretion of material with an angular
momentum vector misaligned to that of the main disk. In models by \citet{jb99} 
and \citet{ss06}, this results in an inclined outer torus in the dark matter halo
that distorts the existing disk and
causes it to become warped. The possibility of a misalignment in the
angular momenta and therefore of the principal planes of
the stellar disk and the dark halo \citep{ds99} has
recently received some observational support from the observations of
\citet{bfos06} of NGC 5055. The \HI\ data suggest that the
inner flat disk and the outer warped part of the \HI\ have different
kinematic centers and systemic velocities, suggesting a dark matter halo
with not only a different orientation, but also an offset with respect
to the disk. The detailed study of the \HI\ in NGC 3718 by \citet{svmsv09}
shows an extensive warping, for which the observed twist can be explained 
as a result of differential precession in a fairly round dark halo with the same orientation as the disk. \citet{bri90} used existing observations and
tilted-ring models in moderately inclined galaxies to define a set of
``rules of behavior for galactic warps''. One was that ``warps
change character at a transition radius near $R_{Ho}$''. The latter
radius is the Holmberg-radius (see footnote~\ref{footnote:Holmberg})
listed for 300 bright galaxies in
\citet{holm58}.

\begin{figure}[t]
\begin{center}
\includegraphics[height=76mm]{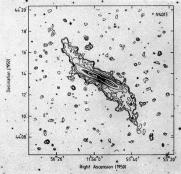}
\includegraphics[height=76mm]{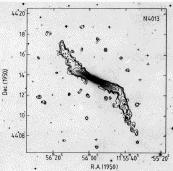}
        \caption{The \HI\ distribution in the edge-on spiral galaxy NGC 4013
\citep{bsvdk87}. On the left the integrated \HI\ map and on the right only
the `high' velocity channel maps. The latter selects the \HI\ along a line
perpendicular to the line-of-sight to the galaxy. The 
warp is less well-defined in the view of the left, since the azimuthal angle in 
the plane of the galaxy, along which the deviation of the warp is largest, is not 
precisely perpendicular to the line-of-sight. The warp starts very
abruptly at almost exactly the truncation radius of the stellar disk. 
\citep[From][]{bot95}}
\label{fig:N4013}
\end{center}
\end{figure}

The onset of \HI\ warps seems to occur very close to the radius of the truncation in
the stellar disk \citep{vdk01}. This can be illustrated with two
archetypal examples. In an edge-on galaxy, the most pronounced warp
known is in NGC 4013 \citep{bsvdk87,bot95,bot96}. The \HI\ warp starts 
very abruptly at the truncation of the stellar disk (see fig.~\ref{fig:N4013}) 
and is accompanied by a significant drop in rotational velocity. The
latter suggests a truncation in the disk mass distribution as well as in
the light. 
A face-on spiral with an \HI-warp is NGC 628 \citep{shvdk84, kb96}. 
The velocity field suggest a warp in the \HI, starting at the edge of the 
optical disk. 

In a extended study, \citet{vdk08} listed the following general characteristics 
of \HI\ warps. Whenever a galaxy has an \HI-disk extended with respect to the 
optical disk, it has an \HI-warp
\citep{grsk02}. Many galaxies, but not all, in this sample 
have relatively sharp truncations in their stellar disks. When an edge-on  
galaxy has an \HI\ warp, the onset occurs just beyond the truncation radius. 
Similarly, in less inclined galaxies, the warp is seen at the boundaries of the 
observable optical disk \citep{bri90}.
In many cases the rotation curve shows a feature at about the
truncation radius \citep{sc83,bot96}, which
indicates that the truncation occurs also
in the mass. The onset of the warp is abrupt and discontinuous and coincides 
in the large majority of cases with a steep drop in the radial \HI\ surface 
density distribution, after which this distribution flattens
off considerably. The inner disks are extremely flat (both stellar disks as
well as in gas and dust) and the onset of the warp is {\it abrupt}; beyond that, 
according to \citet{bri90}, the warp defines a ``new reference frame''.

\begin{figure}[t]
\begin{center}
\includegraphics[width=150mm]{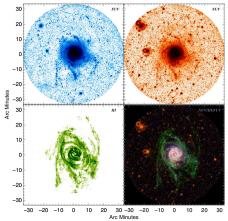}
        \caption{Images of M83, Top: GALEX far-UV and near-UV; Bottom: HI map
          from THINGS and a combination of the three with the HI smoothed.
\citep[From][]{bls10}}
\label{fig:M83bigiel}
\end{center}
\end{figure}

These findings suggest that the inner flat disk and the outer warped
disk are distinct components with different formation histories,
probably involving different epochs. The inner disk
forms initially (either in a monolithic process in a short period or
hierarchically on a somewhat more protracted timescale) and the warped outer disk
forms as a result of much later infall of gas with a higher angular
momentum in a different orientation. This is also consistent
with an origin of the disk truncations that is related
to the maximum specific angular momentum available during its formation,
since then the truncation is also in the disk mass, giving rise to
the abruptness of the onset of the warps.

The misalignment of the material in the inner disk and the outer disk and
warp has been modelled by \citet{rdbqbgdw10} as a result of the interaction
between infalling cold gas and a misaligned hot gaseous halo. The gas that 
forms the warp is
torqued and aligned with the hot gaseous halo rather than the inner disk. 
In this model the outer accreted gas thus
responds differently to the halo than the gas that has formed the inner disk. 
Although this may be consistent with the observed correlation of the
onset of the warp and the truncation or break in the stellar disk, the
abruptness of the onset of the warp may not follow naturally. Furthermore,
this mechanism requirs the existence of a hot
gaseous halo and the question is whether these are present in galaxies with
smaller rotation amplitudes; the presence of warps in M33-size
galaxies would then be another argument against this hypothesis.

To what extent has star formation and chemical enrichment
taken place in these gaseous warps?  The early work of
\citet{fwgh98,fgw98} established the presence of star formation in regions beyond
two $R_{25}$ in a few galaxies through the detection of faint
\HII-regions, while emission line ratios indicated very low abundances
of oxygen and nitrogen in these regions. Since then much work has been
done and more is in progress, in particular in the {\it ACS Nearby Galaxy 
Survey Treasury}  \citep[ANGST,][]{detal09}. 
Early results of this project have been published for M81
\citep{wds09,gdw09} and NGC 300 \citep{gdw10}, but those do not yet refer to
distances beyond the visible spiral structure. However, the disks do
contain mostly old stars at these radii, in agreement with the view that
the disk inside the truncation region and the radial onset of the \HI-warp
forms at an early epoch in the galaxy's evolution. In NGC 2976
\citep{wds10}, populations of old ages are found at all radii, also
beyond the break in the luminosity profile, but star formation does not
appear now to extend into this outer zone. This galaxy may have been in
interaction recently with the M81-group. 

Shostak \&\ van der Kruit (1984; see their fig.~3) pointed out
that in NGC 628 (M74) the extended, warped
\HI\ displays spiral structure that shows a smooth continuation of the
prominent spiral structure in the main disk, right through the onset of the 
warp. Particularly interesting for studying star formation and its history
in disks beyond the
truncation is the comparison of outer \HI\ with UV imaging from GALEX. 
\citet{bls10}  
compare the \HI\ from the THINGS (\HI) survey with
the far UV data from GALEX and conclude that, although star formation
does clearly take place in the outer \HI, its efficiency is extremely low 
compared to that within the optical radius (truncation). A detailed 
comparison of images for M83 \citep[][see fig.~\ref{fig:M83bigiel}]{blw10} 
shows a clear
correlation between star formation and \HI\ surface density. 
The conclusion  that star formation is proceeding in this galaxy in the
very outer parts also
follows from the work of \citet{tjd10}, comparing GALEX imaging to deep 
optical (Gemini) images of the outer regions of M83.

\subsection{Dustlanes in Disks}

We will not review in detail the distributions and properties of dust in
disks of galaxies. However, we noted above that dustlanes are very straight,
such as illustrated in fig.~\ref{fig:edge-ons},
at least in disks of massive galaxies and indicate that galaxy disks are 
extraordinarily flat. It has been known for a long time
that dustlanes sometimes are less well-defined
in galaxies of later type and dwarfs. A good example is NGC 4244, which is a 
late-type, relatively small, pure-disk galaxy \citep{vdks81a}, where the 
dustlane is not sharply outlined against the stellar disk.  
\citet{dyb04} studied a sample of
edge-on galaxies and found that systems with maximum rotation
velocities larger than 120 \kms\ have well-defined dustlanes, while in those
with smaller rotations the scaleheight of the dust is systematically
larger and the distribution much more diffuse. Indeed, NGC 4244 has a maximum
rotation velocity of about 115 \kms. This finding may have
important implications for our understanding of the star formation
history and evolution of disks.

\citet{dyb04} suggest that the transition at 120 \kms\ marks
the rotation speed above which the disks become gravitationally
unstable, whereby instabilities in the disk lead to fragmentation of the
gas component with high density during a collapse which then gives rise
to thin dustlanes. In a study of the vertical distributions in a number
of low-mass edge-on galaxies, \citet{sddj05} find that not only does the
dynamical heating of the stellar population appear to occur at a much reduced rate
compared to the Galactic disk in the Solar Neighborhood, but in these
low-mass systems the vertical distribution of the young stellar population and
of the dust layer is thicker than those in the Milky Way. This is consistent
with a cold interstellar medium in slowly rotating galaxies
that has a larger scaleheight and therefore with an absence of
well-defined dustlanes in such systems.
\citet{sddj05} fit the distributions with an isothermal sheet
($n=1$ in eqn.~(\ref{eqn:vdksdisk}) with $z_{\circ}=2h_{\rm
z}$).  In NGC 4144 (rotation velocity 67 \kms), the dust has a scaleheight 
$z_{\circ}$ of about 0.5 kpc or an exponential scaleheight of half that.
For comparison, in the Milky Way the three-dimensional distribution of dust 
has been modelled by \citet{ds01}: their flaring dust disk has a similar 
exponential scaleheight at the solar radius of about 0.2 kpc. \

We may examine more massive edge-on galaxies with rotation velocities well over
200 \kms\ to compare with the low-rotation systems. For example, three
such galaxies with obvious dustlanes are NGC 4565, 891 and 7814 for
which minor axis profiles (in blue light) are available in the papers of
\citet{vdks81a,vdks81b,vdks82b}. When we determine the height above the
symmetry plane at which the effect of 
the dust extinction is about one magnitude compared to the extrapolated
minor axis profile of the stellar light, we get values of respectively
0.9, 0.7 and 0.6 kpc. These are undoubtedly overestimates when used as
indications for the dust scaleheights; \citet{kb87}
for example find that the dust scaleheight in NGC 891 is 0.22 kpc,
compared to the stellar scaleheight of 0.5 kpc \citep{vdks81b}. 
\citet{whf89} determined in IC 2531 (an NGC 891-like edge-on galaxy)
that the scaleheight of the old disk stars is about 0.5 kpc, while that of
the dust is a quarter of that. A similar determination of the height at
which the minor-axis profile indicates an absorption of one magnitude
compared to the extrapolated profile yields about 0.5 kpc. On this basis
the exponential scaleheights of 
the dust layers in the three systems just mentioned are of the order of
3 or 4 times less than those of the old stellar disks. The important
inference is that the diffuse dust layers in slow
rotators have, in absolute measures, thicknesses that are not very different 
from those in massive galaxies with high rotation velocities.

In early-type edge-on galaxies such as NGC 7814 and NGC 7123, where the 
spheroids dominate
the light (and presumably stellar mass) distributions, the situation
appears different. The dust distributions in these systems have
scaleheights comparable to those of the stellar disks \citep{whf89},
which are of the order of 0.5 to 1 kpc. This may be the result of much longer
dissipation times due to the lower gas content, so that the scaleheights of 
the stars (from which the dust comes) and the dust are similar.

In the sample of late-type, gas-rich dwarf edge-on galaxies 
studied by \citet {obfvdk10c}, the mean \HI\
velocity dispersion increases as a function of the maximum
rotational velocity of the \HI\ disk from about 5 to 8
\kms\ for rotation velocities of 70 to 120 \kms. 
The scaleheight $h$ of a Gaussian dust layer is
related to the ISM velocity dispersion $\sigma$ by
\begin{equation}
\frac{\sigma^2}{h^2} = - \frac{\partial K_z}{\partial z}
= 4\pi G \rho_\circ,
\end{equation}
where $\rho_\circ$ is the mid-plane total density. The density 
$\rho_\circ=\Sigma/2h_{z}$ where $\Sigma$ is the typical surface density of a
disk.  $\Sigma$ and the stellar scaleheight are both approximately
proportional to the maximum circular velocity $V_c$ \citep{gur10,kvdkdg02,kvdkf05},
so the typical value of $\rho_\circ$
is independent of $V_c$ and we expect the scaleheight of the dust
layer to be directly proportional to the ISM velocity dispersion
$\sigma$. In the O'Brien sample of galaxies, the galaxy with the largest
$V_c$ (IC2531) shows a clean well-defined dustlane as expected
from the Dalcanton et al. observations. However there is no evidence
for a decrease in $\sigma$ as $V_c$ increases in the O'Brien et al. sample,
so those data do therefore not 
support the variable turbulence explanation for the change in dust
lane morphology with rotation velocity.

\section{CHEMICAL EVOLUTION AND ABUNDANCE\\
GRADIENTS}

\begin{figure}[t]
\begin{center}
\includegraphics[width=130mm]{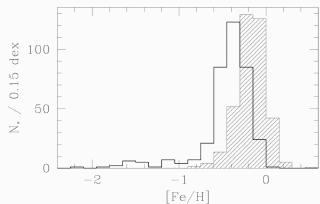}
\caption{Metallicity distribution function for stars in the bar of the 
LMC, compared to the solar neighborhood MDF from \citet{macw90} (hatched).
The solar neighborhood MDF is about 0.2 dex more metal-rich. 
\citep[From ][]{cole05}. }
\label{fig:cole05_9}
\end{center}
\end{figure}

The stars and gas in galactic disks have a mean metallicity which depends 
on the luminosity of the galaxy \citep[\eg][]{tre04} and often shows a 
radial gradient.  In a large galaxy like the Milky Way, the typical disk 
metallicity [Fe/H] is near that of the sun.  For example, in the solar 
neighborhood, the metallicity of the disk stars has a mean of about -0.2 
and ranges from about +0.3 to -1.0 \citep[\eg][]{macw90}.  
Fig.~\ref{fig:cole05_9} compares the stellar metallicity distribution 
functions (MDFs) of the solar neighborhood and the bar of the LMC.  
As expected from the lower luminosity of the LMC, the LMC MDF is 
displaced towards lower metallicities, but the shapes of the two MDFs 
are similar.

The metallicity distribution function (MDF) is believed to come from the 
local chemical evolution of the stars and ISM, including the effect of 
infall of gas from outside the Galaxy. The presence of a fairly tightly 
defined radial abundance gradient in the stars and in the gas in many disk 
galaxies suggests that the chemical evolution of the disk is determined 
mainly by local chemical evolution with limited radial exchange of 
evolution products. Much of the theory of chemical evolution of disk 
galaxies is based on this assumption \citep[\eg][]{chiap97}; this view is 
however currently under challenge.

\begin{figure}[t]
\begin{center}
\includegraphics[width=130mm]{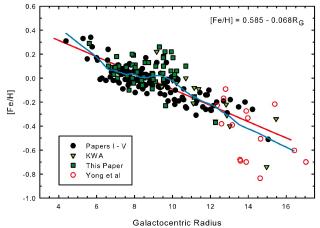}
\caption{The radial abundance distribution 
for Cepheids in the Galactic disk. \citep[From ][]{luck06}. }
\label{fig:luck06_10}
\end{center}
\end{figure}

From the basic theory of chemical evolution via the continued formation 
and evolution of stars and enrichment of the interstellar medium, one 
might expect that the metallicity of disk stars would gradually increase 
with time.  \citet{macw97} pointed out in his review that it was not clear 
observationally at the time that any age-metallicity relation (AMR) exists 
among the stars of the solar neighborhood. This is still the situation today.
\citet{edv93} found evidence for a weak 
decrease of [Fe/H] with age among the disk stars, but later work 
\citep[\eg][]{nord04} is less conclusive. From the white dwarf luminosity 
function and the directly measured ages of disk stars 
\citep[\eg][]{khh99,edv93,san03}, the age of the Galactic thin disk is 
about $8-10$ Gyr.   The open cluster NGC 6791 has an age of about $8-10$ 
Gyr and an abundance [Fe/H]$=+0.2$, indicating that enrichment to 
solar level occurred very quickly in the Galactic disk, on a Gyr 
timescale. An early AMR figure by \citet{se69}, based partly on open 
cluster ages and metallicities, shows a very rapid early evolution of 
the galactic abundances up to near-solar abundances and then little 
further change; this is still a fair representation of the current state 
of knowledge.

The rapidly rising and then flat AMR in the disk of the Milky Way contrasts 
with the situation in the LMC. \citet{dol00} derived the star formation 
history and AMR for two fields in the LMC disk.  The star formation rate 
shows early and late phases of star formation, with a very slow period 
between about $3$ and $7$ Gyr ago, and the AMR shows a smooth rise from 
below $-1.5$ at 10 Gyr ago to the present metallicity of [Fe/H]$=-0.4$.  
A smoothly rising AMR is seen also in the outer regions of M33 
\citep{bar07b}. This difference in the morphology of the AMR between the 
Milky Way and the smaller galaxies is likely to come from the different 
star formation histories of larger and smaller disk systems (the 
downsizing phenomenon).

\subsection{Gas-phase Abundance Gradients}

Many galaxies show a clear radial gradient in their gas-phase abundances. 
\citet{zar94} assembled data on oxygen abundance gradients in 39 disk 
galaxies covering a range of luminosities.  They found the now-familiar 
correlations between oxygen abundance and luminosity, circular velocity 
and morphological type.  The size of the abundance gradients (in 
dex/isophotal radius) did not correlate with luminosity or type.  
The presence of a bar appears to flatten or even erase the abundance 
gradient \citep[see also][]{all81}, probably due to the non-circular 
motions which the bar induces in the gas of the disk.

\citet{mag07} modelled the chemical evolution of M33, assuming that the 
galaxy is accreting gas from an external reservoir. A model with an 
infall rate of about $1$ \msun\ yr$^{-1}$ reproduces the observational 
constraints, including the relatively high star formation rate and the 
shallow abundance gradient. The model indicates that the metallicity 
in the disk has increases with time at all radii, and the abundance 
gradient has continuously flattened over the past $8$ Gyr 
\citep[see also][for a comparison with the evolution of the somewhat 
similar disk galaxy NGC 300]{gdw10}.

For the Milky Way, \citet{sha83} combined radio and optical spectroscopy 
to measure abundances for \HII-regions between about $3$ and $14$ kpc from 
the galactic center.
\citet{fich91} extended the observations to a radius of about $18$ kpc. 
A well defined gradient in the oxygen and nitrogen abundances of $-0.07$ 
to $-0.08$ dex kpc$^{-1}$ is found.  \citet{denn81} see a comparable 
gradient in nitrogen in the \HII-regions of the disk of M31.

\subsection{Stellar Abundance Gradients}

The abundance gradient for relatively young stars in the disk of the 
Milky Way is nicely delineated by the Cepheids \citep{luck06}, shown in 
fig.~\ref{fig:luck06_10}.  The gradient is about $-0.06$ dex kpc$^{-1}$, 
in good agreement with the gas-phase gradient derived by \citet{sha83}. 
The two-dimensional distribution of the Cepheid abundances over the 
Galactic plane shows some localized departures from axisymmetry at the 
level of about $0.2$ dex in abundance. These departures may come from 
radial gas flows associated with the spiral structure.

\begin{figure}[t]
\begin{center}
\includegraphics[width=130mm]{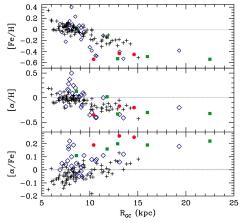}
\caption{{\it Upper panel}: The radial abundance distribution for Cepheids 
(crosses) and older clusters (blue diamonds and green squares) and red giants 
(red circles) in the Galactic disk. {\it Lower panels}: the radial 
[$\alpha$/H] and [$\alpha$/Fe] distributions for the same objects. 
\citep[From ][]{car05}. }
\label{fig:car05_5}
\end{center}
\end{figure}

For the older stars in the outer disk (open clusters and red giants), 
the abundance gradient appears to be somewhat steeper, as seen in the 
study by \citet{car05}. The abundances fall to about $-0.5$ at $11$ kpc 
but then stay approximately constant at this level out to radii beyond 
$20$ kpc (fig.~\ref{fig:car05_5}). The figure also compares the 
$\alpha$/Fe ratio for the older objects and the Cepheids. It indicates 
that the abundance gradient in the outer disk is flatter now than it was 
a few Gyr ago, and also that the [$\alpha$/Fe] ratio is nearer the solar 
value now than it was at the time of formation of the outer old clusters.  
These observations suggest that the chemical evolution of the disk 
gradually flattens the abundance gradient and reduces the [$\alpha$/Fe] 
ratio.  In the outer regions, episodic accretion
 of gas may trigger 
bursts of star formation which could erase the abundance gradient and 
produce the $\alpha$-enriched abundances seen in the older objects. A 
similar bottoming out of the abundance gradient beyond about 15 kpc is 
seen by \citet{wor05} in the outer disk of M31.  We note that the 
environment of the outer disk in M31 has had a complex star formation 
history, with much evidence for an extended period of accretion of smaller 
galaxies \citep[\eg][]{macc09} which complicates the interpretation of 
radial gradients in this system.

\citet{cio09} used the radial change of the ratio of C and M-type AGB 
stars in the LMC and M33 to evaluate their stellar abundance gradients. 
Both show a radial abundance gradient. This is particularly clearly seen 
in M33: the gradient in the disk persists to a radius of about 8 kpc, 
which is the radius at which the radial truncation of the disk occurs.  
At larger radii, the abundance gradient becomes much flatter.  
\citet{bar07a} also found evidence that the radial abundance gradient of 
M33 flattens in the outer regions.

The abundance gradients of disk galaxies may not only flatten at large 
radii but can also reverse.  The galaxy NGC 300 is similar to M33 in 
appearance and absolute magnitude, and it also has a negligible bulge 
component.  The two galaxies do however differ in the details of the 
structure of their disks.  M33 shows a very well defined truncation of 
its disk, at a radius of a few scalelengths \citep{fer07}, while the 
disk of NGC 300 has an unbroken exponential surface density distribution 
extending to a radius of at least $10$ scalelengths \citep{bh05}.  
The negative abundance gradient in NGC 300 persists to a radius of about 
$10$ kpc and then appears to reverse, with the metal abundance increasing 
with radius. \citet{vla09} present two possible scenarios for the reversal 
of the abundance gradient.  One is associated with accretion and local 
chemical evolution.  The other scenario involves radial mixing driven 
by resonant scattering of stars {\it via}  transient spiral disturbances, 
while preserving their near-circular orbits, as discussed by \citet{sb02} 
and \citet{rdsqkw08}. In this picture, the stars in the outermost disk did 
not form {\it in situ} but were scattered from the inner galaxy into the 
outer disk. The scattering occurs from near the co-rotation radius of the 
individual spiral disturbance. The radial extent of the scattering 
depends on the strength of the transient spiral wave, and the radius 
from which the scattering occurs depends on its pattern speed. If the 
inner disk of the galaxy has the usual negative abundance gradient, then 
the abundance distribution in the outer disk (which is populated by stars 
scattered from the inner disk) would depend on the distribution of 
pattern speeds and pattern strengths in the inner disk. A reversal in the 
current abundance gradient could come from an epoch of spiral disturbances 
whose strength increases with their pattern speed: more metal-rich stars 
from smaller radii are then more strongly scattered radially.  Such an 
epoch of spiral disturbances could also explain the difference in 
structure between the truncated disk of M33 and the exponential disk of 
NGC 300.  In this scenario, the outer disk of NGC 300 is more strongly 
populated by radial mixing, building up the continued exponential. For 
an alternative view, see \citet{gdw10}.

The chemical evolution of disks is usually calculated assuming that each 
annulus in the disk evolves independently, with infall of gas from 
intergalactic space but with no radial exchange of processed gas or 
stars \citep[\eg][]{chiap97}. In this picture, one might expect a 
relationship between the age of stars and their metallicity. The 
apparent lack of a well-defined AMR in the solar neighborhood has 
motivated alternative views of the chemical evolution of disks, 
involving radial flows of gas and radial mixing of stars. \citet{schbin09a} 
have constructed such a theory which seems to fit very well the observed 
stellar distribution of thin disk stars in the [$\alpha$/Fe] - [Fe/H] 
plane, and also the existing AMR data.  We caution however that the AMR 
data in the solar neighborhood is still quite uncertain.

\section{SCALING LAWS FOR DISK GALAXIES}
\label{sect:scaling}

Disk galaxies demonstrate several scaling laws, \ie\ relations of observable 
parameters to the luminosity or stellar mass of the galaxies.
Some scaling laws involve parameters associated with the stellar component 
of the galaxies: \eg\ the stellar mass - metallicity relation and the 
luminosity - radius relation. Other relate stellar properties to dark matter 
properties: \eg\ the Tully-Fisher relation between the stellar luminosity 
$L$ (or stellar mass or baryonic mass) and the rotation speed of the 
galaxy, and the scaling relations between the the luminosity of the stellar 
component and the central density of the dark matter halos. These scaling 
laws provide insight into the formation and evolution of disk galaxies.  

When examining the scaling laws and other relations between observables for
disk galaxies, it is useful to ask how many independent parameters there
are in the ensemble of information.  Principal component analysis (PCA) can
provide insight, although it may be difficult to identify which of the 
parameters are fundamental \citep{bro73}. Recently, \citet{disney08} combined 
data from the HIPASS survey of \HI\ in galaxies \citep{hipass04} with SDSS 
data and derived six parameters: two optical radii (containing 50
and 90\%\ of the light), luminosity, \HI\ mass, dynamical mass and 
$(g - r)$ color. PCA shows that the correlations are dominated by only one 
significant principal component. This indicates a somewhat unexpected 
organizational uniformity in galaxy properties. 

\subsection{The Tully-Fisher Law}

This law, in its original form of absolute blue magnitude $M_B$ {\it vs} 
the velocity width of the integrated \HI\ profile, was discovered by 
\citet{tf77}. For their profile width, they used $W_{20}$, the line width 
at 20\% of the peak intensity, which has become widely but not universally 
used in Tully-Fisher relation (TFR) studies. The TFR quickly became an 
important tool for measuring the absolute magnitudes and distances of
galaxies from their \HI\ profile width.  The magnitudes need to be be 
corrected for Galactic and internal absorption, and the velocity widths 
corrected for the inclination of the disk and turbulence in the galaxy's 
ISM. The slope of the TFR depends on wavelength; see for example 
\citet{sak00}. The slope of the $\log L - \log W_{20}$ relation goes from 
about $3.2$ at $B$ to about $4.4$ at $H$, and the scatter about the TFR is 
smaller at longer wavelengths.
 
{ The TFR is an important and complex constraint on galaxy formation theory. Each of 
the variables in the TFR (luminosity and profile width) is 
itself the product of many complex and interacting processes involved in the formation 
and evolution of disk galaxies, and these processes contribute to the slope, zero-point 
and scatter of the TFR. The luminosity measures the integrated star formation history 
and evolution of the baryons.  The \HI\ profile width is a measure of the maximum 
value of $R \partial \Phi/\partial R$ in the plane of the \HI\ disk, where $R$ is the 
radius and $\Phi$ the potential.   This potential comes partly from the stellar 
component and partly from the dark matter halo. It 
therefore depends on the lengthscale of the stellar component (\ie\ on how much the 
baryons have contracted during the formation of the galaxy) and 
hence on the angular momentum of the stellar component. 

The dimensionless 
spin parameter \citep{pee71}
\begin{equation}
\lambda=\frac{J|E|^{1/2}}{GM^{5/2}},
\end{equation}
 (where $J$ is the 
system's spin angular momentum, $E$ its binding energy and $M$ its mass) 
is relevant here. $\lambda$ is a measure of how far a system is from 
centrifugal equilibrium. Cosmological simulations predict 
the distribution of $\lambda$ for dark halos which form in a CDM universe 
(\eg\ \citet{ej79}). The distribution of $\lambda$ is typically lognormal, 
with a mean value of about $0.06$.  In systems with higher values of $\lambda$, the 
baryons settle into disks which are more extended, more slowly rotating and 
of lower surface brightness in the mean. See \eg\ \citet{dal97} for more details.
Furthermore, the parameters involved in the TFR are likely to evolve as the galaxy 
grows: the stellar mass will increase according to the star formation 
history, the baryonic mass will be affected by feedback and accretion, and the 
rotational velocity is likely to change as the stellar mass increases and 
the dark matter halo is gradually built up.  We can expect all of these changes 
to continue to the present time.}

Although the brighter disk galaxies lie on a well-defined luminosity-velocity 
relation, the fainter galaxies with circular velocities less than about 
100 \kms\ are observed to have luminosities that lie below the relation 
defined by the brighter galaxies.  Many of these fainter disk galaxies are 
gas (\HI) rich, so the stellar mass is only a fraction of their baryonic 
content. If the baryonic (stellar + \HI) mass is used instead of the 
stellar luminosity or mass, the fainter disk systems move up to lie on 
the same baryonic TFR as the brighter galaxies \citep{kcf99, mcg00}, 
with a relatively small scatter of 0.33 mag. This suggests that the TFR 
is really a relation between the rotational velocity and the total baryonic 
mass $M_{bar}$ { \citep[see also ][]{mg05}}. The change in the slope of the $L - W_{20}$ relation with 
wavelength is partly due to the way in which the $L/M_{bar}$ ratio changes with 
$M_{bar}$ at different wavelengths, which in turn probably reflects the 
different star formation histories of disk galaxies of different masses. 
In the mean, the less massive galaxies are now more affected by current 
star formation (downsizing), and we can expect more scatter in their 
$L/M_{bar}$ ratio.

There is still some disagreement about the slope of the baryonic 
Tully-Fisher relation (BTFR). Various authors find slopes between 
$M_{bar} \propto V^3$ and $M_{bar} \propto V^4$ where $V$ is the rotational 
velocity \citep[\eg][]{gur10,tra09,sta09}.   CDM theory predicts a slope 
closer to 3, so it is important to settle this question observationally. The
studies quoted estimate the stellar masses from $M/L$ ratios based on various
models. In a more direct approach, \citet{kvdkf05} use stellar disk 
masses derived from stellar dynamical analysis; after adding the \HI\ content, 
they find a slope for the BTFR of $3.33 \pm 0.37$ over 2 dex, with a scatter
of 0.21 dex.

One potentially important difference in approach comes from the way in 
which different authors measure the rotational velocity 
\citep[see][for more discussion]{can07}. Some authors use the profile 
width $W_{20}$ as their velocity measure; the relationship between this 
quantity and the asymptotic (flat) rotational velocity of the galaxy 
becomes less well defined for lower-mass galaxies, in which the \HI\ often 
does not extend to the flat part of the rotation curve.  Other authors 
restrict their samples to galaxies in which the rotation curve is observed 
to reach the flat region, and use this flat level of the rotation curve as 
their velocity measure. Although the flat level of the rotation curve 
provides a consistent estimate of the rotation, this is itself a selection 
effect, restricting the sample of galaxies to those in which the \HI\ 
extends to the flat rotation curve region.  The selected galaxies are 
therefore biased towards systems in which either the \HI\ is intrinsically 
more extended, or the dark matter halos are more centrally concentrated with 
relatively smaller scalelengths and larger concentrations. As one might 
expect from the above discussion, analyses based on $W_{20}$ appear to 
give lower (flatter) slopes for the BTFR than those based on the flat 
region of the rotation curve. Ideally, to relate the baryonic content of 
disk galaxies to the properties of the dark halos, we would like to use 
the circular velocity of the dark halo at the virial radius, but for most
galaxies this quantity cannot be measured at present 
\citep[see][for a useful discussion]{cou07}. 

The interpretation of the BTFR is not straightforward. For normal high 
surface brightness disk galaxies, the rotation speed depends on the 
contributions to the gravitational potential from both the stellar 
distribution and the dark matter, including any effects of the stellar 
distribution on the dark matter (baryonic contraction). This in turn 
involves the evolution of the stellar disk itself.  In gas-rich and 
low surface brightness galaxies, the contribution of the luminous matter 
to the potential field is relatively small but, even in this simpler 
situation, the BTFR involves the ratio of baryon to dark matter mass, 
the structure of the dark matter halo, and the radial extent or 
maximum angular 
momentum of the \HI\ gas.  \citet{zwa95} and \citet{spr95} constructed 
a TFR law for low surface brightness (LSB) galaxies: these galaxies 
have surface brightnesses that are typically at least a magnitude fainter 
than the normal galaxies.  They found that the TFR for LSB and normal disk 
galaxies were very similar in slope and zero-point. Although this result 
may seem surprising, it makes sense if interpreted in the context of the 
BTFR.  The stellar luminosity is used as a proxy for the baryonic mass, 
with some scatter and bias depending on the gas fraction (and adopted 
stellar $M/L$ ratio).  The velocity for these galaxies reflects the circular 
velocity of the dark matter potential, because the contribution to the 
gravitational field from the stellar component is relatively insignificant.

 \begin{figure}[t]
\begin{center}
\includegraphics[width=150mm]{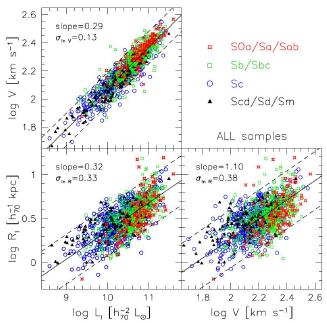}
\caption{Scaling relations between the rotation velocity $V$, the
scalelength $R$ and the luminosity $L$, the latter two in the 
$I$-band. \citep[From ][]{cou07}. }
\label{fig:scaling}
\end{center}
\end{figure}

The following clever argument of \citet{cr99} makes use of the scatter in the 
Tully-Fisher relation. The amplitude of the rotation curve of the self-gravitating
exponential disk is
\begin{equation}
V_{\rm disk} \propto \sqrt{h \Sigma_{\circ }} \propto
\sqrt{M_{\rm disk} \over h}.
\end{equation}
For fixed disk-mass $M_{\rm disk}$ we then get by differentiation
\begin{equation}
{{\D \log V_{\rm max}} \over {\D \log h}} = -0.5.
\end{equation}
So at a given absolute magnitude (or mass) a lower scalelength
disk should have a higher rotation velocity. If all galaxies were 
maximum disk, then this anticorrelation should be visible in the scatter of the
Tully-Fisher relation. It is however not observed and the inference 
is that on average $V_{\rm disk} \sim 0.6 V_{\rm total}$ and galaxies
in general do not have maximal disks. We note that this argument ignores the
contribution of the gas to the total baryonic mass, which can be significant
even for relatively bright galaxies \citep[\eg][]{gur10}.

\subsection{Scaling Laws involving the Galaxy Diameter}

In their pioneering paper, \citet{tf77} also considered the relationship 
between galaxy optical diameter and rotational velocity.  \citet{cou07}
made an extensive study of the distributions of $L$ and stellar 
scalelength $h$ (in the I and K bands) and $V$ for a sample of 1300 
galaxies.  The scalelength reflects the angular momentum distribution of 
the star-forming baryons but probably includes further complications of 
the baryonic dissipation and settling to equilibrium. 

The scatter in the relations involving $h$ is larger than for the $L-V$ 
relation (see fig.~\ref{fig:scaling}), 
as one might expect from the additional complexity of the 
underlying physics and also the difficulty of measuring disk scalelengths 
accurately.  The mean relations found by \citet{cou07} are 
\begin{equation}
V \propto L^{0.29},~h \propto L^{0.32}~{\rm and}~ h \propto V^{1.10}
\label{eqn:courteau}
\end{equation}
in the $I$-band.  The relations involving $h$ show some morphological 
dependence, but no significant morphological dependence is seen in the 
$V-L$ relation.

\subsection{The Mass-Metallicity Relation}

The mass-metallicity relation pertains to the baryonic component of the 
galaxies. The relationship between stellar mass or luminosity and 
metallicity in galaxies goes back at least to \citet{leq79}. A study of a 
large sample of SDSS galaxies by \citet{tre04} gives references and 
demonstrates a tight correlation between the stellar mass and the gas-phase 
oxygen abundance extending over 3 orders of magnitude in stellar mass and a 
factor of 10 in oxygen abundance. \citet{kir08} show that the relation 
between luminosity and metallicity continues down to the faintest known 
dwarf spheroidal galaxies with absolute V magnitudes fainter than $-4$.  
The deeper potential wells of the more massive galaxies are believed to 
retain more effectively the metal-enriched ejecta of supernovae 
\citep{lar74, ds86}, which can be lost via galactic winds. \citet{tas08} 
have argued however that this kind of supernova feedback may not be 
essential for generating the mass-metallicity relation; the increasingly 
inefficient conversion of gas into stars in the lower mass galaxies may 
be responsible. 

\subsection{The Surface Density-Mass Relation}

This relation again pertains to the baryonic component of the galaxies; 
as for the mass-metallicity relation, it provides some constraints on the 
theory of the evolution of the baryonic component. In the mean, the surface 
density of galaxies appears to increase with increasing stellar mass or 
luminosity, as shown { first by \citet{kor85} and followed up by}
\citet{ds86} for a sample of nearby ellipticals 
and irregular galaxies with stellar masses less than about $10^{10.5}$ \msun.  
For disk galaxies, \citet{gur10} found a similar relation: the baryonic 
(stellar + \HI) surface density is observed to increase approximately linearly 
with $W_{20}$.  Again, this scaling law can be interpreted in terms of 
supernova-driven loss of gas or as due to increasingly inefficient star 
formation for systems of lower masses.

\subsection{Scaling Laws for Dark Matter Halos}

The properties of the dark matter halos of disk galaxies appear 
to scale with the luminosity of their stellar component. \citet{jkkf04} 
analysed the rotation curves of spirals and dwarf irregular galaxies to 
estimate parameters for their dark matter halos. They modelled the 
dark halo density distributions as isothermal spheres with a central core 
of density $\rho_\circ$ and a core radius $r_c$. Estimates for the dark  
halos of dwarf spheroidal galaxies were also included, using the velocity 
dispersion profiles of these systems.  \citeauthor{jkkf04} found that the 
core density $\rho_\circ$ decreases with luminosity,  as 
$\rho_\circ \propto L^{-0.28}$ : the core densities increase from about 
$10^{-2.5}$ \msun\ pc$^{-3}$ for the brighter spirals to a rather high value 
of about $10^{-0.5}$ \msun\ pc$^{-3}$ for the fainter dwarf irregular and 
spheroidal galaxies. The core radius $r_c$ increases with luminosity, 
as $r_c \propto L^{+0.32}$, so the surface density $\rho_\circ r_c$ of the 
dark matter halos is approximately independent of the luminosity of the stellar 
component.  This remarkable observational result was confirmed by 
\citet{don09} with more recent data for the dark halos of the dwarf 
galaxies. The surface density $\rho_\circ r_c$ is constant at about 
$140$ \msun\ pc$^{-2}$ over about $15$ mag in stellar luminosity.

The dark halo scaling laws reflect the changing density of the universe as halos of 
different masses are formed, with the less massive halos 
forming earlier.  The difference in halo densities indicates that the smallest 
dwarfs formed about $7$ units of redshift earlier than the largest spirals.
\citet{djor92} showed that protogalactic clumps which 
separate from an evolving density field with power spectrum 
$|\delta_k|^2 \sim k^n$ have a scaling law between density and radius 
$\rho\sim r^{-3(3+n)/(5+n)}$.  The observed scaling laws for dark matter halos 
correspond to $n \approx -2$, close to what is expected for $\Lambda$CDM 
on galactic scales.

\section{THICK DISKS}

Most spirals, including our Galaxy, have a second thicker disk component 
surrounding the thin disk. Thick disks were discovered in other galaxies 
via surface photometry \citep{tsi81, bur79}  and then in the Milky Way 
through star counts \citep{gilr83}.  It appears that thick disks are very 
common in disk galaxies and that they are mostly very old.  The thick 
disk is therefore an important component in understanding the assembly of 
disk galaxies.

\subsection{Statistics of Incidence} 

The photographic surface photometry of \citet{vdks81a, vdks81b, vdks82a} 
showed that thick disks were common in disk galaxies but perhaps not 
ubiquitous \citep[see also][]{fry98}.  At the time, evidence indicated 
that the thick disk was associated with the presence of a central bulge.  
A more recent extensive study of a large sample of edge-on galaxies by 
\citet{yd06} showed that thick disks are probably present in all or 
almost all disk galaxies. They found that the ratio of thick disk stars to 
thin disk stars depends on the luminosity or circular velocity of the 
galaxy: it is about 10\% for large spirals like the Milky Way, and rises 
to about 50\% for the smallest disk systems.

Our Galaxy has a thick disk. Star counts by \citet{gilr83} at high 
Galactic latitude showed two vertically exponential components: the 
thin disk and the
more extended thick disk.  Its scaleheight is about 1000 pc, compared 
to about 300 pc for the old thin disk, and its surface brightness is 
about 10\% of the surface brightness of the thin disk, but there is 
still some disagreement about these parameters

\subsection{Structure of Thick Disks}

From within our Galaxy, it is difficult to estimate reliably the 
scaleheight and scalelength of the thick disk. Values for the scaleheight 
between about 0.5 and 1.2 kpc have been reported.  A recent analysis of 
the SEGUE photometry gives a relatively short exponential scaleheight of 
$0.75 \pm 0.07$ kpc and an exponential scalelength of $4.1 \pm 0.4$ kpc 
\citep{dej10}.  For comparison, the scaleheight of the thin disk is about 
250 to 300 pc; the scalelength of the thin disk is poorly determined but 
is probably between 2 and 4 kpc. Their model gives the stellar density 
of the thick disk near the sun to be about 0.0050 \msun\ pc$^{-3}$, about 
7\%\ of the stellar density of the thin disk near the sun (about 0.07 
\msun\ pc$^{-3}$). 

\begin{figure}[t]
\begin{center}
\includegraphics[width=160mm]{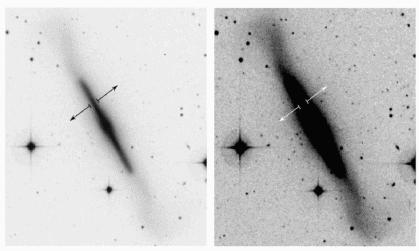}
        \caption{ The S0-galaxy NGC 4762, which has a very bright thick disk,
as was first described by \citet{tsi81}. The $z$-extent indicated by the arrows is where
the thin disk dominates. On the right the outer extent of the thick disk is slanted w.r.t.
the symmetry plane (producing an approximately diamond shape), 
indicative of a double exponential light distribution. These images
were produced with the use of the {\it Digital Sky Survey}.
}
\label{fig:thick_disk}
\end{center}
\end{figure}

{ Thick disks are not easy to see in galaxy images. Fig.~\ref{fig:thick_disk} 
shows that in NGC 4762 \citep[that has a bright thick disk ][]{tsi81}
the outer extent of the faint starlight has an approximate diamond shape, 
indicating the double exponential light distribution of a thick disk. In
fig.~\ref{fig:truncations}, we see the same thing in the outer outline in the bottom
picture for NGC 4565 at the deepest stretch.}
{ The luminosity distribution of}
the edge-on galaxy NGC 891, which is often regarded as
an analog for the Milky Way \citep{vdk84}, { shows after subtraction of the disk
a  light distribution that becomes progressively more flattened at fainter levels
\citep[see fig.~7 in ][]{vdks81b}; in \citet{vdk84} it is shown that this distribution
can be interpreted as a superposition of a thin and thick disk plus a small,
central bulge. Recent studies} 
show that the scaleheight of its thick 
disk is $1.44 \pm 0.03$ kpc and its radial scalelength is $4.8 \pm 0.1$ 
kpc, only slightly longer than that of the thin disk \citep{ib09}. The 
relationship between the scalelengths of the thin and thick disk is an 
important constraint on the various formation mechanisms of thick disks, 
as discussed below.

\begin{figure}[t]
\begin{center}
\includegraphics[width=130mm]{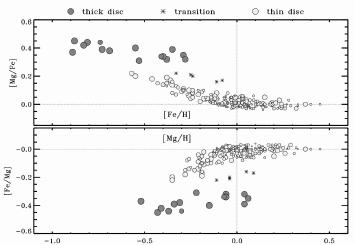}
        \caption{Mg-enrichment of the thick disk relative to the thin disk,
indicating that the thick disk is a chemically distinct population.
\citep[From ][]{fuh08}}
\label{fig:fuhrmann}
\end{center}
\end{figure}

\subsection{Kinematics and Chemical Properties}

Little information is available on the kinematics and chemical properties 
of thick disks in galaxies other than the Milky Way. The larger scaleheight 
of the Galactic thick disk means that its velocity dispersion is higher 
than for the thin disk (about $40$ \kms\ in the vertical direction near the 
sun, compared to about $20$ \kms\ for the thin disk \citep[\eg][]{qg00}. 
The stars of the thick disk are usually identified by their larger 
motions relative to the Local Standard of Rest, but kinematic 
selection is inevitably prone to contamination by the more abundant thin disk
stars. 
Recently it has become clear that the Galactic thick disk is a discrete 
component, kinematically and chemically distinct from the thin disk. It 
now appears that thick disk stars can be more reliably selected by their 
chemical properties. 

Near the Galactic plane, the rotational lag of the thick disk relative 
to the LSR is only about 30 \kms\ \citep{cb00, dam09}, but its rotational 
velocity appears to decrease with height above the plane.  The stars of 
the thick disk are old ($\gt10$ Gyr) and more metal-poor than the thin disk. 
The metallicity distribution of the thick disk has most of the stars with 
[Fe/H] between about $-0.5$ and $-1.0$, with a tail of metal-poor stars 
extending to about $-2.2$.  The thick disk stars are enhanced in 
$\alpha$-elements relative to thin
disk stars of the same [Fe/H] \citep[\eg\ fig.~\ref{fig:fuhrmann} 
from][]{fuh08}, indicating a more rapid history of chemical evolution.  
The thick disk does not show 
a significant vertical abundance gradient \citep{gwj95}. It appears to be 
chemically and kinematically distinct from the thin disk.

A decomposition of the kinematics and distribution of stars near the Galactic 
poles, by \citet{vel08}, nicely illustrates the three main kinematically 
discrete stellar components of the Galaxy: a thin disk with a scaleheight 
of 225 pc and a mean vertical velocity dispersion of about 18 \kms, a thick 
disk with a scaleheight of 1048 pc and mean velocity dispersion of 40 \kms, 
and a halo component with a velocity dispersion of about 65 \kms.

The old thick disk presents a kinematically recognizable relic of
the early Galaxy and is therefore a very significant component for
studying galaxy formation.  Because its stars spend most of their
time away from the Galactic plane, the thick disk is unlikely to
have suffered much secular heating since the time of its formation.
Its dynamical evolution was probably dominated by the changing
potential field of the Galaxy associated with the continuing growth
of the Galaxy since the time at which the thick disk was formed.

For the thick disks of other galaxies, broadband colors suggest that 
the thick disks are again old and metal-poor relative to their thin disks. 
\citet{yd08} studied the rotation of the thick disks of a small sample 
of extragalactic thick disks. In one galaxy { (FGC 227)\footnote{ FGC is
the Flat Galaxy Catalogue of \citet{kkp93}.}}, 
they found that the thick 
disk appeared be counter-rotating relative to its thin disk. If confirmed, 
this difficult observation would have important implications for the 
origin of thick disks.

\subsection{Relation of Thick Disk to the other Galactic Components}

At this time, there is no convincing evidence that the thick disk is in 
any way related to the halo or the bulge of its parent galaxy.
The almost ubiquitous nature of thick disks, 
independent of the bulge to disk ratio, argues against any causal 
relation with the bulge. In the Milky Way, some chemical similarities of 
bulge and thick disk stars are observed.  For example, the [$\alpha$/Fe] 
ratio of the thin disk stars near the sun are somewhat like the enhanced 
[$\alpha$/Fe] ratios seen in bulge stars \citep{mel08} but this is probably 
more a reflection of the rapid star formation history of both components, 
rather than any deeper relation.  The metal-poor tail of the metallicity 
distribution of the thick disk reaches down to metallicities usually 
associated with halo stars ([Fe/H]$\lt -1$), but the kinematics are 
different (the metal-poor thick disk stars are rotating more rapidly 
than the halo stars \citep[\eg][]{car10} and do not hint at any cosmogonic 
relation of thick disk and halo. \citet{gil95} found that in the Galaxy 
the cumulative angular momentum distribution function for stars in the
thick disk is rather similar to that of the thin disk, but distinctly 
different from that of the stellar halo and the bulge.

Thin disks and thick disks do appear to be causally linked: the survey of 
\citet{yd06} shows that all or almost all galaxies selected as having a 
thin disk also have a thick disk. Thick disk formation seems to be a 
normal event in the early formation of thin disks, but the details of how 
disks form are not yet understood. 

\subsection{Thick Disk Formation Scenarios}

Thick disks appear to be old and very common, puffed up relative to their 
parent thin disks, and (at least in the case of the Milky Way) { to have} suffered 
rapid chemical evolution.  Scenarios for their formation include:
\begin{itemize}
\item thick disks come from energetic early star burst events, maybe
associated with gas-rich mergers \citep{samger03, brook04} 
\item thick disks are the debris of accreted galaxies which were
dragged down by dynamical friction into the plane of the parent galaxy
and then disrupted \citep{abad03, wal96}. To provide the observed metallicity 
of the Galactic thick disk ([Fe/H] $\sim-0.7$), the accreted galaxies that 
built up the Galactic thick disk would have been more massive than the SMC 
and would have had to be chemically evolved at the time of their accretion.  
The possible discovery of a counter-rotating thick disk \citep{yd08} { (FGC 
227; see above)} would 
favor this mechanism.
\item the thick disk's energy comes from the disruption of massive 
clusters or star-forming aggregates \citep{krou02b}, possibly like the 
massive clumps seen in the high redshift clump cluster galaxies. Other 
authors have discussed the formation of thick disks through the merging 
of clumps and heating by clumps in clump cluster galaxies \citep[\eg][]{bour09}.
\item the thick disk may be associated with the effects of radial mixing 
of stars and gas in the evolving Galaxy \citep{schbin09a,schbin09b}.
\item the thick disk represent the remnant early thin disk, heated by
accretion events.  In this picture, the thin disk begins to form at a
redshift of 2 or 3, and is partly disrupted and puffed up during the 
active merger epoch. Subsequently the rest of the gas gradually settles
to form the present thin disk \citep[\eg][]{qg86, kcf87}.
\end{itemize}

\citet{sal09} have shown that the predicted distribution of orbital 
eccentricities for nearby thick disk stars is different for several of 
these formation scenarios.  As large samples of accurate orbital 
eccentricities become available for thick disk stars, they can be used 
to exclude some of the proposed formation mechanisms.


\section{FORMATION OF DISKS}

\begin{figure}[t]
\begin{center}
\includegraphics[width=130mm]{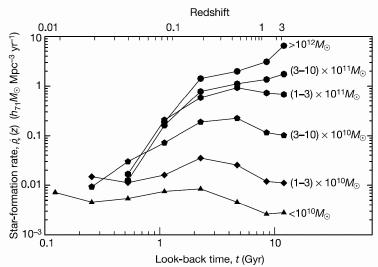}
        \caption{The history of star formation as a function of stellar mass
          of the galaxy. The curves have been offset by 0.5 in the log, except
          for the most massive one, where it has been an additional 1.0. The
          phenomenon of star formation occurring in the early universe mainly
          in large systems while later shifting to smaller systems is known as
          `downsizing'. 
\citep[From ][]{hea04}
}
\label{fig:heavens}
\end{center}
\end{figure}

\subsection{Disk Formation Scenarios}

Stellar disks are close to centrifugal equilibrium, suggesting
dissipation of baryons to a near-equilibrium structure before the
gas became dense enough to initiate the onset of the main epoch of
star formation.  This picture goes back at least to \citet{els62}
and \citet{sfs70} in the pre-dark-matter  era, and was followed by
many innovative landmark papers in the late 1970s and early 1980s
on the cooling of gas and dissipational formation of disks in the
potential of the pre-formed dark matter halo.  \citet{ro77} discussed the
cooling of gas and the thermalising of infalling gas. \citet{wr78}
described the two-stage theory of galaxy formation, in which the
dark matter clustered under the influence of gravity and the gas cooled into
the dark matter halos. \citet{fe80} considered the dissipational collapse
of a disk within a dark halo, conserving its detailed $M(h)$ distribution
as suggested by \citet{mes63}. Assuming that the specific angular
momenta of the dark matter and gas were similar, they showed that 
the gas would collapse by at least a factor of 10.  \citet{gun82} 
showed that in this hypothesis approximately exponential stellar disks 
arise naturally and \citet{vdk87} showed that these would then have 
radial truncations at about 4.5 scalelengths.  \citet{bfpr84}  discussed
the various possible kinds of dark matter (cold, warm, hot) and
their implications for the formation of galaxies by cooling.  Later
versions of this basic picture were discussed by \citet{dal97} and
\citet{mmw98}. { The so-called `rotation curve
conspiracy', that the properties of disk and halo as precisely those necessary
to produce essentially flat, featureless rotation curves and to provide the
Tully-Fisher law was heuristically explained in the context
discussed here by \citet{gunn87} and \citet{rg87}.}

 In summary, this paradigm involves the dark matter
halo forming gravitationally and relaxing to virial equilibrium;
then infalling gas is shock-heated to the halo virial temperature
and then cools radiatively from inside out, gradually building up
the disk and forming stars quiescently. 

The subsequent star formation history of disks is very different
from galaxy to galaxy.  In the S0 galaxies, the star formation has
already come to a halt through gas exhaustion or gas stripping.  In
the Galactic thin disk, star formation began about 10 Gyr ago
\citep{edv93,khh99} and a wide range of stellar age is seen,
indicating that disk formation was a extended process starting about
10 Gyr ago  and continuing to the present time.  The star formation
history of the disk is not very well constrained, but is consistent
with a roughly constant star formation rate \citep[\eg][]{rp00}
with fluctuations of about a factor 2 on Gyr timescales. In the
lower luminosity systems, the main epoch of star formation did not
begin right away, and much of the current star formation in the
universe is now taking place in these smaller disk galaxies.  This
is the phenomenon of downsizing: star formation in the early universe
occurred mainly in the larger systems and has has gradually proceeded
to progressively smaller galaxies (see fig.~\ref{fig:heavens}).

This is nicely shown by \citet{noe07a,noe07b}. They analysed the
star formation rate as a function of stellar mass $M_*$ and redshift.
In galaxies which are forming stars, the star formation rate depends
on the stellar mass of the galaxies. The observed range of star
formation rate remains approximately constant with redshift, but
the sequence moves to higher star formation rate with increasing
redshift.  The dominant mode of evolution since $z=2$ is a gradual
decline of the average star formation rate. At a given mass, the
star formation rate at $z=2$ was larger by a factor $\sim4$ and 
$\sim30$ than that in star forming galaxies at $z=1$ and $0$ 
\citep{da07}.  The star formation drives the growth of disks.  
\citet{tru06} studied the evolution of the luminosity-size and stellar 
mass-size relation in galaxies since $z=3$ and found that, at a 
given luminosity, galaxies were $3$ times smaller at $z=2.5$ than 
now; at a given $M_*$ they are 2 times smaller.  In the Local Group, 
\citet{wi09} directly measured the evolution of the scale length
of M33, which has increased from $1$ kpc $10$ Gyr ago to $1.8$ kpc
at more recent times.

The present high specific star formation rate of many less massive
galaxies reflects the late onset of their dominant star formation
episode, after which the star formation rate gradually declines.
The less massive galaxies appear to have a longer e-folding time
and a later onset of their star formation history.

The color distributions of galaxies reflect this mass dependence
of their star formation history.  At low redshifts, the color
distribution is bimodal,  showing the blue cloud of star forming
galaxies at lower stellar masses and the overlapping red sequence
of more luminous systems in which star formation has now stopped
or is at a very low rate \citep{kau03a, bla03}.  The restframe color
distribution is now known to be bimodal at all redshifts $z\lt2.5$
\citep[\eg][]{bra09}.  The red sequence persists to beyond $z=2$,
indicating that many galaxies had completed their star-forming lives
by this time.  Galaxies in the ``green valley" between the red
sequence and the blue cloud may be in transition between the two
sequences,  as blue cloud galaxies whose star formation is running
down, but also as red sequence systems in which the star formation
has been revitalised \citep[\eg][]{beau10}.  However it seems more
likely now that most of these green valley galaxies are reddened 
star forming systems.

The bimodality manifests itself in several ways.  Blue galaxies,
as defined by their color and their star formation rates,  dominate
the stellar mass function below a transition mass of about $3 \times
10^{10}$ \msun.  It persists to redshifts of at least $2.5$. The
bulge-to-disk ratio shows a related transition from disk-dominated
blue galaxies below the transition mass to spheroid-dominated red
sequence galaxies above the transition mass \citep[\eg][]{kau03b}.
The surface brightness $\mu$ - stellar mass $M_*$ relation changes
from $\mu \propto M_*^{0.6}$ at lower masses to $\mu \sim$ constant
at higher masses.  The mean mass - metallicity relation
also shows a break at the transition mass \citep[\eg][]{tre04}.
\citet{dyb04} observed a  break in the morphology of dustlanes of
edge-on disk galaxies at a similar transition mass, changing from
diffuse for lower-mass disks to sharply-defined for higher-mass
disks.  This could be related to infalling cold gas streams enhancing
star formation and turbulence in the less massive galaxies (see \S\S~
4.4 and 8.3).

The role of mergers in the evolution of disk galaxies remains
uncertain.  The thin disks of disk galaxies are relatively fragile
and are easily puffed up by even minor mergers \citep{qg86, to92}.
CDM simulations show a high level of merger activity on all scales.
These mergers tend to excite star formation, puff up the disks and
build up spheroidal components in the simulations. This makes 
it difficult for CDM simulations to generate large spirals
like the Milky Way, with relatively small spheroidal components.
While there has been discussion by simulators about the unusually
quiescent merger history of the Milky Way, such systems are not
unusual. \citet{kdbc10} have shown that at least 11 of 19 nearby
disk galaxies with circular velocities $\gt150$ \kms\ show no evidence
for a classical bulge. They argue that pure-disk galaxies are far
from rare.  \citet{lo08} studied the evolution of the merger rate
since redshift $z=1.2$.  They find that the fraction of galaxies
involved in mergers is about 10\%, and conclude that the decrease
in star formation rate density since $z=1$ is a result of the
declining star formation rate in disk galaxies rather than a decrease
in major mergers. From their GOODS galaxy sample, \citet{bu09} show
that mergers contribute little to galaxy growth since $z=1.2$ for
galaxies with $M_*\lt3\times10^{10}$ \msun.  For the more massive galaxies,
which are mostly spheroidal systems, mergers are more important;
they estimate that about 30\% have undergone (mostly dry or gas-free)
mergers. The Milky Way will be an important part of assessing the
significance of mergers in building up large spiral galaxies.  The
unique possibility to make very detailed chemical studies of stars
in the Milky Way provides an independent opportunity to evaluate
the merger history of our large disk galaxy via chemical tagging
techniques \citep{fbh02}.

\subsection{Disks at High Redshift}

One of the most fundamental observations that Hubble Space Telescope
made possible is the imaging of galaxies when they were
very young. Studies of the Hubble Deep Fields \citep{hdfn96,
hdfs00, hudf06} showed observationally that few galaxies are present
at redshifts $\gt4$ which resemble present-day spirals or ellipticals.
Simulations indicate that disks should be present at redshifts
around 2 \citep[\eg][]{sl03}: is this consistent with observations?  
If disks are present at these redshifts, what are they like:  how
do their baryonic mass distributions compare with those of disk
galaxies at low redshifts?

\citet{lrf03} observed the Hubble Deep Field-South in the near-infrared
(from the ground) and found six galaxies at redshifts $z=1.4$ to
$3.0$ that have disklike morphologies. The galaxies are regular and
large in the near-infrared (corresponding to rest-frame optical),
with face-on effective radii of $5.0 - 7.5$ kpc, which is comparable
to the Milky Way. The surface brightness proﬁles are consistent
with an exponential law over 2 to 3 effective radii.  The HST
morphologies (rest-frame UV) are irregular and show large complex
aggregates of star-forming regions ($\sim15$ kpc across), symmetrically
distributed around the centers.

\citet{ge06} described the rapid formation of a luminous star-forming
disk galaxy at a redshift $z=2.38$.  Their IFU observations
indicate that a large protodisk is channelling gas towards a growing
central bulge. The high surface density of gas and the high rate
of star formation show a system in rapid assembly, with no obvious
evidence for a major merger.  Integral-field spectroscopy by
\citet{fs09} of several UV-selected galaxies with stellar masses
$\sim3\times10^{10}$ \msun, star formation rates of $\sim70$ \msun yr$^{-1}$
and redshifts between 1.3 and 2.6 provides rotation curves and
indicators of dynamical evolution.  The morphology is typically
clumpy. About 1/3 of the galaxies are turbulent and rotation-dominated,
another 1/3 are compact and dominated by velocity dispersion, while
the rest are interacting or merging systems. The rotation-dominated
fraction is higher for higher masses.

\citet{ee09a, ee09b} investigated the relation to modern spirals
of clumpy high redshift galaxies in the GOODS, GEMS and Hubble UDF.
The clump properties indicate the gradual dispersal of clumps to
form disks and bulges, with little indication of merger activity.
The morphological similarity of these systems to modern dwarf
irregulars suggests that the clumpy morphology comes from gravitational
instability in the turbulent gas. They note that about 50\% of these
clump cluster galaxies have massive red clumps which could be
interpreted as young bulges.  We have already discussed the clump
cluster galaxies in the context of the formation of the thick disk
component (\S~6.5)

\citet{kvdfi09} studied 19 massive galaxies at $z\sim2.3$: nine
of them are compact quiescent systems and 10 are emission line
systems (6 star-forming galaxies and 4 AGNs).  The star forming
galaxies again have clumpy morphologies.  In the rest-frame $(U-B)
- M_*$ plane, the galaxies appear bimodal: the large star-forming
galaxies lie in a blue cloud and the compact quiescent galaxies in
a red sequence.  A bimodal distribution similar to that at lower
redshifts is already in place at redshifts $\gt2$.

In summary, it appears that the formation of disk galaxies is already
well advanced at redshifts $\gt2$, but the systems have mostly not yet
settled to a quiescent disk in rotational equilibrium.
The clumpy structure of the massive star forming systems is likely
to be an important factor in the subsequent evolution of these
systems and in the formation of their thick disks and bulges.  The
role of mergers in building up these disk galaxies may not be as
important as it appears to be from CDM simulations.

\subsection{Baryon Acquisition by Disk Galaxies}

In the scenarios for disk formation that emerged soon after the discovery
of dark matter in disk galaxies,  the gas was shock-thermalised to
the virial temperature and then gradually cooled to form the disk.
Simulations \citep[\eg][]{sl99} suggest that this hot halo is further
populated by gas blown out from the disk via feedback into the hot
halo. This feedback provides a way to reduce the problem of angular 
momentum loss which led to unrealistically low angular momenta for 
disks seen in earlier simulations.  These simulations are successful 
in reproducing the peak in the star formation rate seen within individual 
galaxies at $z\sim2$.

More recent simulations point to the likelihood of cold gas accretion
into disk galaxies.  SPH simulations by \citet{ke05} showed that
typically about half of the gas shock-heats to the virial temperature
of the potential well ($\sim10^6$K in a Milky-Way-like galaxy),
while the other half radiates its gravitational energy at $T\lt10^5$K.
A cold mode of infall is seen for stellar masses $\lt2\times10^{10}$
\msun.  This cold gas often falls in via the cosmic filaments,
allowing galaxies to draw their gas from a large volume. \citet{ke09}
found that most of the baryonic mass is acquired through the
filamentary cold accretion of gas that was never shock-heated to
its virial temperature.   This cold accretion is the main driver
of the cosmic star formation history.  

Hot halos are seen only for dark halo masses $\gt2-3\times10^{11}$
\msun.   \citet{dekb06} ascribed the bimodality of galaxy properties
to the nature of the gas acquisition.  Galaxies with stellar masses
$\lt3\times10^{10}$ \msun\ are mostly ungrouped star-forming disk
systems, while the more massive galaxies are mostly grouped old red
spheroids.  They argue that the bimodality is driven by the thermal
properties of the inflowing gas.  In halos with masses $\lt10^{12}$
\msun, the disks are built by cold streams, giving efficient early
star formation regulated by supernova feedback.  In the more massive
halos, the infalling gas is shock-heated and is further vulnerable
to AGN feedback, shutting off the gas supply and leading to red and
dead spheroids at redshift $z\sim1$. Simulations by \citet{dek09}
showed that the evolution of massive disk systems is governed by
interplay between smooth and clumpy cold streams, disk instability
and bulge formation. The streams maintain an unstable gas-rich disk,
generating giant clumps which can migrate into the bulge in a few
dynamical times.  The streams prolong this clumpy phase for several
Gyr.  The clumps form stars in dense subclumps and each clump
converts to stars in $\sim0.5$ Gyr.  The star forming disk is
extended because the incoming streams keep the outer disk dense and
unstable, and also because of angular momentum transport by secular
processes within the disk \citep[\eg][]{kk04}. Observationally, the 
large chemical tagging surveys which will soon begin (HERMES, APOGEE) 
will be able to evaluate the role of giant clumps in the formation of 
the thin and thick disks of the Milky Way \citep[\eg][]{jbh10}. The 
debris of the dispersed giant clumps should be very apparent from the 
chemical tagging analysis.

The Milky Way is surrounded by a system of infalling high velocity \HI\ 
clouds (HVCs) whose nature is not yet fully understood.  The associated 
infall rate is estimated at about $0.2$ \msun\ yr$^{-1}$ 
\citep[\eg][]{pps08}, an order of magnitude smaller than the current 
star formation rate of the Milky Way. \citet{mb04} proposed that the 
cooling of the Galactic hot corona is thermally unstable and generates 
pressure-confined HVCs with masses $\sim5\times10^6$ \msun, which 
contribute to fueling the continued star formation of the disk.  
\citet{bnf09} argued however that thermal instability of the hot halo 
is unlikely to be the source of the Galactic HVC system.

Some disk galaxies \citep[\eg\ NGC 891:][]{ofs07} show thick \HI\ layers 
surrounding their galactic disks, which is lagging in rotation relative to 
the gas in the disk. This gas is accompanied by ionized gas (observed in
\Ha) that shares the lag
in rotation velocity \citep{hrbb07,pketal07a} and by dust \citep{pketal07b}.
In more face-on systems this \HI\ appears associated with regions of star
formation \citep[\eg][]{ksvdh91}, indicating that at least a part of it
may have originated in the disk. \citet{fb08} and \citet{mb10} suggest 
that these layers are associated with gas that has been swept up from the 
hot corona by galactic fountain clouds ejected from the disk by star 
formation.  In this way, the star formation in the disk is self-fueling, 
through the gas brought down from the hot corona.

Deep \HI\ images of other disk systems show a very extended rotating \HI\ 
distribution. M83 (see \S~\ref{sec:warps} and 
fig.~\ref{fig:M83bigiel}, and  www.atnf.csiro.au/people/bkoribal/m83/m83.html) 
is an example, with \HI\ extending far beyond the optical extent of the 
system. Its outermost \HI\ shows spectacular \HI\ arms and filaments, some 
of which are forming stars at a low level \citep{bls10}.  It seems 
unlikely that this structure should be interpreted as spiral structure in 
an extended \HI\ disk, because the density of the \HI\ is so low. It may 
represent a slow filamentary infall of \HI\ into the disk.  The observed 
star formation in this outer \HI\ disk indicates that the outer disk is 
still in the process of construction.  NGC 6946 is another more orderly 
example of such a very extended \HI\ disk \citep{bof08}.


\section{S0 GALAXIES}
\label{sect:S0}

\citet{hubble36} introduced S0 galaxies as the transition type between
elliptical galaxies and spirals. \citet{sb51} described them as spiral
galaxies without arms and suggested they were spirals stripped of their
dust, gas and arms as a result of collisions in clusters of galaxies. 
\citet{sfs70} argued that the morphological type of a galaxy is
defined at the time of the formation of the old disk stars, and S0
galaxies are those in which at the time of the completion of the
formation of the disk there was little gas left for star formation.
Subsequently, the observation of a high proportion of S0 galaxies in
clusters \citep{oem74}, the evolution of blue galaxies populations in
clusters with redshift \citep{bo78} and the increasing ratio of S0s
compared to spiral in regions of higher galaxy density \citep{dress80}
led to the general acceptance that S0 galaxies in clusters are stripped
spirals. \citet{ltc80} reconsidered the issue. They remarked that the
apparent rate of consumption of gas by star formation
leads to the paradoxical situation that
spirals will exhaust their gas supplies on a timescale `considerably less than
the Hubble time'. The solution they suggested to this was that spirals
constantly replenish their gas content from a reservoir in a gaseous envelop
remaining from their formation and therefore can sustain star formation over a
much longer timescale, while
S0 galaxies would have lost these envelopes early on and therefore do run
out of gas on a relatively short timescale. 

One way to address the issue of whether S0s are stripped spirals 
further is to investigate the
structure, and in particular the kinematics of disks in S0s in order to
investigate whether or not there are differences in the stellar
dynamics. We will not fully review the work on S0 galaxies, but 
concentrate on this aspect and refer for comprehensive reviews of S0 
galaxies to those presented by \citet{qmb00} and \citet{fva04} {
and also point out that the importance of ram-pressure stripping in 
environments like Virgo has convincingly been demonstrated by \citet{cvgkcv09}.}

Although S0s have a relatively bright surface brightness, measurements of 
their kinematics remained difficult and this prevented for a long time 
a detailed understanding of their dynamics. 
It is illustrative in this context to
consider the historical development of the subject. One of the earliest
identified and most easily accessible S0s is NGC 3115. 
\citet{oort40} { already as early as during the thirties(!)} 
considered its dynamics; his motivation was to study issues of
stability as he was interested in the origin and maintenance of spiral
structure (the first part of the paper concerns the origin of the
deviation of the vertex as caused by spiral arms). Although his photometry
shows evidence for a disk component (`concentration of light near the
major axis') he does not treat it as a separate component. The
velocity data (only a few points of the rotation curve
by Humason as reported in the annual report of the Mount
Wilson Observatory) was insufficient for a significant treatment.
Oort concludes that, if these data are correct, 
the distribution of mass does not correspond
to that of the light and quotes mass-to-light ratios of order 250. It
took two decades before some advancement took place. \citet{mink60}
reported in a review at meeting on {\it ``Les Recherches Galactiques et 
Extragalactiques et la Photografie \'Electronique''} in Paris in 1959 a
new rotation curve, which showed an initial rise, then a secondary
minimum followed be another strong rise. In the discussion after 
Minkowski's paper, Oort reported that he was able to reproduce this 
behavior on the assumption of a
proportionality of mass to light and a large velocity dispersion, which was
reported also by Minkowski.\footnote{Oort was, however, not satisfied
with his solution and never published it. He did illustrate it during
his lectures on `Stellar Dynamics' in Leiden, 
as the notes of one of us --PCK is a student of Oort-- show, 
and mentioned there that he felt that the
rotation curve might very well be wrong and needed confirmation.}

Oort urged Maarten Schmidt to remeasure the rotation of NGC 3115 and
together they took spectra in 1968 on the 200-inch Hale Telescope. It
took until 1974 before \citet{wil75} reduced the data. Oort never used
this for a detailed dynamical study, his interests having turned to
problems of galactic nuclei and cosmology \citep{oort77,oort81,oort83}.
In the mean time, measurements of the light distribution improved
\citep{mp68, ssjmtt77, sswr78}, but although color information seemed to
support the existence of color variations and the rotation curve
allowed an estimate of the mass of the disk as a fraction of the total
($\simlt 0.4$), no comprehensive dynamical model was possible without better
spectroscopy. The detailed surface photometry study of \citet{tsi79} constituted
the first attempt to separate photometric components. { The most accurate
measurements of the stellar kinematics, confirming the flat shape of the 
stellar rotation and providing also evidence for a supermassive black hole 
in the center, has been presented by \citet{kr92}.}

The kinematic data necessary for a detailed dynamical modelling
started to become available only in the eighties, first in the central
regions \citep{rpf80} and then over a more extended region \citep{is82}.
In the latter study, the kinematics of the disk were estimated from a
decomposition of the contributions to the observed velocities and
velocity dispersions. The main result of the study was the important
deduction, that later was proved to be more general, that in bulges of disk
galaxies rotation plays a bigger role in supporting its shape and
density distribution than in ellipticals.

\citet{kor84a,kor84b} was the first to measure the velocity dispersion
in the disk of an S0 galaxy. His main aim was to estimate the \citet{too64}
$Q$-parameter for local stability. Both in NGC 1553 and in the barred S0
NGC 936 he was able to estimate $Q$ and found it to be well above unity 
(more like 2 or 3). He therefore concluded 
that S0 galaxies may differ from spirals in that
their stellar disks are too hot (in addition to suffering from lack of
gas, which would lower the overall $Q$) to form small-scale structure.

\begin{figure}[t]
\begin{center}
\includegraphics[height=95mm]{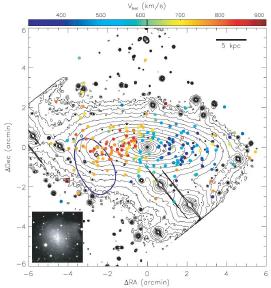}
\includegraphics[height=95mm]{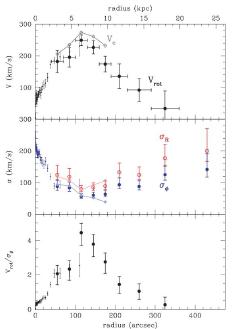}
\caption{Kinematics of the S0 galaxy NGC 1023 from velocity
measurements of planetary nebulae. On the left the distribution over the
face of the system and in color the pattern of rotation. On the right
the results of an analysis of the measurement, with from top to bottom
the rotation velocities, the dispersions (red radial and blue
tangential), and the ratio between rotation and velocity dispersion.
the rotation velocities. \citep[From][]{nmc08}}
\label{fig:N1023}
\end{center}
\end{figure}

For NGC 3115 the observations of photometry and kinematics
have become more sophisticated in recent years
\citep[\eg][]{cvh88,sbtj89,cchv93,mic07}, but the issue has become
much more complicated. A simple answer to the question on whether an S0
originated as a spiral that was swept of its gas or a system that
formed its disk with little gas that was not replenished, has not
emerged. GMOS multi-object long-slit
spectroscopy by \citet{nsk06} shows that there is a
difference between the [$\alpha $/Fe] in the bulge ($\sim$0.3) and the
disk (close to solar), while the average age of stars in the disk (5--8 
Gyr) is significantly less than in the bulge (10--12 Gyr). The fact that
the disk is bluer than the bulge would then primarily be an age
difference. Star formation in the disk of this archetypal S0 has
proceeded at least for some time after the formation of the bulge.

We note that the advent of new techniques has greatly improved the
observational possibilities. SAURON can measure kinematic data using
integral field spectroscopy, such as in the study of the S0 galaxy NGC 7332
by \citet{fbp04}. In this system, the stellar populations in the
disk (but also in the bulge) are again relatively young, so that star
formation must have proceeded in the disk until fairly recently. In
addition there is evidence for a strong influence by a bar. 

Another recent development is the use of planetary nebulae (PNe) to measure the
kinematics. This has the advantage that velocities and dispersions can be 
measured in faint outer parts of galaxies \citep[\eg][]{coc09}. 
Comparing photometry, absorption-line and PNe kinematics shows that there 
is good agreement between the PNe number density distribution and the stellar 
surface brightness and also a good agreement between PNe and absorption line
kinematics. An application of this technique to an S0 galaxy is the study of 
NGC 1023 by \citet{nmc08}; fig.~\ref{fig:N1023}
shows that PNe can be measured with a very good distribution over the
face of the system. They show a clear pattern of rotation. The left-hand 
panel shows the accuracy with which velocities and velocity dispersions can be
measured. With these data, it is possible to test whether
an S0 could result from a minor merger such that its kinematics become
dominated by random motions. The inner parts are fitted quite well with
a disk that is rotationally supported, but the outer parts would suggest
a minor merger event. Information like this on more systems is required
to answer the basic questions. 

Finally, an interesting approach is that of \citet{asbm06} and \citet{bbasmb07},
who use globular clusters to measuring the fading of S0 galaxies. They
do this by comparing the specific frequency of globular clusters (their
number per unit luminosity, $S_{\rm N}$) between S0s and normal spirals. This
innovative and powerful method has led to the picture in which the disk
has faded by a factor of three or so. S0s with younger ages have values
for $S_{\rm N}$ that are more like those of
spirals. This is consistent with the view
that S0s are formed as a result of removal of gas from normal spirals. 

Although the kinematics of S0s would suggest that their
disks have been present on a long timescale, the evidence for
relatively recent star formation indicates that their history is more
complicated than previously thought.

\vspace{0.5cm}

{\small
{\bf Acknowledgments.} 
A major part of this review was written during work visits by PCK to Mount
Stromlo Observatory. PCK thanks the Research School for Astronomy and 
Astrophysics of the Australian National University and its directors for
hospitality and facilities over the years in support of such visits.
He is grateful to the Governing Board of the University of Groningen 
for the appointment as distinguished Jacobus C. Kapteyn Professor of
Astronomy, and the Faculty of Mathematics and Natural Sciences for an
accompanying annual research grant that supported among other things 
his travels to scientific meetings and work visits. Additional financial 
support was provided from an annual research grant as member
of the Area Board for Exact Sciences of the Netherlands Organisation for 
Scientific Research (NWO). KCF is very grateful to many colleagues for
discussions of galactic disks, and particularly to G\'erard de Vaucouleurs 
and Allan Sandage. PCK acknowledges many stimulating
collaborations, of which he wishes to acknowledge here especially that with
Leonard Searle.}


\vspace{0.5cm}

{\small

\addcontentsline{toc}{section}{References}
\bibliographystyle{apalike}

}

\end{document}